\documentclass[prb,aps,english,twocolumn,notitlepage,superscriptaddress]{revtex4-2}
\usepackage{graphicx}
\usepackage{dcolumn}
\usepackage{rotating}
\usepackage{lipsum}
\usepackage{xcolor}
\usepackage{bm}
\usepackage[T1]{fontenc}  
\usepackage[latin1]{inputenc}
\usepackage{multirow,amsmath,amssymb}
 \usepackage{babel} 
 \usepackage{txfonts}
 \usepackage{epsfig,epsf,psfrag} 
\usepackage{graphicx} 
\usepackage{subfigure}
\usepackage{pslatex}
\usepackage{epic,eepic} 
\usepackage{color,pstcol}
\usepackage{pstricks} 
\usepackage{fancyhdr} 
\usepackage{tabularx}
\usepackage{physics}
\usepackage{url}
\usepackage{listings}
\usepackage{color}
\usepackage{caption}
\begin{document}

\title{Finite temperature quantum noise correlations as a probe for topological helical edge modes}
\author{Sachiraj Mishra}%
\email{sachiraj29mishra@gmail.com}

\author{Colin Benjamin}%
\email{colin.nano@gmail.com}
\affiliation{School of Physical Sciences, National Institute of Science Education and Research, HBNI, Jatni-752050, India}
\affiliation{Homi Bhabha National Institute, Training School Complex, AnushaktiNagar, Mumbai, 400094, India }

\begin{abstract}
 The distinction between chiral, trivial helical, and topological helical edge modes can be effectively made using quantum noise measurements at finite temperatures. Quantum noise measurements consist of mainly two components. The first is thermal noise, whose provenance is thermal fluctuations, and the second is shot noise, whose origin is the quantum nature of charge particles. Studying these edge modes at finite temperatures is important as it more accurately reflects the conditions in real-world experiments. Additionally, we have verified that our results for finite temperature quantum noise correlations are valid at finite frequencies too.
\end{abstract}

\maketitle

\section{\label{sec:level1}Introduction}
The discovery of the integer quantum Hall effect \cite{PhysRevLett.45.494} and subsequent development of Landauer-Buttiker formula \cite{PhysRevB.38.9375, datta_1995} were critical milestones in our understanding of electron transport in 2DEGs. The concept of chiral edge modes, which are responsible for quantized Hall
resistance has since been extended to other systems, including the quantum spin Hall (QSH) effect in materials such as HgTe/CdTe heterostructures \cite{zhang}, wherein the edge
modes are helical. The discovery of the QSH effect \cite{Konig, Roth_2009} was important because it demonstrated that spin-orbit scattering could be used to control the direction of spin
polarization in edge modes, opening up new avenues for
spintronics and quantum computing.

Chiral and helical edge modes are topologically protected and are not affected by the presence of impurities along the edge channels \cite{PhysRevB.38.9375, datta_1995}. Helical and chiral edge states can be distinguished by different methods, such as a conductance test using the Landauer-Buttiker approach \cite{Shen_2017}. However, a recent experiment showed that the helical edge states may not always be topologically protected \cite{Nichele_2016}; instead, trivial helical edge modes can arise. In this case, it is essential to distinguish between the different kinds of helical edge states, a topological one and another non-topological or trivial. Of course, we can do a conductance measurement using the Landauer-Buttiker approach, but this cannot effectively show any distinction between trivial and topological helical edge modes \cite{Mani_2017}. In this case, it is wiser to rely upon the noise correlations \cite{BLANTER20001}, which can also be performed experimentally.

At zero temperature in a multiterminal conductor and in the presence of inelastic scattering, transport via chiral and helical edge modes can be distinguished by the nonlocal charge shot noise ``HBT" correlations \cite{Mani_2017}. For the transport via chiral edge modes, ``HBT" correlations are always negative, while it is positive for transport via topological helical edge states \cite{Mani_2017}. The most significant advantage of this is that it works as a probe for helicity as it turns negative for trivial helical edge modes due to in the presence of inelastic scattering \cite{Mani_2017}. Experiments involving nonlocal shot noise correlations can be instrumental in distinguishing different modes of quantum transport. However, this work \cite{Mani_2017} has certain limitations. Firstly, this was at zero temperature, where only shot noise matters. Secondly, this does not distinguish between trivial and topological helical edge modes for low spin-flip probabilities, which is a disadvantage for experiments in clean samples where spin-flip scattering is low. Thirdly, this setup does not provide any information regarding the distinction via shot noise autocorrelation. In this work, we remove these limitations by showing the differentiation between the chiral and topological helical edge modes and between the helical, trivial, and topological edge modes for any arbitrary value of spin-flip scattering by studying both cross and autocorrelation of the quantum noise at finite temperatures.

{Many other groups have also explored this topic. A recent experimental study \cite{10.1063/1.5111626} presented noise measurements  on InAs/Ga(In)Sb Corbino structures at different temperatures and current ranges, which deals with the distinction of trivial and topological helical edge modes. The same experimental technique can be extended to QH and QSH Hall bars, which is our proposal in this paper. In our work, we have shown the distinction between chiral, topological helical as well as trivial helical by both thermal noise as well as shot noise. In Ref. \cite{PhysRevLett.115.036803}, tuning between trivial and topological phases is done through electrical gating and it has shown the
discrimination between trivial and topological phases via Hall resistance and longitudinal resistance, but with a large magnetic and electric field. In a recent study \cite{PhysRevB.104.045144}, it was seen that a technique called ``Resistively detected Nuclear Magnetic Resonance (RDNMR)" can be used as a probe for helicity in QSH samples. Further in a recent theoretical work \cite{PhysRevB.104.195307}, it is shown that applying an external electric field and inter-edge coupling can lead to oscillations in energy spectra verifying the presence of topologically protected helical edge modes. Further, one can perform the noise experiments at very low temperatures, such as $50nK$ and $10mK$, as reported in QSH experiments in~\cite{Nichele_2016, PhysRevLett.115.036803}.}

 The total noise in mesoscopic conductors at finite temperature is called ``quantum noise". At finite temperatures, {both thermal and quantum fluctuations} are significant. {Thermal fluctuations are responsible for thermal noise, whereas
quantum fluctuations result in shot noise} \cite{BLANTER20001}. In this work, we have used a familiar setup used in the context of chiral edge modes, see Ref. \cite{PhysRevB.46.12485}. We see distinct behaviour for both chiral and helical edge modes as well as topological and trivial helical edge modes. The {thermal noise-like} contribution cannot distinguish between chiral and topological helical, but distinguishes trivial helical from topological helical via its magnitude, while the {shot noise-like contribution, distinguishes between the aforesaid edge modes via both its magnitude and nature}. The most striking result was that we could observe the distinct behavior of trivial helical edge modes as opposed to topological helical edge modes by measuring both the quantum noise autocorrelation and cross-correlations.

This paper is organized as follows. In the next section, we explore the behaviour of quantum noise in the presence
of chiral edge states by calculating cross-correlations between different terminals and autocorrelation in a four-terminal Hall bar. The terminals are connected to reservoirs at differing chemical potentials. Section \ref{section III} extends our study to helical edge mode transport and
calculates the quantum noise correlations both for topological as well as trivial helical edge modes. We also compare the results obtained for chiral and topological helical edge modes and find that they are distinct. In section~\ref{section III.B}, we study transport via trivial helical edge modes and
find distinct behaviour for the quantum noise
correlations as compared to topological helical edge modes and the results are further analyzed via tables in  section~\ref{section IV}: Analysis. In the same section, we have discussed the validity of our results at finite frequencies. {In Section \ref{section V}, we provide the experimental realization and conclusion to this work by explaining how to distinguish the thermal and shot noise experimentally using a two-terminal quantum Hall sample with chiral edge modes and we further discuss the possible analysis via $\Delta_T$ noise.} The paper ends with an exhaustive Appendix which gives a historic context of quantum noise calculation related with edge modes, then we derive the quantum noise and the individual thermal and shot noise contributions for chiral, helical and trivial helical setups. 

\section{Quantum noise in quantum Hall setup} \label{Section. II}
\subsection{Theory} \label{section II A}
The current-current correlation between metallic contacts $\alpha$ and $\beta$ is given by \cite{BLANTER20001},
\begin{equation} \label{eq:1}
S_{\alpha \beta}(t-t') = \frac{1}{2}\langle \Delta \hat{I}_{\alpha}(t)\Delta\hat{I}_{\beta}(t')+\Delta\hat{I}_{\beta}(t')\Delta\hat{I}_{\alpha}(t)\rangle,
\end{equation}
where $\Delta \hat{I}_{\alpha}(t)$ is the fluctuation in current from the average value in contact $\alpha$ and is given by $\Delta \hat{I}_{\alpha}(t) = \hat{I}_{\alpha} - \langle \hat{I}_{\alpha}\rangle$.
For a multiterminal system, the current in contact $\alpha$ is given by \cite{BLANTER20001},
\begin{equation} \label{eq:2}
\begin{split}
\hat{I}_{\alpha}(t) = \frac{e}{h} \sum_{\beta \gamma}\sum_{mn}\int dE dE' e^{i(E-E')t/\hbar}\\\times\hat{a}^{\dagger}_{\beta m}(E)A_{\beta \gamma}^{mn}(\alpha;E,E')\hat{a}_{\gamma n}(E'),
\end{split}
\end{equation}
where summation over $\beta, \gamma$ runs over the contacts except $\alpha$, summation over $m, n$ runs over number of edge channels. $\hat{a}_{\beta m}^{\dagger}(E)$ is the creation operator, which creates a particle (Boson or Fermion) with energy $E$ in the terminal $\beta$ in the edge channel $m$. Similarly, $\hat{a}_{\gamma n}(E')$ is the annihilation operator which annihilates a particle (Boson or Fermion) with energy $E'$ in terminal $\gamma$ in the edge channel $n$. Using these two operators, we can find the occupation number of the incident charge carriers in the terminal $\beta$ in the edge channel $m$, given by,
\begin{equation} \label{eq:3}
\hat{n}_{\beta m}(E) = \hat{a}_{\beta m}^{\dagger}(E)\hat{a}_{\beta m}(E),
\end{equation}
In Eq. (\ref{eq:1}), $A_{\beta \gamma}^{mn}(\alpha;E,E')$ is given by \cite{BLANTER20001},
\begin{equation} \label{eq:4}
A_{\beta \gamma}^{mn}(\alpha;E,E') = \delta_{mn}\delta_{\alpha \beta} \delta_{\alpha \gamma}-\sum_{k} s_{\alpha \beta;m k}^{\dagger}(E) s_{\alpha \gamma;k n}(E').
\end{equation}
where $s_{\alpha \beta; mk}(E)$ is the amplitude for an electron to scatter from terminal $\beta$ in the edge channel $k$ to terminal $\alpha$ in the edge channel $m$ with energy $E$.
Similarly, average current in terminal $\alpha$ is given by,
\begin{equation} \label{eq:5}
\langle I_{\alpha} \rangle = \frac{2e^2}{h}\sum_{\beta}V_{\beta}\int dE \left(-\frac{\partial f}{\partial E}\right) \bigg[N_{\alpha}\delta_{\alpha \beta}-Tr(s_{\alpha \beta}^{\dagger}s_{\alpha \beta})\bigg].
\end{equation}
where $V_{\beta}$ is the applied voltage to contact $\beta$ with $\beta \ne \alpha$ {and $N_{\alpha}$ is the number of edge modes in terminal $\alpha$. $s_{\alpha \beta}$ is the amplitude for an electron to scatter from terminal $\beta$ to terminal $\alpha$}. The extra factor of 2 in Eq. (\ref{eq:5}) has been taken due to the spin degeneracy. We can calculate the current fluctuation using, Eq. (\ref{eq:2}) and, Eq. (\ref{eq:5}) and derive the quantum noise.
The Fourier transform of Eq. (\ref{eq:1}) is called quantum noise, which is given by,
\begin{equation} \label{eq:6}
\begin{split}
2\pi \delta(\omega+\omega') S_{\alpha \beta}^q(\omega) = \frac{1}{2}\langle \Delta \hat{I}_{\alpha}(\omega)\Delta\hat{I}_{\beta}(\omega')\\+\Delta\hat{I}_{\beta}(\omega')\Delta\hat{I}_{\alpha}(\omega)\rangle.
\end{split}
\end{equation}
We can derive an expression for quantum noise \cite{BLANTER20001} using Eqs. (\ref{eq:2}, \ref{eq:5}, and \ref{eq:6}) and the Fermi-Dirac statistics, we get,
\begin{equation} \label{eq:7}
\begin{split}
S_{\alpha \beta}^q(\omega) =\frac{2e^2}{h}\sum_{\gamma,\delta} \sum_{m,n}\int dE  [A_{\gamma \delta}^{mn}(\alpha;E,E+\hbar \omega)\\\times
A_{\delta \gamma}^{nm}(\beta; E +\hbar \omega,E)](f_{\gamma}(E)[1-f_{\delta}(E+\hbar \omega)]\\+[1-f_{\gamma}(E)]f_{\delta}(E+\hbar \omega)).
\end{split}
\end{equation}
Our focus will be on zero frequency quantum noise, and for simplicity, we consider single edge mode occupation, which reduces Eq. (\ref{eq:7}) to,
\begin{equation} \label{eq:8}
\begin{split}
S_{\alpha \beta}^q =\frac{2e^2}{h}\sum_{\gamma,\delta}\int dE [A_{\gamma \delta}(\alpha)
A_{\delta \gamma}(\beta)]\\ (f_{\gamma}(E)[1-f_{\delta}(E)]+[1-f_{\gamma}(E)]f_{\delta}(E)).
\end{split}
\end{equation}
where, $A_{\beta \gamma}(\alpha) = I_{\alpha}\delta_{\alpha \beta}\delta_{\alpha \gamma} - s_{\alpha \beta}^{\dagger}s_{\alpha \gamma}$, and $s_{\alpha \beta}$ is an element of the s-matrix of the setup, describing the amplitude for an electron to scatter from terminal $\beta$ to $\alpha$ with single edge mode only. $s_{\alpha \beta}$ is the scattering amplitude for an electron to scatter from terminal $\beta$ to terminal $\alpha$. $f_{\gamma}(f_{\delta})$ is the Fermi Dirac distribution function for contact $\gamma(\delta)$ i.e., $f_{\gamma}(f_{\delta}) = (1+e^{(E-{\mu}_{\gamma(\delta)})/k_B\mathcal{T}})^{-1}$, $\mathcal{T}$ is the temperature of the system, $\mu_{\gamma(\delta)}$ is the chemical potential of contact $\gamma(\delta)$.
Quantum noise, as in, Eq. (\ref{eq:8}), {contains two parts}: (i) {the thermal noise-like contribution ($S_{\alpha \beta}^{th}$), and} (ii) {the shot noise-like contribution ($S_{\alpha \beta}^{sh}$)}.
The expression for both {thermal noise and shot noise-like contributions}  can be derived separately, and the full calculation is given in {\ref{Appendix B.1} and \ref{Appendix B 2} respectively}. Thus, {$S_{\alpha \beta}^{q} = S_{\alpha \beta}^{th} + S_{\alpha \beta}^{sh}$}. {For the quantum noise cross-correlation ($\alpha \ne \beta$), we have,}
\begin{equation} \label{eq:9}
\begin{split}
{S_{\alpha \beta}^{th} = -\frac{4e^2}{h}\int dE [T_{\alpha \beta}f_{\beta}(1-f_{\beta}) + T_{\beta \alpha}f_{\alpha}(1-f_{\alpha})]},\\
{S_{\alpha \beta}^{sh} = -\frac{4e^2}{h} \sum_{\gamma,\delta}\int dE (f_{\gamma}-f_a)(f_{\delta} - f_b)Tr(s_{\alpha \gamma}^{\dagger}s_{\alpha \delta}s_{\beta \delta}^{\dagger}s_{\beta \gamma})}.
\end{split}
\end{equation}

 {  $T_{\alpha \beta}$ is the transmission probability for an electron to scatter from terminal $\beta$ to terminal $\alpha$. $f_a$ and $f_b$ are energy-dependent functions, whose significance has been explained in Appendix~\ref{Appendix A.2}. As an aside, we provide a historical perspective to quantum noise calculation by considering a two-terminal quantum Hall sample with chiral edge modes in the presence of a constriction in Appendix~\ref{Appendix A}. In that setup, we  calculate the quantum noise cross-correlation $S_{12}^{q}$ between terminals 1 and 2, via wavepacket approach~\cite{LANDAUER1991167, PhysRevB.45.1742, imry} and also by the scattering matrix approach of Buttiker~\cite{PhysRevB.46.12485, BUTTIKER1991199} using Eq.~ (\ref{eq:8}). After deriving the quantum noise cross-correlation via the wavepacket approach, we derive the same quantum noise correlation via Buttiker's approach using Eq.~(\ref{eq:8}), which unfortunately does not match with that of the wavepacket approach. Although the thermal noise-like contribution is similar to that of the wavepacket approach, it is the shot noise-like contribution which doesn't match. Further, at zero temperature and zero applied voltage bias, shot noise does not  vanish which is a unphysical result. Therefore, Buttiker~\cite{PhysRevB.46.12485, BUTTIKER1991199} introduced a modification to the derivation of shot noise-like contribution, we explain this modification in Appendix~\ref{Appendix A.2}, Eq. (\ref{eq:9}) such that the problems with deriving shot noise using scattering matrix approach are corrected.}

We can also derive the {thermal and shot noise-like contributions separately for the autocorrelation ($\alpha = \beta$)}. The {thermal noise-like contribution to the autocorrelation is given as,}
\begin{equation} \label{eq:10}
{S_{\alpha \alpha}^{th}} = \frac{8e^2}{h}\int dE f_{\alpha}(1-f_{\alpha})(M_{\alpha}-R_{\alpha \alpha}).
\end{equation}
where $M_{\alpha}$ is the number of edge modes in terminal $\alpha$ and $R_{\alpha \alpha}$ is the probability for an electron to reflect from terminal $\alpha$ to itself.
{Similarly, the shot noise-like contribution is,}
\begin{equation} \label{eq:11}
\begin{split}
{S_{\alpha \alpha}^{sh}} = \frac{4e^2}{h}\int dE \bigg(\sum_{\gamma}T_{\alpha \gamma}(f_{\gamma}-f_{\alpha})+M_{\alpha}f_{\alpha}^2\\-\sum_{\gamma, \delta}f_{\gamma}f_{\delta}Tr(s_{\alpha \gamma}^{\dagger}s_{\alpha \delta}s_{\alpha \delta}^{\dagger}s_{\alpha \gamma})\bigg).
\end{split}
\end{equation}

We consider a 4-terminal system with QH edge modes. We can apply voltages at any of the contacts and measure the current-current correlation between any contacts based on a system as in Fig.~(\ref{fig:1}). {In our work, we have considered a variety of setups with different biasing conditions to verify that quantum noise correlations are the best probe to distinguish the different modes of transport irrespective of any situation.} We have three separate setups: setup 1 with $\mu_2 = \mu_3 = \mu_0$ and $\mu_1 = \mu_4 = \mu$, setup 2 with $\mu_2 = \mu_4 = \mu_0$ and $\mu_1 = \mu_3 = \mu$, setup 3 with $\mu_3 = \mu_4 = \mu_0$ and $\mu_1 = \mu_2 = \mu$.  In all these setups, $\mu - \mu_0 = eV$, where $V$ is the applied voltage bias and we will be working in the regime $\mu, \mu_0 \gg k_B \mathcal{T}$, where $\mathcal{T}$ is the temperature of each setup. In each setup, we have a constriction with scattering amplitudes $r, t$ respectively, and the general s-matrix for this quantum Hall bar, which is valid for each of the three setups, is given as,
\begin{equation} \label{eq:12}
s = \begin{pmatrix}
0 & te^{i\theta} & 0 & -ire^{i\theta}\\
0 & 0 & e^{i\phi_2} & 0\\
0 & -ire^{i\theta} & 0 & te^{i\theta}\\
e^{i\phi_1} & 0 & 0 & 0
\end{pmatrix}.
\end{equation}
The phases $\phi_2$ and $\phi_1$ are the propagating phases connecting different contacts due to the path difference of the electron. $\theta$ is the phase acquired when there is scattering due to the constriction. There will be an extra phase $\pi$/2 whenever there is a reflection due to the constriction such that the s-matrix is unitary. {From the s-matrix, we can find the transmission probability $T_{\alpha \beta}$ by taking the modulus square $|s_{\alpha \beta}|^2$ and since the s-matrix element is energy-independent, the transmission probability $T_{\alpha \beta}$ will also be energy independent.}

{In each of our setups, the thermal noise-like contributions $S_{\alpha \beta}^{th}$ both in quantum noise cross and auto-correlations are identical in the regime $\mu, \mu_0\gg k_B \mathcal{T}$. We  explain this in the following way in the context of cross-correlation ($\alpha \ne \beta$)- since the transmission probability $T_{\alpha \beta}$ is energy independent, $S_{\alpha \beta}^{th}$ as shown in Eq.~(\ref{eq:9}) only involves the integrations of $f_{\beta}(1-f_{\beta})$ and $f_{\alpha}(1-f_{\alpha})$. Now, in the regime $\mu_{\beta}, \mu_{\alpha} \gg k_B \mathcal{T}$, we see,}
\begin{equation} \label{eq:13}
   {\int_0^{\infty}dE f_{\beta}(1-f_{\beta}) = \int_0^{\infty}dE f_{\alpha}(1-f_{\alpha}) = k_B \mathcal{T}}.
\end{equation}
{where, $f_{\alpha} = \frac{1}{1+e^{\frac{E-\mu_{\alpha}}{k_B \mathcal{T}}}}$ and $f_{\beta} = \frac{1}{1+e^{\frac{E-\mu_{\beta}}{k_B \mathcal{T}}}}$. The thermal noise-like contribution, thus reduces to,}
\begin{equation} \label{eq:14}
    {S_{\alpha \beta}^{th}=-\frac{4e^2}{h}k_B \mathcal{T}(T_{\alpha \beta}+T_{\beta \alpha})}.
\end{equation}
{As seen in Eq.~(\ref{eq:14}), the thermal noise-like contribution is only dependent upon the transmission probabilities $T_{\alpha \beta}$ and $T_{\beta \alpha}$ and temperature $\mathcal{T}$ of the setup, and not on the applied voltage biases. So, this will be identical in each of the setups under different biasing conditions. Similarly, for the auto-correlation too, we can verify that the thermal noise-like contribution ($S_{\alpha \alpha}^{th}$) is same irrespective of the voltage biases  applied. Using Eq.~(\ref{eq:10}), the thermal noise-like contribution reduces to,}
\begin{equation} \label{eq:15}
   {S_{\alpha \alpha}^{th}=\frac{8e^2}{h}k_B \mathcal{T}(M_{\alpha}-R_{\alpha \alpha})}.
\end{equation}
Herein below, we analyze the thermal and shot noise-like contributions and the quantum noise calculation for each setup.
\subsection{Quantum noise due to chiral edge modes} \label{section IIB}
\subsubsection{Setup 1 [$\mu_2 = \mu_3 = \mu_0$ and $\mu_1 = \mu_4 = \mu$]} \label{section IIB.1}
\begin{figure}
\centering
\includegraphics[width=1.00\linewidth]{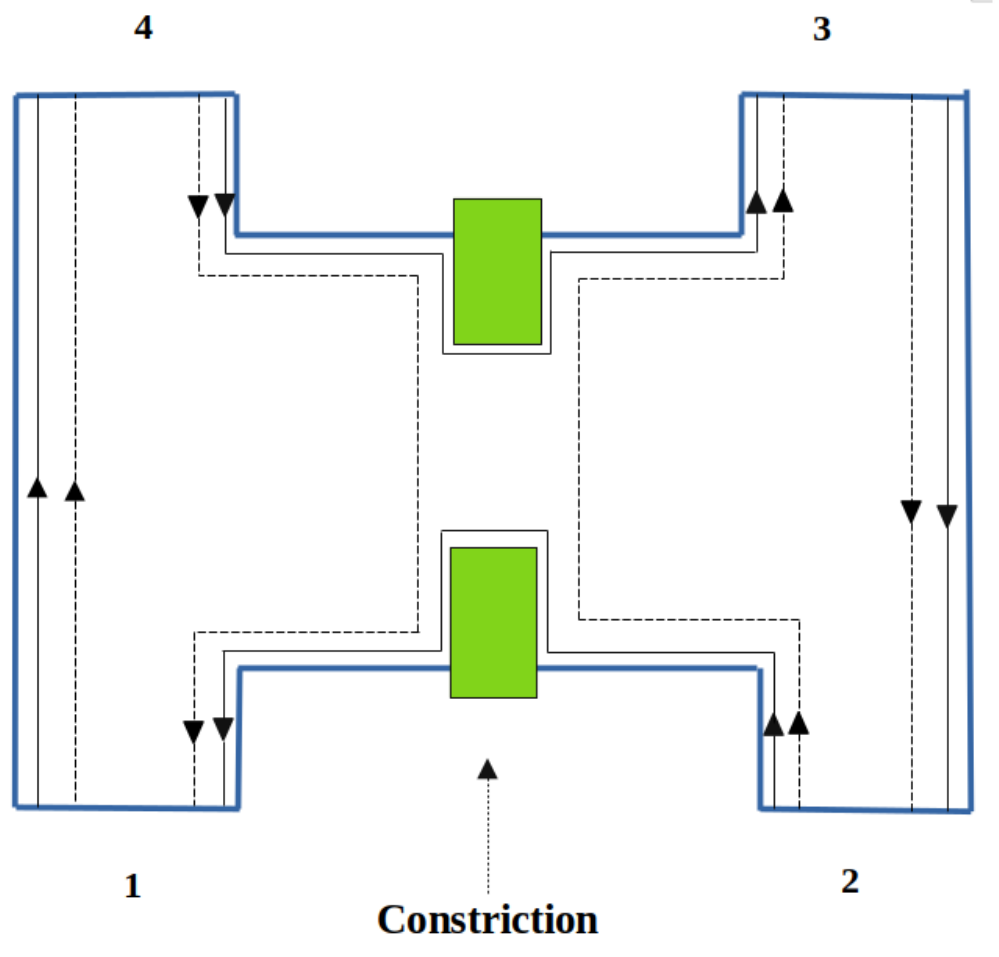}
\caption{Four terminal quantum Hall sample with chiral edge modes and a constriction. The black solid (dashed)line depicts the edge mode for an electron, which can transmit (reflect) via constriction}
\label{fig:1}
\end{figure}

Fig. \ref{fig:1} shows a quantum Hall setup, which is characterized by $\mu_2 = \mu_3 = \mu_0$ and $\mu_1 = \mu_4 = \mu = \mu_0 + eV$, where $V$ is the applied voltage.
In this setup, all contacts are assumed to be ideal and, we analyze all current correlations between different contacts. {First, we will calculate the quantum noise cross-correlation between terminals 1 and 3, i.e., $S_{13}^q$. The general expression for $S_{13}^q$ according to Eq. (\ref{eq:8}) is given as,}
\begin{equation} \label{eq: 16}
    \begin{split}
{S_{13}^q =\frac{2e^2}{h}\sum_{\gamma,\delta}\int dE [A_{\gamma \delta}(1)
A_{\delta \gamma}(3)]}\\ \times {(f_{\gamma}(E)[1-f_{\delta}(E)]+[1-f_{\gamma}(E)]f_{\delta}(E)).}
\end{split}
\end{equation}
{where $A_{\gamma \delta}(1) = \delta_{1 \gamma}\delta_{1 \delta}-s_{1 \gamma}^{\dagger}s_{1 \delta}$ and $A_{\delta \gamma}(3)= \delta_{3 \delta}\delta_{3 \gamma}-s_{3 \delta}^{\dagger}s_{3 \gamma}$. The thermal noise-like contributions {($S_{13}^{th}$)} using Eq. ~(\ref{eq:9}) vanishes, since an electron cannot directly tunnel to terminal 3 from terminal 1 and vice-versa,} and only {shot noise-like contribution ($S_{13}^{sh}$) is finite}. We can see this below,
\begin{equation} \label{eq:17}
{S_{13}^{sh}}=-\frac{4e^2}{h} \sum_{\gamma,\delta = 1,4}\int dE (f_{\gamma}-f_a)(f_{\delta}-f_b)Tr(s_{1 \gamma}^{\dagger}s_{1 \delta}s_{3 \delta}^{\dagger}s_{3 \gamma}),
\end{equation}

$f$ and $f_0$ are the Fermi-Dirac distribution functions for terminals $1$, $4$ and $2$, $3$. Choosing $f_a$ and $f_b$ to be $f_0$, which is similar to the assumption taken in \cite{PhysRevB.46.12485} and using the s-matrix given in Eq.~(\ref{eq:12}), we get
\begin{equation} \label{eq:18}
S_{13}^{q}={S_{13}^{sh}}=-\frac{4e^2}{h}\int dE R T (f-f_0)^2,
\end{equation}
where $T=|s_{12}|^2$ and $R=|s_{14}|^2$ are the transmission and reflection probabilities of the constriction. The integration~\cite{LANDAUER1991167} in Eq.~(\ref{eq:18}) is
\begin{equation} \label{eq:19}
\int_0^{\infty}dE(f-f_0)^2 = eV \coth \left[\frac{eV}{2k_B \mathcal{T}}\right]-2k_B \mathcal{T},
\end{equation}
where, $\mu-\mu_0 = eV$ and we have assumed the condition $\mu, \mu_0\gg k_B \mathcal{T}$ and taking the lower limit in Eq.~(\ref{eq:18}) to be $0$ and assuming it to be the energy at the bottom of the conduction band. So, the quantum noise correlation between the terminals $1$ and $3$ is,
\begin{equation} \label{eq:20}
S_{13}^{q}={S_{13}^{sh}}=-\frac{4e^2}{h}RT\left(eV\coth \left[\frac{eV}{2k_B \mathcal{T}}\right]-2k_B\mathcal{T}\right).
\end{equation}

{In this setup, we observe that quantum noise correlation ($S_{13}^{q}$) between terminals 1 and 3 is non-zero. $S_{13}^q$ is the current-current correlation $\langle \Delta I_{1} \Delta I_{3} \rangle$, which in principle should have contributions from thermal as well as shot noise. In all of our setups, the thermal noise-like contribution ($S_{13}^{th}$) is zero, since there is no direct tunneling between terminals 1 and 3, i.e., $T_{13} = |s_{13}|^2 = T_{31} = |s_{31}|^2 = 0$, which can be directly verified via $s$-matrix as in Eq.~(\ref{eq:12}). Thus, in $S_{13}^q$, only the shot noise-like contribution ($S_{13}^{sh}$) is finite. This can also be verified from the $s$-matrix as has been done in Eq.~(\ref{eq:17}). Shot noise-like contribution arises due to the constriction as well as applied voltage bias. If either the constriction is absent i.e., $R = 0, T = 1$ or voltage bias is zero, then $S_{13}^{sh}$ vanishes since for $eV \rightarrow 0$, the expression inside the parenthesis in Eq.~(\ref{eq:20}) is zero.} 

{We can have three distinct regimes. In the regime, $eV \gg k_B \mathcal{T}$,}
\begin{equation} \label{eq:21}
  {S_{13}^q = S_{13}^{sh} = -\frac{2e^2}{h}2RT (eV-2k_B \mathcal{T}) \simeq \frac{-4e^2}{h}RT eV,}
\end{equation}
 as for $eV \gg k_B \mathcal{T}$, $\coth\left[\frac{eV}{2k_B \mathcal{T}}\right] = 1$.
 
{In the regime $eV \ll k_B \mathcal{T}$, $S_{13}^q = 0$,} {and in the regime $eV = k_B \mathcal{T}$,}
\begin{equation} \label{eq:22}
    S_{13}^q = \frac{-4e^2}{h} 0.16 RT  k_B \mathcal{T}.
\end{equation}

The quantum noise correlation between contacts 1 and 4 denoted by $S_{14}^q$ is given as,
\begin{equation} \label{eq: 23}
    \begin{split}
{S_{14}^q =\frac{2e^2}{h}\sum_{\gamma,\delta}\int dE [A_{\gamma \delta}(1)
A_{\delta \gamma}(4)]}\\ {(f_{\gamma}(E)[1-f_{\delta}(E)]+[1-f_{\gamma}(E)]f_{\delta}(E)).}
\end{split}
\end{equation}
The {thermal noise-like contribution ($S_{14}^{th}$)} to the quantum noise correlation is given by,
\begin{equation} \label{eq: 24}
{S_{14}^{th}} = -\frac{4e^2}{h}\int dE [T_{14}f_4(1-f_4) + T_{41}f_1(1-f_1)],
\end{equation}
now, using $\int_0^{\infty}dE f_{\alpha}(1-f_{\alpha}) = k_B \mathcal{T}$, where $\alpha$ = 1,2,3,4 with $\mathcal{T}$ being the temperature {in the limit $\mu, \mu_0 \gg k_B \mathcal{T}$}, we get,
\begin{equation} \label{eq: 25}
{S_{14}^{th}} = -\frac{4e^2}{h} k_B \mathcal{T} (T_{14} + T_{41}) = -\frac{4e^2}{h} k_B \mathcal{T}(1+R).
\end{equation}
Similarly, we can calculate the {shot noise-like} contribution, which is
\begin{equation} \label{eq: 26}
{S_{14}^{sh}} = -\frac{4e^2}{h}\sum_{\gamma,\delta = 1,4}\int dE(f_{\gamma}-f_a)(f_{\delta} - f_b)Tr(s_{1 \gamma}^{\dagger}s_{1 \delta}s_{4 \delta}^{\dagger}s_{4 \gamma}).
\end{equation}
{Because each term in the summation in Eq.~(\ref{eq: 26}) vanishes due to s-matrix elements, see Eq. (\ref{eq:12}), vanishing. The shot noise contribution {$S_{14}^{sh}$} vanishes.} So, the quantum noise correlation between terminals $1$ and $4$ is entirely due to thermal noise. Thus,
\begin{equation} \label{eq: 27}
S_{14}^q = -\frac{4e^2}{h} k_B \mathcal{T}(1+R).
\end{equation}
Similarly, when we calculate the quantum noise correlation between terminals 1 and 2, 3 and 4, and 2 and 3, we see that only the thermal noise contributes, i.e.,
\begin{equation} \label{eq: 28}
 S_{12}^q = S_{34}^q = \frac{1-R}{1+R}S_{14}^q, \quad  S_{23}^q = S_{23}^{th} = S_{14}^q.
 \end{equation}
{Here also, the shot noise-like contributions vanish because of the elements of s-matrix as in Eq. (\ref{eq:12}).} There is one particular case as well, i.e., the quantum noise correlation between terminals 2 and 4 which is precisely zero since both the {thermal noise-like} and {shot noise-like} contributions vanish. {The thermal noise-like contribution ($S_{24}^{th}$) is zero, since there is no direct tunneling process between terminals 2 and 4.} 

{The shot noise-like contribution is given as,}
\begin{equation}
{S_{24}^{sh} = \frac{-4e^2}{h} \sum_{\gamma, \delta = 1,2,3,4} \int dE (f_{\gamma} - f_0)(f_{\delta} - f_0)Tr(s_{2 \gamma}^{\dagger}s_{2 \delta}s_{4 \delta}^{\dagger}s_{4 \gamma}).}
\end{equation}

{The summation in the above equation for shot noise-like contribution ($S_{24}^{sh}$) vanishes due to one or more of the $s$-matrix elements being zero, as in Eq.~(\ref{eq:12}). Therefore, the quantum noise cross-correlation between terminals 2 and 4 is given as,}
\begin{equation} \label{eq: 29}
   {S_{24}^{q} = S_{24}^{th} + S_{24}^{sh} = 0}.
\end{equation}

{Similarly, we can calculate autocorrelation for each of the terminals. For this purpose, we  use the formulae given in Eqs.~(\ref{eq:10}) and (\ref{eq:11}) to calculate thermal noise and shot noise-like contributions.} 
{The thermal noise-like contribution ($S_{11}^{th}$) to quantum noise autocorrelation ($S_{11}^q$) in terminal 1 is given as,}
\begin{align} \label{eq:166}
\begin{split}
{S_{11}^{th}} & {= \frac{8e^2}{h} \int dE f(1-f)(M_1 - R_{11})} \\ &= {\frac{8e^2}{h}\int dE f(1-f)}.
\end{split}
\end{align}

{Using the $s$-matrix as in Eq.~(\ref{eq:12}), $R_{11} = |s_{11}|^2$ = 0 and number of edge modes in terminal 1 ($M_1$) to be 1, we get}
\begin{equation} \label{eq: 30}
  S_{11}^{th} = \frac{8e^2}{h} \int dE f(1-f)
\end{equation}
{The shot noise-like correlation ($S_{11}^{sh}$) is given as,}
\begin{equation} \label{eq: 31}
\begin{split}
   {S_{11}^{sh} = \frac{4e^2}{h}\int dE \bigg (T(f_0 - f) + f^2}\\ { - \sum_{\gamma, \delta}f_{\gamma}f_{\delta}Tr(s_{1\gamma}^{\dagger}s_{1 \delta}s_{1 \delta}^{\dagger}s_{1 \gamma})\bigg)}.
    \end{split}
\end{equation}

{In the summation in the above equation, the surviving terms arise when $\gamma$ and $\delta$ are either 2 or 4, which can be seen in the $s$-matrix as in Eq. (\ref{eq:12}).}
{Using the $s$-matrix given in Eq.~(\ref{eq:12}) and adding Eqs.~(\ref{eq: 30}),~(\ref{eq: 31}), we get,}
\begin{equation}
\begin{split}
{S_{11}^q = S_{11}^{th} + S_{11}^{sh} = \frac{4e^2}{h}\int dE (2f(1-f) + T(f_0 - f) + f^2} \\ {- R^2 f^2 - T^2 f_0^2 - 2 RT f f_0))}
\end{split}
\end{equation}

{Simplifying further, we get the quantum noise autocorrelation ($S_{11}^q$) to be,}
\begin{equation} \label{eq: 32}
\begin{split}
S_{11}^q = \frac{4e^2}{h}\int dE [f(1-f)+Rf(1-f)+Tf_0(1-f_0)\\+RT(f-f_0)^2].
\end{split}
\end{equation}
Similarly, the autocorrelation in contact 3: $S_{33}^q$ is the same as $S_{11}^q$.
{Doing the integration in, Eq.~(\ref{eq: 32}) taking the limit from $0$ to $\infty$ and assuming $\mu, \mu_0 \gg k_B \mathcal{T}$, we get,}
\begin{equation} \label{eq: 33}
\begin{split}
S_{11}^q = S_{33}^q =\frac{8e^2}{h}k_B \mathcal{T} \\+\frac{2e^2}{h} 2RT\left(eV\coth \left[\frac{eV}{2k_B \mathcal{T}}\right]-2k_B \mathcal{T}\right).
\end{split}
\end{equation}
Eq. (\ref{eq: 33}) has both the {thermal} and shot noise-like contributions. They are given as,
\begin{equation}   \label{eq:34}
       {S_{11}^{th}} {= S_{33}^{th} = \frac{8e^2}{h}k_B \mathcal{T},}
        \end{equation}
        \begin{equation} \label{eq:35}
            {S_{11}^{sh}}  {= S_{33}^{sh} = \frac{4e^2}{h}RT \left
             (eV \coth \left[\frac{eV}{2k_B \mathcal{T}}\right]-2k_B \mathcal{T}\right).}
  \end{equation}
{Here, we can see that the quantum noise auto-correlation has contributions from equilibrium (thermal noise) and transport (shot noise) fluctutions. The thermal noise-like contribution in terminal 1 arises effectively due to transport with probability 1.  The shot noise-like contribution comes due to both constriction and applied voltage bias. If there is no constriction, i.e., $R = 0, T = 1$ or at zero applied voltage bias, i.e., $eV \rightarrow 0$, then the shot noise-like contribution vanishes and what remains is only the thermal noise-like contribution.}

{Here, again we have three distinct regimes. For $eV\gg k_B \mathcal{T}$, the quantum noise autocorrelation is,}
  \begin{equation} \label{eq:36}
   {S_{11}^q = S_{33}^q = \frac{4e^2}{h}RT \left(eV -2k_B \mathcal{T}\right)},
  \end{equation}
 { for the regime $eV \ll k_B \mathcal{T} $,}
\begin{equation} \label{eq:37}
  {S_{11}^q = S_{33}^q = \frac{8e^2}{h}k_B \mathcal{T}.}
\end{equation}
{and, for the regime $eV = k_B \mathcal{T}$,}
\begin{equation}\label{eq:38}
    S_{11}^q = S_{33}^q = \frac{2e^2}{h}4 k_B \mathcal{T} + \frac{2e^2}{h}0.32 RT k_B \mathcal{T}.
\end{equation}

We can also calculate auto-correlation in other terminals as well.{The thermal noise-like contribution in terminal 2 is given as,}
\begin{equation} \label{eq:39}
   {S_{22}^{th} = \frac{8e^2}{h}\int dE f_0 (1-f_0),}
\end{equation}
{and the shot noise-like contribution is given as,}
\begin{equation} \label{eq:40}
{S_{22}^{sh} = \frac{4e^2}{h}\int dE \bigg( f_0^2} { - \sum_{\gamma, \delta}f_{\gamma}f_{\delta}Tr(s_{2\gamma}^{\dagger}s_{2 \delta}s_{2 \delta}^{\dagger}s_{2 \gamma})\bigg)} = 0.
    \end{equation}
   {Now, using the s-matrix as given in Eq.~(\ref{eq:12}), we see that the trace in Eq.~ (\ref{eq:40}) only survives only if $\gamma = \delta = 3$, which makes the total shot noise-like contribution zero.}
The noise auto-correlation in terminal 2 is only due to thermal noise and it is the same as that of terminal 4, which is given as,
\begin{equation} \label{eq:41}
\begin{split}
S_{22}^q = {S_{22}^{th}} = S_{44}^q = {S_{44}^{th}} = \frac{8e^2}{h}k_B \mathcal{T}.
\end{split}
\end{equation}
It should be noted that the quantum noise autocorrelations $S_{22}^q$ and $S_{44}^q$ are independent of constriction.

\subsubsection{setup 2 [$\mu_2 = \mu_4 = \mu_0$ and $\mu_1 = \mu_3 = \mu$]} \label{section IIB.2}
In this subsection, we are concerned with the outcomes for setup 2, i.e., $\mu_2 = \mu_4 = \mu_0$ and $\mu_1 = \mu_3 = \mu$. Using Eq. (\ref{eq:9}), we can find the thermal and shot noise-like contributions to total quantum noise. {As shown in Eq. (\ref{eq:14}) and (\ref{eq:15}) in Sec.~\ref{section II A}}, the { thermal noise}-like contribution to any cross-correlation between terminals will remain unchanged in any setup since it only depends on the {tunneling probability} for the electron to scatter from one terminal to another and not on how voltage biases are applied. The {shot noise}-like contribution may change depending on the voltage bias applied across the terminals. Below we provide a list of cross-correlations, {where only thermal noise contributes}.
\begin{align} 
S_{14}^q &= {S_{14}^{th}} = S_{23}^q = {S_{23}^{th}} = -\frac{4e^2}{h}k_B \mathcal{T}(1+R), \label{eq:42}\\
\text{and}, S_{12}^q &= {S_{12}^{th}} = S_{34}^q = {S_{34}^{th}} =
\frac{1-R}{1+R}S_{14}^q. \label{eq:43}
\end{align}
In all these cases, the {shot noise-like} contribution vanishes.  Like we explained in Eq.~(\ref{eq: 26}), the shot noise-like contributions here too vanish due to the $s$-matrix given in Eq.~(\ref{eq:12}) with its elements vanishing.

Further, the quantum noise correlation between terminals 1 and 3, and between terminals 2 and 4 is zero. The quantum noise correlations between terminals 1 and 3, 2 and 4 for setup 2 are,
\begin{align}
{S_{13}^{q}} &{= S_{13}^{th} = S_{13}^{sh} = 0,} \quad
{S_{24}^{q}}  {= S_{24}^{th} = S_{24}^{sh} = 0}. \label{eq:44}
\end{align}

{We also see that $S_{13}^q$ vanishes, because the chemical potential in terminals 1 and 3 are now the same, which implies absence of any applied bias voltage, which drives the shot noise-like contribution to zero. Further, thermal noise-like contribution ($S_{13}^{th}$) is zero, since there is no direct tunneling between terminals 1 and 3.}
The autocorrelations in each terminal are the same, which is only {thermal noise-like} and {is given as,}
\begin{equation}
\begin{split}
S_{ii}^q={S_{ii}^{th}} = \frac{8e^2}{h}k_B \mathcal{T}, \quad i \in \{1,2,3,4\} \label{eq:45}
\end{split}
\end{equation}

{This can be understood using the $s$-matrix as in Eq.~(\ref{eq:12}). Using Eq.~(\ref{eq:10}), the thermal noise-like contribution in terminal 1 is given as,}
\begin{equation}
{S_{11}^{th}  = \frac{8e^2}{h}\int_0^{\infty}dE f(1-f) = \frac{8e^2}{h}k_B \mathcal{T}.}
\end{equation}
{As proved previously in Eq.~(\ref{eq:15}), the thermal noise-like contribution is independent of setups. The only thing which can change is the shot noise-like contribution.}
{Using Eq.~(\ref{eq:11}), we get the shot noise-like contribution is given as,}
\begin{equation}
\begin{split}
{S_{11}^{sh} = \frac{4e^2}{h}\int dE \bigg(\sum_{\gamma}T_{1 \gamma}(f_{\gamma} - f_1) + f_1^2} \\ { -\sum_{\gamma, \delta}f_{\gamma}f_{\delta} Tr(s_{1 \gamma}^{\dagger}s_{\alpha \delta}s_{\alpha \delta}^{\dagger}s_{\alpha \gamma})\bigg)}
\end{split}
\end{equation}

{Using the $s$-matrix as in Eq.~(\ref{eq:12}), we get} 
\begin{equation}
{S_{11}^{sh} = \frac{4e^2}{h} \int_0^{\infty} dE (f_0(1-f_0) - f(1-f)) = 0}
\end{equation}
{where, we used Eq.~(\ref{eq:13}) in the limit $\mu, \mu_0 \gg k_B \mathcal{T}$, which makes the shot noise to vanish. It effectively means the constriction does not play any role even though there is an applied bias voltage. That is why the quantum noise auto-correlation in terminal 1 is only thermal noise-like. The quantum noise auto-correlation in terminal 3 ($S_{33}^q$) is identical to $S_{11}^q$. Similarly, we can find quantum noise autocorrelation in terminal 2 is also thermal noise-like and we can derive it using the $s$-matrix. The thermal noise-like contribution in terminal 2 is,}

\begin{equation}
{S_{22}^{th} = \frac{8e^2}{h} \int_0^{\infty}dE f_0(1-f_0) = \frac{8e^2}{h} k_B \mathcal{T},}
\end{equation}
{and, the shot noise-like contribution is given as,}
\begin{equation}
{S_{22}^{sh} = \frac{4e^2}{h} \int_0^{\infty}dE \bigg (f(1-f) - f_0(1-f_0) \bigg) = 0.}
\end{equation}
{The shot noise-like contribution is also zero at terminal 2. The quantum noise auto-correlation in terminal 4 ($S_{44}^q$) is identical to $S_{22}^q$. Therefore, we conclude that the quantum noise autocorrelations in this setup are only thermal noise-like.}

\subsubsection{setup 3 [$\mu_1 = \mu_2 = \mu$ and $\mu_3 = \mu_4 = \mu_0$]} \label{section IIB.3}
In this setup, we see that the quantum noise correlation between terminals 1 and 3 are {shot noise}-like only. The quantum noise correlation for all other terminals are {thermal noise-like}. Below, we derive expressions of all the possible quantum noise cross-correlations.
\begin{equation} \label{eq:46}
\begin{split}
S_{14}^q = {S_{14}^{th}} = S_{23}^q= {S_{23}^{th}} = -\frac{4e^2}{h}k_B \mathcal{T}(1+R),\\
S_{12}^q = {S_{12}^{th}} = S_{34}^q = {S_{34}^{th}} = \frac{(1-R)}{1+R}S_{14}^q,\\
S_{13}^{q} = {S_{13}^{sh}} = -\frac{4e^2}{h}RT \left(eV\coth \left[\frac{eV}{2k_B \mathcal{T}}\right]-2k_B \mathcal{T}\right),\\
{S_{24}^q = S_{24}^{th} = S_{24}^{sh} = 0.}
\end{split}
\end{equation}
{The reason behind the vanishing shot noise-like contribution can be understood from the $s$-matrix as in Eq.~(\ref{eq:12}). Since the elements involved in the summation in the shot noise-like contributions vanish, the shot noise-like contribution automatically vanishes.}
For the autocorrelations, we get,
\begin{equation} \label{eq:47}
\begin{split}
S_{22}^q = {S_{22}^{th}} = S_{44}^q = {S_{44}^{th}} = \frac{8e^2}{h}k_B \mathcal{T}, \text{with}\quad S_{22}^{sh} = S_{44}^{sh} = 0,\\
S_{33}^q = S_{11}^q = S_{11}^{th} + {S_{11}^{sh}}.
\end{split}
\end{equation}
and,{$S_{11}^{sh} = S_{33}^{sh} = \frac{4e^2}{h}RT\bigg(eV\coth \left[\frac{eV}{2k_B \mathcal{T}}\right]-2k_B \mathcal{T}\bigg)$.}

{We observe that the quantum noise-correlations $S_{11}^q$ and $S_{33}^q$ are a combination of both equilibrium (thermal noise) and transport (shot noise) fluctuations, whereas in case of $S_{22}^q$ and $S_{44}^q$, only equilibrium (thermal noise) fluctuation matters. } 
{The reason behind this can be derived and explained in a similar fashion as we did in setup-1 below Eqs. (\ref{eq: 33}) and (\ref{eq:38}) using the $s$-matrix.}

{In the following section, we will calculate the quantum noise correlations in the presence of helical edge modes.}
\section{Quantum noise in quantum spin Hall setup} \label{section III}
\subsection{Theory} \label{section III.A}
The current-current correlation between the metallic contacts $\alpha$ and $\beta$ due to spin polarised current is given by,
\begin{equation} \label{eq:48}
S_{\alpha \beta}^{\sigma \sigma'}(t-t') = \frac{1}{2}\langle \Delta \hat{I}_{\alpha}^{\sigma}(t)\Delta\hat{I}_{\beta}^{\sigma'}(t')+\Delta\hat{I}_{\beta}^{\sigma'}(t')\Delta\hat{I}_{\alpha}^{\sigma}(t)\rangle,
\end{equation}
where $\Delta\hat{I_{\alpha}^{\sigma}}$ is the fluctuation in current from the average value in contact $\alpha$ with spin $\sigma$ and is given by $\Delta \hat{I_{\alpha}^{\sigma}} = \hat{I_{\alpha}^{\sigma}} - \langle \hat{I_{\alpha}^{\sigma}} \rangle$. The current due to a charged particle with spin $\sigma$ flowing through terminal $\alpha$ is
\begin{equation} \label{eq:49}
\begin{split}
\hat{I_{\alpha}^{\sigma}}(t) = \frac{e}{h}\sum_{n=1}^M \int \int dE dE' e^{i(E-E')t/\hbar}\\\times (\hat{a}_{\alpha n}^{\sigma \dagger}(E)\hat{a}_{\alpha n}^{\sigma}(E') - \hat{b}_{\alpha n}^{\sigma \dagger}(E)\hat{b}_{\alpha n}^{\sigma}(E')),
\end{split}
\end{equation}
where $\hat{a}_{\alpha n}^{\sigma \dagger}(E)$ is the creation operator, which creates an incoming electron in terminal $\alpha$ with energy $E$ in edge channel $n$ and $\hat{a}_{\alpha n}^{\sigma}(E)$ is the annihilation operator, which annihilates an incoming electron in the terminal $\alpha$ with energy $E'$ in edge channel $n$. Similarly, $\hat{b}_{\alpha n}^{\sigma \dagger}(E)$ and $\hat{b}_{\alpha n}^{\sigma}(E')$ are the creation and annihilation operators, which create and annihilate an outgoing electron from the terminal $\alpha$ with spin $\sigma$ with energy $E$ and $E'$ respectively.
The expression for finite temperature quantum noise correlation in quantum spin Hall set up due to the spin polarised currents $I^{\uparrow}$ and $I^{\downarrow}$ see in Fig. \ref{fig:2} is given by,
\begin{equation} \label{eq:50}
S_{\alpha \beta}^{q} = S_{\alpha \beta}^{\uparrow \uparrow, q} + S_{\alpha \beta}^{\uparrow \downarrow, q} + S_{\alpha \beta}^{\downarrow \uparrow, q} + S_{\alpha \beta}^{\downarrow \downarrow, q},
\end{equation}
the individual spin polarised contributions are given as \cite{PhysRevB.75.085328},

\begin{equation} \label{eq:51}
\begin{split}
S_{\alpha \beta}^{\sigma \sigma', q} =\frac{e^2}{h}\sum_{\rho, \rho' = \uparrow, \downarrow}\sum_{\gamma,\delta}\int dE Tr[A_{\gamma \delta}^{\rho \rho'}(\alpha,\sigma, E, E + \hbar \omega)\\ \times
A_{\delta \gamma}^{\rho' \rho}(\beta, \sigma', E + \hbar \omega, E)](f_{\gamma}(E)[1-f_{\delta}(E + \hbar \omega)]\\+[1-f_{\gamma}(E + \hbar \omega)]f_{\delta}(E)),
\end{split}
\end{equation}

Our focus is on zero-frequency quantum noise, which reduces Eq.~(\ref{eq:51}) to,
\begin{equation} \label{eq:52}
\begin{split}
S_{\alpha \beta}^{\sigma \sigma', q} =\frac{e^2}{h}\sum_{\rho, \rho' = \uparrow, \downarrow}\sum_{\gamma,\delta}\int dE Tr[A_{\gamma \delta}^{\rho \rho'}(\alpha,\sigma)
A_{\delta \gamma}^{\rho' \rho}(\beta, \sigma')]\\(f_{\gamma}(E)[1-f_{\delta}(E)]+[1-f_{\gamma}(E)]f_{\delta}(E)),
\end{split}
\end{equation}
where, $A_{\gamma \beta}^{\rho \rho'}(\alpha, \sigma) = \delta_{\alpha \gamma}\delta_{\alpha \beta}\delta_{\sigma \rho}\delta_{\sigma \rho'} - s_{\alpha \gamma}^{\sigma \rho \dagger}s_{\alpha \beta}^{\sigma \rho'}$, and $\sigma, \sigma'$ denotes the spin of electrons.
Eq.~(\ref{eq:52}) has both the thermal noise-like contribution and shot noise-like contribution to the quantum noise. We will first look into the cross-correlation, i.e., $\alpha \ne \beta$. The {thermal noise-like} contribution is given by,
\begin{equation} \label{eq: 53}
{S_{\alpha \beta}^{th} = S_{\alpha \beta}^{\uparrow \uparrow, th} + S_{\alpha \beta}^{\uparrow \downarrow, th} + S_{\alpha \beta}^{\downarrow \uparrow, th} + S_{\alpha \beta}^{\downarrow \downarrow, th}},
\end{equation}
and the individual spin polarised contributions to the {thermal noise-like contributions} are given as,
\begin{equation} \label{eq:54}
\begin{split}
{S_{\alpha \beta}^{\sigma \sigma', th}} = -\frac{2e^2}{h}\int_0^{\infty}dE[T_{\alpha \beta}^{\sigma \sigma'}f_{\beta}(1-f_{\beta}) + T_{\beta \alpha}^{\sigma \sigma'}f_{\alpha}(1-f_{\alpha})].
\end{split}
\end{equation}
see Appendix \ref{Appendix C 1} for the derivation of Eq. (\ref{eq:54}).
$T_{\alpha \beta}^{\sigma \sigma'}$ is the transmission probability for an electron to scatter from terminal $\beta$ with initial spin $\sigma'$ to terminal $\alpha$ with final spin $\sigma$.
$f_{\beta}$ is the Fermi-Dirac distribution function at contact $\beta$ connected to a reservoir given by the expression $f_{\beta} = (1+e^{(E-\mu_{\beta})/k_B \mathcal{T}})$, where $\mathcal{T}$ is the temperature of the system, $\mu_{\beta}$ is the chemical potential of the contact $\beta$.

{From the s-matrix, we can find the transmission probability $T_{\alpha \beta}^{\sigma \sigma'}$ by taking the modulus square $|s_{\alpha \beta}^{\sigma \sigma'}|^2$ and since the s-matrix element in our setups is energy-independent, the transmission probability $T_{\alpha \beta}^{\sigma \sigma'}$ will also be energy independent.}

{In each of our setups, the thermal noise-like contributions $S_{\alpha \beta}^{\sigma \sigma', th}$ are identical in the regime $\mu, \mu_0 \gg k_B \mathcal{T}$. We can explain this in the following way. Since the transmission probability $T_{\alpha \beta}^{\sigma \sigma'}$ is energy independent, $S_{\alpha \beta}^{\sigma \sigma', th}$ as shown in Eq. (\ref{eq:54}) only involves the integrations of $f_{\beta}(1-f_{\beta})$ and $f_{\alpha}(1-f_{\alpha})$. Now, in the regime $\mu_{\beta}, \mu_{\alpha} \gg k_B \mathcal{T}$, we see}
\begin{equation} \label{eq:55}
    {\int_0^{\infty}dE f_{\beta}(1-f_{\beta}) = \int_0^{\infty}dE f_{\alpha}(1-f_{\alpha}) = k_B \mathcal{T}}.
\end{equation}
{so, the thermal noise-like contribution reduces to,}
\begin{equation} \label{eq:56}
   {S_{\alpha \beta}^{\sigma \sigma',th} = -\frac{2e^2}{h}k_B \mathcal{T}(T_{\alpha \beta}^{\sigma \sigma'} + T_{\beta \alpha}^{\sigma \sigma'})}
\end{equation}
{As seen in Eq. (\ref{eq:56}), the thermal noise-like contribution is only dependent upon the transmission probabilities $T_{\alpha \beta}^{\sigma \sigma'}$ and $T_{\beta \alpha}^{\sigma \sigma'}$ and temperature $\mathcal{T}$ of the setup, not on the applied voltage biases. So, this will be identical in each of the setups under different biasing conditions. }

Similar to, Eq. (\ref{eq: 53}), the {shot noise-like contribution} is
\begin{equation} \label{eq:57}
{S_{\alpha \beta}^{sh} = S_{\alpha \beta}^{\uparrow \uparrow, sh} + S_{\alpha \beta}^{\uparrow \downarrow, sh} + S_{\alpha \beta}^{\downarrow \uparrow, sh} + S_{\alpha \beta}^{\downarrow \downarrow, sh},}
\end{equation}
the individual spin-polarised contributions are
\begin{equation} \label{eq:58}
\begin{split}
{S_{\alpha \beta}^{\sigma \sigma', sh}} = -\frac{2e^2}{h}\int dE \sum_{\gamma, \delta}\sum_{\rho \rho' = \uparrow, \downarrow}(f_{\gamma}-f_a)(f_{\delta}-f_b)\\\times Tr(s_{\alpha \gamma}^{\sigma \rho^{\dagger}}s_{\alpha \delta}^{\sigma \rho'}s_{\beta \delta}^{\sigma' \rho'^{\dagger}}s_{\beta \gamma}^{\sigma' \rho} ).
\end{split}
\end{equation}
See Appendix \ref{Appendix C 2} for the derivation of, Eq. (\ref{eq:58}).

Similarly, the quantum noise autocorrelation ($\alpha = \beta$) is
\begin{equation} \label{eq:59}
S_{\alpha \alpha}^{q} = S_{\alpha \alpha}^{\uparrow \uparrow, q} + S_{\alpha \alpha}^{\uparrow \downarrow, q} + S_{\alpha \alpha}^{\downarrow \uparrow, q} + S_{\alpha \alpha}^{\downarrow \downarrow, q},
\end{equation}
where the individual spin-polarised correlations are
\begin{equation} \label{eq:60}
\begin{split}
S_{\alpha \alpha}^{\sigma \sigma', q} =\frac{e^2}{h}\sum_{\rho, \rho' = \uparrow, \downarrow}\sum_{\gamma,\delta}\int dETr[A_{\gamma \delta}^{\rho \rho'}(\alpha,\sigma)
A_{\delta \gamma}^{\rho' \rho}(\alpha, \sigma')]\\\times(f_{\gamma}(E)[1-f_{\delta}(E)]+[1-f_{\gamma}(E)]f_{\delta}(E)),
\end{split}
\end{equation}
Eq. (\ref{eq:60}) also has both thermal noise-like and shot noise-like contributions. For the autocorrelation, i.e., $\alpha = \beta$, the {thermal noise-like contribution} is given by:
\begin{equation} \label{eq:61}
{S_{\alpha \alpha}^{th} = S_{\alpha \alpha}^{\uparrow \uparrow, th} + S_{\alpha \alpha}^{\uparrow \downarrow, th} + S_{\alpha \alpha}^{\downarrow \uparrow, th} + S_{\alpha \alpha}^{\downarrow \downarrow, th}},
\end{equation}
where the spin-polarised contributions, as derived in Appendix \ref{Appendix C 1}, are given as,
\begin{equation} \label{eq:62}
{S_{\alpha \alpha}^{\sigma \sigma', th}} = \frac{4e^2}{h}\int dE f_{\alpha}(1-f_{\alpha})(M_{\alpha}\delta_{\sigma \sigma'} - R_{\alpha \alpha}^{\sigma \sigma'}).
\end{equation}
Similar to {thermal noise-like contribution, shot noise-like contribution} to the autocorrelation is,
\begin{equation} \label{eq:63}
{S_{\alpha \alpha}^{sh} = S_{\alpha \alpha}^{\uparrow \uparrow, sh} + S_{\alpha \alpha}^{\uparrow \downarrow, sh} + S_{\alpha \alpha}^{\downarrow \uparrow, sh} + S_{\alpha \alpha}^{\downarrow \downarrow, sh}},
\end{equation}
where,
the spin polarised contribution, as derived in Appendix \ref{Appendix C 2}, is given as,
\begin{equation} \label{eq:64}
\begin{split}
{S_{\alpha \alpha}^{\sigma \sigma', sh}} = \frac{2e^2}{h}\int dE \bigg(\sum_{\gamma}T_{\alpha \gamma}^{\sigma \sigma'}(f_{\gamma} - f_{\alpha})+M_{\alpha}\delta_{\sigma \sigma'}f_{\alpha}^2\\-\sum_{\gamma \delta}\sum_{\rho \rho' = \uparrow, \downarrow}f_{\gamma}f_{\delta}Tr(s_{\alpha \gamma}^{\sigma \rho^{\dagger}}s_{\alpha \delta}^{\sigma \rho'}s_{\alpha \delta}^{\sigma' \rho'^{\dagger}}s_{\alpha \gamma}^{\sigma' \rho})\bigg).
\end{split}
\end{equation}
Here $M_{\alpha}$ is the number of edge modes from contact $\alpha$.
{Similarly, for auto-correlations, the thermal noise-like contributions are identical irrespective of voltage biases applied. Using Eq. (\ref{eq:62}), we get,}
\begin{equation} \label{eq:65}
    S_{\alpha \alpha}^{\sigma \sigma', th} = \frac{4e^2}{h}k_B \mathcal{T}(M_{\alpha} \delta_{\sigma \sigma'}-R_{\alpha \alpha}^{\sigma \sigma'}).
\end{equation}
\subsection{Results for topological helical edge modes} \label{section III.B}
\begin{figure}
\centering
\includegraphics[width=1.00\linewidth]{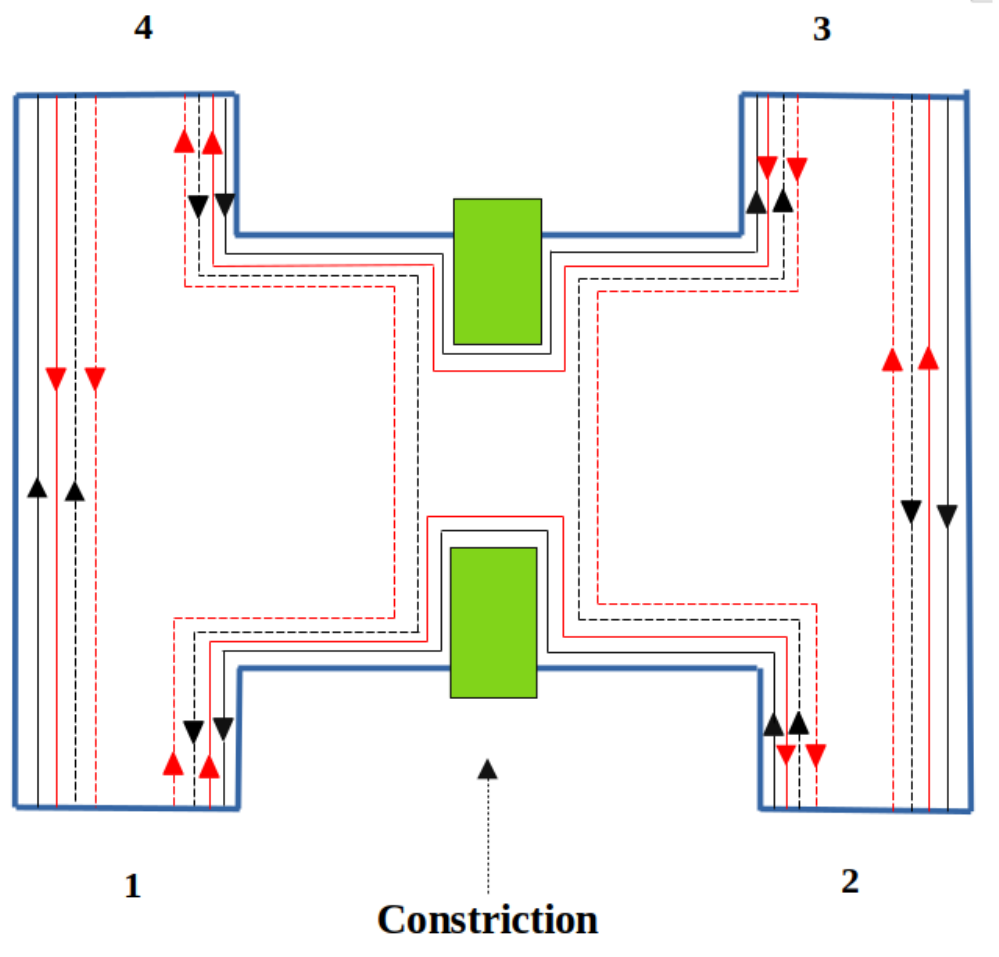}
\caption{Four terminal quantum Hall sample with a constriction, showing Helical edge modes. The Black (Red) solid (dashed)line depicts the edge mode for a spin-up (down) electron, which can, transmit (reflect) via constriction.}
 \label{fig:2}
\end{figure}

. 
This section considers a 4-terminal system with QSH edge modes, which is topological, and the electron motion is helical, as shown in Fig. \ref{fig:2}. We can apply voltages at any contact and measure the current-current correlation between any of those contacts based on a system as in Fig. \ref{fig:2}. There we have a constriction with reflection and, transmission amplitudes $r$, $t$, respectively. The general $s$-matrix for this quantum spin Hall bar setup is,
\begin{equation} \label{eq:66}
s=
\begin{pmatrix}
0 & 0 & te^{i\theta} & 0 & 0 & 0 & -ire^{i\theta} & 0\\
0 & 0 & 0 & 0 & 0 & 0 & 0 & e^{i\phi_1}\\
0 & 0 & 0 & 0 & e^{i\phi_2} & 0 & 0 & 0\\
0 & te^{i\theta} & 0 & 0 & 0 & -ire^{i\theta} & 0 & 0\\
0 & 0 & -ire^{i\theta} & 0 & 0 & 0 & te^{i\theta} & 0\\
0 & 0 & 0 & e^{i\phi_2} & 0 & 0 & 0 & 0\\
e^{i\phi_1} & 0 & 0 & 0 & 0 & 0 & 0 & 0\\
0 & -ire^{i\theta} & 0 & 0 & 0 & te^{i\theta} & 0 & 0
\end{pmatrix}
\end{equation}
{The phase $\phi_1$ is determined by the path difference for the electron reaching contact 1 from contact 4 and vice-versa. Similarly, phase $\phi_2$ is determined by path difference for electron reaching contact 2 from contact 3 or vice-versa without scattering.}
$\theta$ is the phase acquired during, transmission via the constriction. An extra phase $\pi$/2 occurs whenever there is reflection due to the constriction. The $s$-matrix is unitary as it is. We have three setups. Setup 1 is defined by parameters such that $\mu_2 = \mu_3 = \mu_0$ and $\mu_1 = \mu_4 = \mu$ condition holds. Similarly, Setup 2 is defined by $\mu_2 = \mu_4 = \mu_0$ and $\mu_1 = \mu_3 = \mu$ and setup 3 by $\mu_3 = \mu_4 = \mu_0$ and $\mu_1 = \mu_2 = \mu$. {In all setups, $\mu - \mu_0 = eV$ and $\mu, \mu_0 \gg k_B \mathcal{T}$.}
We herein analyze the thermal noise-like, shot noise, and quantum noise correlations in each setup using the $s$-matrix given in Eq. (\ref{eq:66}).
\subsubsection{setup 1 [$\mu_1 = \mu_4 = \mu$ and $\mu_2 = \mu_3 = \mu_0$]} \label{section IIIB.1}
From the $s$ matrix, we can calculate the thermal and thermal noise-like contributions to the total quantum noise. We will examine the quantum noise cross correlation between the probes $1$ and $4$. {The quantum noise correlation $S_{14}^q$ between terminals 1 and 4 is given as,}
\begin{equation} \label{eq:67}
{S_{14}^{q} = S_{14}^{\uparrow \uparrow, q} + S_{14}^{\uparrow \downarrow, q} + S_{14}^{\downarrow \uparrow, q} + S_{14}^{\downarrow \downarrow, q},}
\end{equation}
{the individual spin polarised contributions using Eq. (\ref{eq:52}) are given as,}
\begin{equation} \label{eq:68}
\begin{split}
{S_{14}^{\sigma \sigma', q} =\frac{e^2}{h}\sum_{\rho, \rho' = \uparrow, \downarrow}\sum_{\gamma,\delta}\int dE Tr[A_{\gamma \delta}^{\rho \rho'}(1,\sigma)
A_{\delta \gamma}^{\rho' \rho}(4, \sigma')]}\\{(f_{\gamma}(E)[1-f_{\delta}(E)]+[1-f_{\gamma}(E)]f_{\delta}(E)),}
\end{split}
\end{equation}

{Eq. (\ref{eq:68}) has both the thermal noise-like contribution ($S_{14}^{\sigma \sigma',th}$) and shot noise-like contribution ($S_{14}^{\sigma \sigma',sh}$)}.
The {total thermal noise-like} {contribution} to the quantum noise {correlation ($S_{14}^{q}$)} is given by,
\begin{equation} \label{eq:69}
{S_{14}^{th} = S_{14}^{\uparrow \uparrow, th} + S_{14}^{\uparrow \downarrow, th} + S_{14}^{\downarrow \uparrow, th} + S_{14}^{\downarrow \downarrow, th}}.
\end{equation}
We first calculate {$S_{14}^{\uparrow \uparrow, th}$ can be calculated using Eq. (\ref{eq:54}) and is given as,} 
\begin{equation} \label{eq:70}
{S_{14}^{\uparrow \uparrow, th}} = -\frac{2e^2}{h}\int dE [T_{14}^{\uparrow \uparrow}f_4(1-f_4) + T_{41}^{\uparrow \uparrow}f_1(1-f_1)].
\end{equation}
Making use of the integral $\int_{0}^{\infty}dE f_{\alpha}(1-f_{\alpha}) = k_B \mathcal{T}$ for $\mu, \mu_0 \gg k_B \mathcal{T}$, we get
\begin{equation} \label{eq:71}
{S_{14}^{\uparrow \uparrow, th}} = -\frac{2e^2}{h}k_B \mathcal{T}[T_{14}^{\uparrow \uparrow} + T_{41}^{\uparrow \uparrow}]= -\frac{2e^2}{h}k_B \mathcal{T}(1+R).
\end{equation}
Similar calculation shows, {$S_{14}^{\uparrow \downarrow, th}$} = 0, {$S_{14}^{\downarrow \uparrow, th}$ = 0,} and {$S_{14}^{\downarrow \downarrow, th} = -\frac{2e^2}{h}k_B \mathcal{T}(1+R)$}.
Thus the {thermal noise-like contribution} to the quantum noise is given by {$S_{14}^{th} = -\frac{4e^2}{h}k_B \mathcal{T}(1+R)$}. $R$ is the reflection probability for an electron to reflect from the constriction. Here, we observe that the thermal noise-like contribution is twice that of the chiral quantum Hall case.
The {shot noise-like contribution} to the quantum noise is given by,
\begin{equation} \label{eq:72}
{S_{14}^{sh} = S_{14}^{\uparrow \uparrow, sh} + S_{14}^{\uparrow \downarrow, sh} + S_{14}^{\downarrow \uparrow, sh} + S_{14}^{\downarrow \downarrow, sh}},
\end{equation}
The {shot noise-like contribution $S_{14}^{\sigma \sigma', sh}$} can be calculated from, Eq. (\ref{eq:58}),
\begin{equation} \label{eq:73}
\begin{split}
{S_{14}^{\sigma \sigma', sh}} = -\frac{2e^2}{h}\int dE \sum_{\gamma, \delta}\sum_{\rho \rho' = \uparrow, \downarrow}(f_{\gamma}-f_a)(f_{\delta}-f_b)\\\times Tr(s_{1 \gamma}^{\sigma \rho^{\dagger}}s_{1 \delta}^{\sigma \rho'}s_{4 \delta}^{\sigma \rho'^{\dagger}}s_{4 \gamma}^{\sigma \rho} ).
\end{split}
\end{equation}
 Once again, we consider the energy-dependent functions $f_a$ and $f_b$, equal to $f_0$, which is the Fermi-Dirac distribution of contacts $2$ and $3$.
Using the $s$-matrix as in Eq. (\ref{eq:66}), we get {$S_{14}^{\uparrow \uparrow, sh}$ = 0}. Similarly other {contributions} such as {$S_{14}^{\uparrow \downarrow, sh}$, $S_{14}^{\downarrow \uparrow, sh}$ and $S_{14}^{\downarrow \downarrow, sh}$} are also zero.
Thus, the {shot noise-like contribution} to the quantum noise $S_{14}^q$ vanishes. The quantum noise, therefore, is
\begin{equation} \label{eq:74}
S_{14}^q = {S_{14}^{th}} = -\frac{4e^2}{h}k_B \mathcal{T}(1+R).
\end{equation}
Like the chiral case, the total contribution to the quantum noise comes from thermal noise only. Similarly, correlations between terminals 1 and 2, 3 and 4, 2 and 3, and 1 and 4 show similar nature where only the thermal noise contributes to the quantum noise. These quantum noise correlations are given as
\begin{equation} \label{eq:75}
S_{12}^q = S_{34}^q =\frac{1-R}{1+R}S_{14}^q, \quad
S_{23}^q = S_{14}^q.
\end{equation}
We can see that the quantum noise correlation between terminals $2$ and $4$ is now finite, which was zero in the quantum Hall case. The noise correlation between these two terminals is {shot noise-like. It is a significant result since it distinguishes chiral edge modes from helical ones.}

{The shot noise-like contribution ($S_{24}^{sh}$) between terminals 2 and 4 is given as,}
 \begin{equation} \label{eq:76}
    {S_{24}^{sh} = S_{24}^{\uparrow \uparrow, sh} + S_{24}^{\uparrow \downarrow, sh} + S_{24}^{\downarrow \uparrow, sh} + S_{24}^{\downarrow \downarrow, sh}},
 \end{equation}
 where individual spin-polarised components are 
 \begin{equation} \label{eq:77}
 \begin{split}
     {S_{24}^{\sigma \sigma', sh}}  {= \langle \Delta I_{2}^{\sigma} \Delta I_{4}^{\sigma'} \rangle^{sh}}, \quad \text{{where}} \quad
{\sigma, \sigma' = \uparrow /\downarrow}\\
  {= -\frac{2e^2}{h}\int dE \sum_{\gamma, \delta}\sum_{\rho \rho' = \uparrow, \downarrow}(f_{\gamma}-f_a)(f_{\delta}-f_b)}\\{\times Tr(s_{2 \gamma}^{\sigma \rho^{\dagger}}s_{2 \delta}^{\sigma \rho'}s_{4 \delta}^{\sigma \rho'^{\dagger}}s_{4 \gamma}^{\sigma \rho})}
     \end{split}
 \end{equation}

{The component $S_{24}^{\uparrow \uparrow, sh}$ is given as,}
\begin{equation}
\begin{split}
{S_{24}^{\uparrow \uparrow, sh} = \frac{-2e^2}{h} \int dE \sum_{\gamma, \delta} \sum_{\rho, \rho' = \uparrow, \downarrow} (f_{\gamma} - f_a)(f_{\delta} - f_b)}\\ {\times Tr (S_{2 \gamma}^{\uparrow \rho \dagger} s_{2 \delta}^{\uparrow \rho' \dagger} s_{4 \delta}^{\uparrow \rho' \dagger}s_{4 \gamma}^{\uparrow \rho})}.
\end{split} 
\end{equation} 

{Each term in the summation is zero, which can be seen from $s$-matrix as in Eq.~(\ref{eq:66}). Similarly, we can calculate other spin-polarized components. $S_{24}^{\downarrow \downarrow sh}$ is given as, }
\begin{equation}
\begin{split}
{S_{24}^{\downarrow \downarrow, sh} = \frac{-2e^2}{h} \int dE \sum_{\gamma, \delta} \sum_{\rho, \rho' = \uparrow, \downarrow} (f_{\gamma} - f_a)(f_{\delta} - f_b)}\\ {\times Tr (S_{2 \gamma}^{\downarrow \rho \dagger} s_{2 \delta}^{\downarrow \rho' \dagger} s_{4 \delta}^{\downarrow \rho' \dagger}s_{4 \gamma}^{\downarrow \rho})}.
\end{split} 
\end{equation} 

{The summation in the above equation survives only when $\gamma, \delta = 1,3$. Now, using the appropriate $s$-matrix elements as in Eq.~(\ref{eq:66}), we get,}
\begin{equation} \label{eq:78}
 \begin{split}
     {S_{24}^{\downarrow \downarrow, sh}} = {-\frac{2e^2}{h}RT \bigg(eV\coth \left[\frac{eV}{2k_B \mathcal{T}}\right] - 2k_B \mathcal{T}\bigg)}.
       \end{split}
 \end{equation}

{The other components such as $S_{24}^{\uparrow \downarrow}$ and $S_{24}^{\downarrow \uparrow}$ are zero and may become significant in the presence of spin-flip scattering. For helical edge modes, we see that the shot noise-like contribution $S_{24}^{sh}$ arises due to edge modes via spin-down electrons and this is a result of both constriction and the applied voltage bias. In absence of the constriction, i.e., $R = 0, T =1$, or at zero applied voltage bias, i.e., $eV \rightarrow 0$, the shot-noise vanishes.}
 {The thermal noise-like contribution ($S_{24}^{th}$) is given as,}
 \begin{equation} \label{eq:79}
     {S_{24}^{th} = S_{24}^{\uparrow \uparrow, th} + S_{24}^{\uparrow \downarrow, th} + S_{24}^{\downarrow \uparrow, th} + S_{24}^{\downarrow \downarrow, th}} .
 \end{equation}
We first calculate {$S_{24}^{\uparrow \uparrow, th}$ can be calculated using Eq. (\ref{eq:54}) and is given as,} 
\begin{equation} \label{eq:80}
{S_{24}^{\uparrow \uparrow, th}} = -\frac{2e^2}{h}\int dE [T_{24}^{\uparrow \uparrow}f_4(1-f_4) + T_{42}^{\uparrow \uparrow}f_1(1-f_1)].
\end{equation}
{Since there is no direct tunneling for an electron to tunnel from contact 2 to 4 is zero and vice-versa, we have $T_{24}^{\uparrow \uparrow}$ and $T_{42}^{\uparrow \uparrow}$ is zero, which makes $S_{24}^{\uparrow \uparrow, th}$ is zero. Similarly, the other spin-polarised components are zero.     So, the total thermal noise-like contribution is given as,}
\begin{equation} \label{eq:81}
    {S_{24}^{th} = 0}
\end{equation}

 {The total quantum noise correlation $S_{24}^q$ is only shot-noise like, i.e.,}
 \begin{equation} \label{eq:82}
 {S_{24}^q = S_{24}^{sh} = -\frac{2e^2}{h}RT \bigg(eV\coth \left[\frac{eV}{2k_B \mathcal{T}}\right] - 2k_B \mathcal{T}\bigg)}.
 \end{equation}

\begin{widetext}

\begin{figure}[h!]
\includegraphics[scale=0.60]{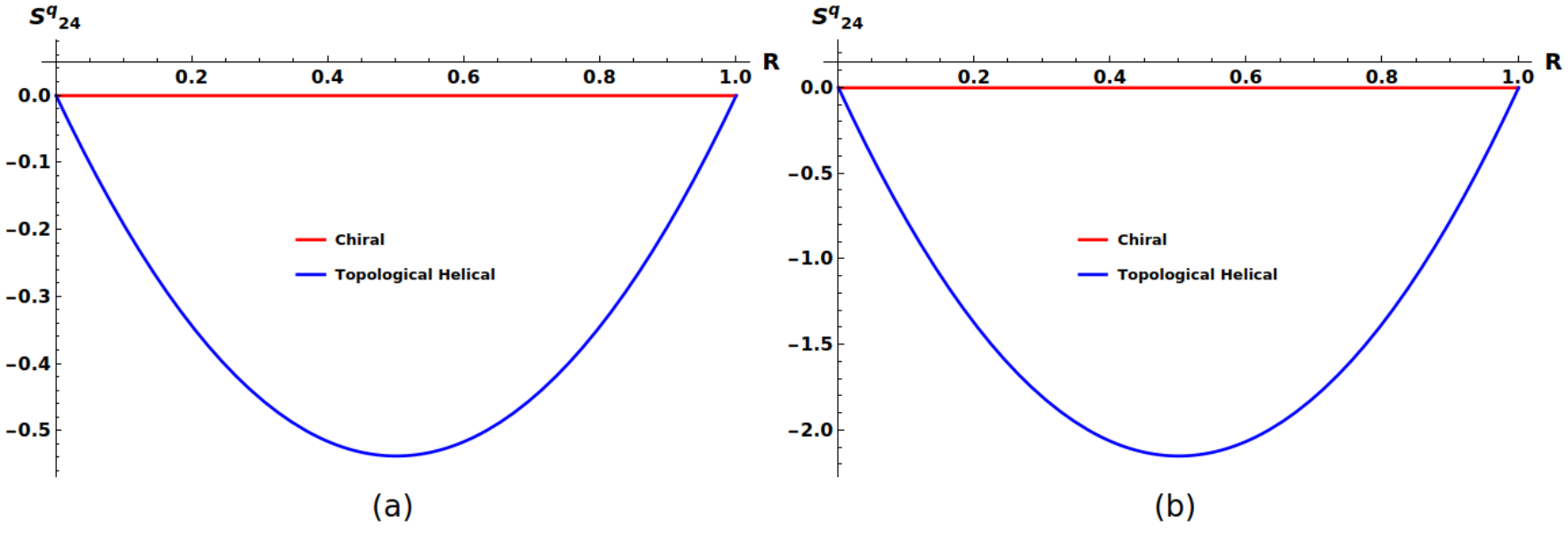}

\caption{(a) Behaviour of quantum noise cross correlation between terminals $2$ and $4$ in setup 1 (Results are identical in setup 3 as well), and (b) quantum noise cross correlation between terminals $2$ and $4$ in setup 2 in units of $\frac{2e^2}{h} k_B \mathcal{T}$ versus $R$ with the voltage bias ($eV = 4$$k_B \mathcal{T}$) with $\mu_0 = 100 k_B \mathcal{T}$ . This is in the regime $eV \gg k_B \mathcal{T}$.}
\label{fig:3}
\end{figure}
\end{widetext}

\begin{widetext}

\begin{figure}[h!]
\includegraphics[scale=0.60]{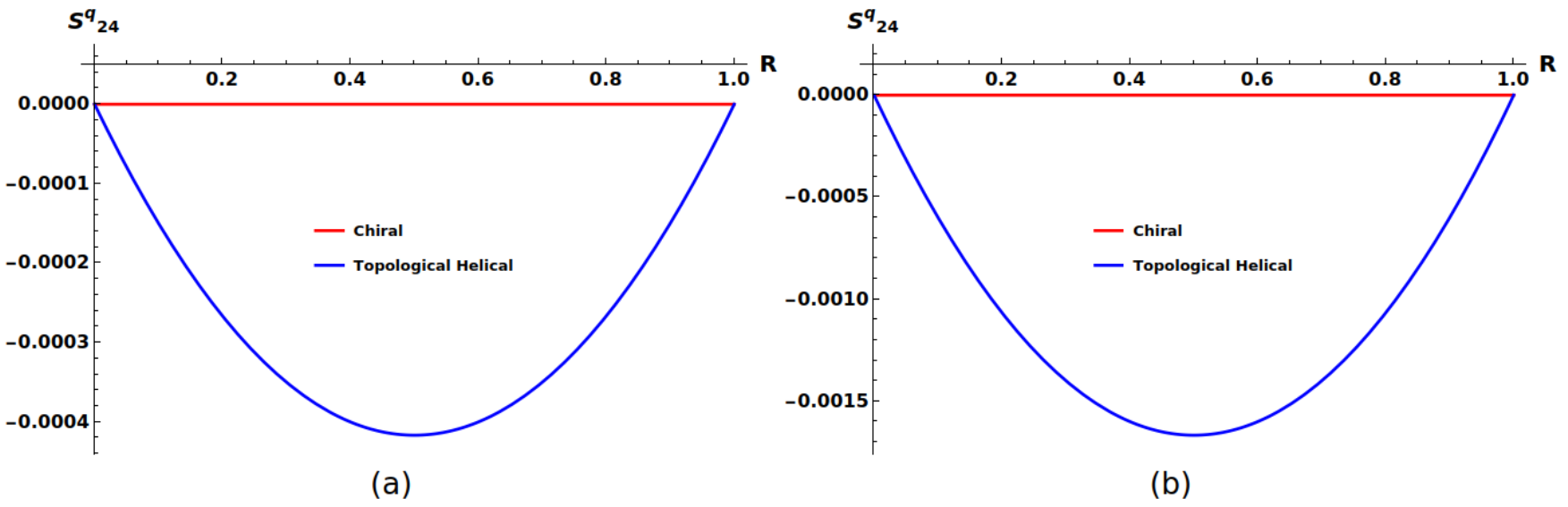}

\caption{(a) Behaviour of quantum noise cross correlation between terminals $2$ and $4$ in setup 1 (Results are identical in setup 3 as well), and (b) quantum noise cross correlation between terminals $2$ and $4$ in setup 2 in units of $\frac{2e^2}{h} k_B \mathcal{T}$ versus $R$ with the voltage bias ($eV = 0.01$$k_B \mathcal{T}$) with $\mu_0 = 100 k_B \mathcal{T}$ . This is in the regime $eV \ll k_B \mathcal{T}$.}
\label{fig:4}
\end{figure}
\end{widetext}

{Here, we observe that the shot noise-like contribution is non-zero for the helical edge mode transport, while it is zero for the chiral edge mode transport. This gives the distinction between chiral and helical edge modes. }
{Here, we have two distinct regimes. In the regime $eV \gg k_B \mathcal{T}$, the quantum noise correlation ($S_{24}^q$) becomes}
\begin{equation} \label{eq:83}
{S_{24}^q = -\frac{2e^2}{h}RT (eV - 2k_B \mathcal{T}) ,}
\end{equation}
{In this regime, the chiral and topological helical edge modes can be distinguished and has been shown in Fig. \ref{fig:3}(a)} {and in the regime $k_B \mathcal{T} \gg eV$, $S_{24}^q$ is very small but still helps in distinguish chiral and topological helical edge modes as shown in Fig. \ref{fig:4}(a).}

\begin{widetext}

\begin{figure}[h!]
\includegraphics[scale=0.60]{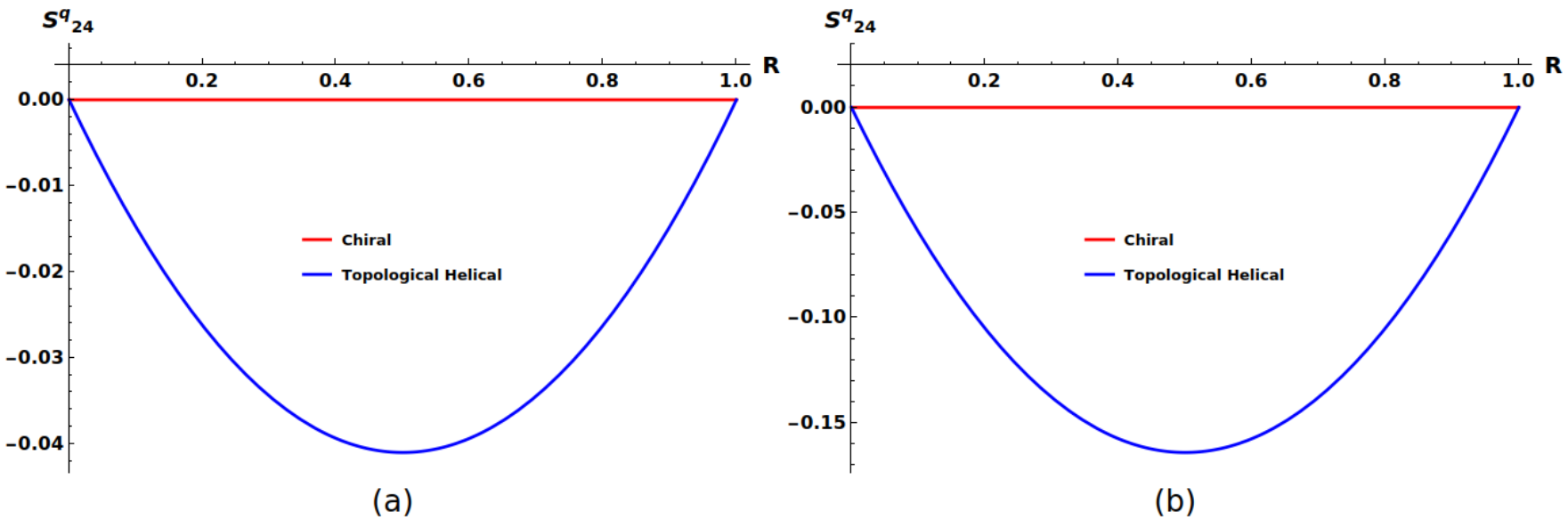}

\caption{(a) Behaviour of quantum noise cross correlation between terminals $2$ and $4$ in setup 1 (Results are identical in setup 3 as well), and (b) quantum noise cross correlation between terminals $2$ and $4$ in setup 2 in units of $\frac{2e^2}{h} k_B \mathcal{T}$ versus $R$ with the voltage bias ($eV = $$k_B \mathcal{T}$) with $\mu_0 = 100 k_B \mathcal{T}$ . This is in the regime $eV = k_B \mathcal{T}$.}
\label{fig:5}
\end{figure}
\end{widetext}

{In the regime $eV = k_B \mathcal{T}$ also, we can see the distinction between these two edge modes., which has been shown in Fig. \ref{fig:5}(a). In this regime, $S_{24}^q$ is given as,}
\begin{equation} \label{eq:84}
    S_{24}^q = \frac{-2e^2}{h}RT\left(\coth \left[\frac{1}{2}\right]-2\right)k_B \mathcal{T}.
\end{equation}

The quantum noise correlation between terminals 1 and 3 is only due to the {shot noise-like contribution} since the thermal noise-like contribution vanishes, which is the same for the chiral case. {The reason behind $S_{13}^q$ being only shot noise-like can be understood in a similar way one understands why $S_{24}^q$ is shot noise-like. }

The quantum noise correlation between terminals $1$ and $3$ gives us,
\begin{equation} \label{eq:85}
S_{13}^q = {S_{13}^{sh}} = -\frac{2e^2}{h}RT \bigg(eV\coth \left[\frac{eV}{2k_B \mathcal{T}}\right]-2k_B \mathcal{T}\bigg).
\end{equation}

Further, we can calculate the autocorrelations as well. The autocorrelation in terminal 1 is,
\begin{equation} \label{eq:86}
S_{11}^q = S_{11}^{\uparrow \uparrow, q}+S_{11}^{\uparrow \downarrow, q}+S_{11}^{\downarrow \uparrow, q} + S_{11}^{\downarrow \downarrow, q}.
\end{equation}
The {thermal noise-like contributions} to $S_{11}^{\uparrow \uparrow, q}$ can be calculated from Eq. {(\ref{eq:62})} and is given as,
\begin{equation} \label{eq:87}
{S_{11}^{\uparrow \uparrow, th}} = \frac{4e^2}{h}\int_0^{\infty} dE f(1-f) (M_{1}-R_{11}^{\uparrow \uparrow})= \frac{4e^2}{h}\int_0^{\infty} dE f(1-f)
\end{equation}

{where, using the $s$-matrix as in Eq.~(\ref{eq:66}), one can get $R_{11}^{\uparrow \uparrow} = |s_{11}^{\uparrow \uparrow}|^2 = 0$ and the number of edge modes ($M_1$) is 1 due to spin-up electrons in terminal 1.}

Similarly, the {shot noise-like contribution} can be derived from, Eq. {(\ref{eq:64})} and is given by,
\begin{equation} \label{eq:88}
\begin{split}
{S_{11}^{\uparrow \uparrow, sh}} = \frac{2e^2}{h}\int dE \bigg(T_{12}^{\uparrow \uparrow}(f_{0} - f)+\\f^2-\sum_{\gamma \delta}\sum_{\rho \rho' = \uparrow, \downarrow}f_{\gamma}f_{\delta}Tr(s_{1 \gamma}^{\sigma \rho^{\dagger}}s_{1 \delta}^{\sigma \rho'}s_{1 \delta}^{\sigma' \rho'^{\dagger}}s_{1 \gamma}^{\sigma' \rho})\bigg).
\end{split}
\end{equation}

{In Eq.~(\ref{eq:88}), the summation survives only if $\gamma$, $\delta$ = 2, 4 as can be seen by the $s$-matrix as in Eq. (\ref{eq:66}). $S_{11}^{\uparrow \uparrow, sh}$ is therefore,}
\begin{equation} \label{eq:170}
\begin{split}
    S_{11}^{\uparrow \uparrow, sh} = \frac{2e^2}{h}\int dE (T(f_0 - f) + f^2 \\
    - f^2 R^2 - f_0^2 T^2 - 2 f f_0 RT)
    \end{split}
\end{equation}

Adding Eqs. (\ref{eq:87}), (\ref{eq:170}) and simplifying further the quantum noise $S_{11}^q$ is therefore,
\begin{equation} \label{eq:89}
\begin{split}
S_{11}^{\uparrow \uparrow, q} = \frac{2e^2}{h}\int_0^{\infty} dE[f(1-f)+Rf(1-f)\\+Tf_0(1-f_0)+RT(f-f_0)^2],
\end{split}
\end{equation}

On integration, Eq. (\ref{eq:89}) becomes,
\begin{equation} \label{eq:90}
S_{11}^{\uparrow \uparrow, q} = \frac{4e^2}{h}k_B \mathcal{T}+\frac{2e^2}{h}RT\left(eV\coth\bigg[\frac{eV}{2k_B\mathcal{T}}\bigg]-2k_B \mathcal{T}\right).
\end{equation}
{Similarly, $S_{11}^{\downarrow \downarrow, sh}$ can also be calculated from Eqs.~(\ref{eq:62}) and (\ref{eq:64}). The thermal noise-like contribution $S_{11}^{\downarrow \downarrow, th}$ is,}
\begin{equation} \label{eq:171}
    {S_{11}^{\downarrow \downarrow, th} = \frac{4e^2}{h}\int_0^{\infty}dE f(1-f)},
\end{equation}
{and, the shot noise-like contribution is given as,}
\begin{equation}\label{eq:172}
\begin{split}
    {S_{11}^{\downarrow \downarrow, sh} = \frac{2e^2}{h}\int dE \bigg(T_{12}^{\downarrow \downarrow}(f_0 - f) +} \\ {f^2 - \sum_{\gamma \delta} \sum_{\rho, \rho' = \uparrow, \downarrow}f_{\gamma}f_{\delta}Tr(s_{1\gamma}^{\downarrow \rho \dagger}s_{1 \delta}^{\downarrow \rho'}s_{1\delta}^{\downarrow \rho' \dagger}s_{1 \gamma}^{\downarrow \rho})\bigg).}
    \end{split}
\end{equation}
{Here, $T_{12}^{\downarrow \downarrow} = |s_{12}^{\downarrow \downarrow}| = 0$ from the $s$-matrix as in Eq.~(\ref{eq:66}), and performing the summation using the same $s$-matrix, $S_{11}^{\downarrow \downarrow, sh}$ goes to zero. It implies that $S_{11}^{\downarrow \downarrow, q}$ is only thermal noise-like, which is given as,} 
\begin{equation} \label{eq:91}
 S_{11}^{\downarrow \downarrow, q} = \frac{4e^2}{h}k_B\mathcal{T},
\end{equation}

{Similarly, other spin-polarized components such as $S_{11}^{\uparrow \downarrow,q}$ and $S_{11}^{\downarrow \uparrow,q}$ can also be calculated and all of them vanish since there is no spin-flip scattering. However, these may become significant in the presence of spin-flip scattering. }
Thus, the quantum noise autocorrelation in terminal 1 is, therefore,
\begin{equation} \label{eq:92}
S_{11}^q = \frac{2e^2}{h} 4k_B \mathcal{T} +\frac{2e^2}{h} RT\left(eV\coth \left[\frac{eV}{2k_B \mathcal{T}}\right]-2k_B \mathcal{T}\right).
\end{equation}

{In the regime $eV \gg k_B \mathcal{T}$, $S_{11}^q = \frac{2e^2}{h}RT (eV - 2k_B \mathcal{T} )$, in $eV \ll k_B \mathcal{T}$, $S_{11}^q = \frac{2e^2}{h}4k_B \mathcal{T}$ and in the regime, $eV = k_B \mathcal{T}, S_{11}^q = \frac{2e^2}{h}4k_B \mathcal{T} + \frac{2e^2}{h}RT 0.16 k_B \mathcal{T}$. While in the regime $eV \ll k_B \mathcal{T}$, the quantum noise autocorrelation is thermal noise-like and it exactly matches that of the chiral result, which does not help in distinguishing these two edge modes. }

\begin{widetext}

\begin{figure}[h!]
\includegraphics[scale=0.60]{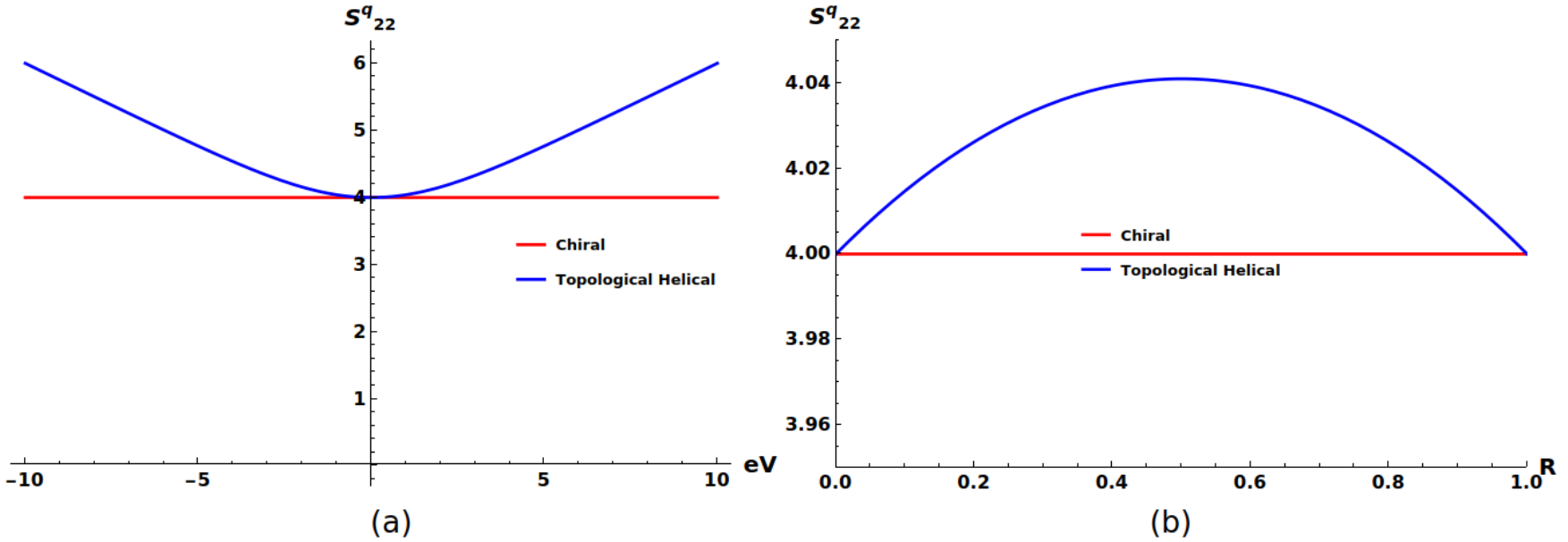}

\caption{(a) Quantum noise autocorrelation $S_{22}^q = S_{44}^q$ in setup 1 (Results are identical in setup 3 as well) in units of $\frac{2e^2}{h}k_B \mathcal{T}$ versus the voltage bias ($eV$) in units of $k_B \mathcal{T}$ with $R=0.5$ and $\mu_0 = 100 k_B \mathcal{T}$. (b) Quantum noise autocorrelation $S_{22}^q = S_{44}^q$ in setup 1 (Results are identical in setup 3 as well) in units of $\frac{2e^2}{h}k_B \mathcal{T}$ versus reflection probability ($R$) in units of $k_B \mathcal{T}$ with $eV = k_B \mathcal{T}$ and $\mu_0 = 100 k_B \mathcal{T}$. This is in the regime $eV = k_B \mathcal{T}$.}
\label{fig:6}
\end{figure}
\end{widetext}

{We can also calculate the other autocorrelations by a similar method used to calculate $S_{11}^q$ and we find all of them to be same, i.e.,}
\begin{equation} \label{eq:93}
\begin{split}
S_{11}^q = S_{22}^q = S_{33}^q = S_{44}^q.
\end{split}
\end{equation}
which differentiates helical from the chiral case. For chiral,  $S_{11}^q = S_{33}^q \ne S_{22}^q = S_{44}^q$. Again, we observe another profound difference. For chiral edge modes, the quantum noise $S_{22}^q = S_{44}^q$ is pure {thermal noise-like}, but for helical edge modes, {shot noise-like contribution} also contributes to autocorrelation. This distinguishes chiral from helical edge modes, {which is shown in Figs.~\ref{fig:6}(a) and (b) in the regimes $eV\gg k_B \mathcal{T}$ and $eV = k_B \mathcal{T}$ respectively.}

\subsubsection{setup 2 [$\mu_1 = \mu_3 = \mu$ and $\mu_2 = \mu_4 = \mu_0$]} \label{section IIIB.2}
As explained earlier, the {thermal noise-like contributions} does not change for different setups {as explained in Eqs. (\ref{eq:56}) and (\ref{eq:65}) in Sec. \ref{section III.A}}. The quantum noise cross correlation between different terminals is,
\begin{align}
S_{14}^q &= {S_{14}^{th}} = S_{23}^q = {S_{23}^{th}} = -\frac{4e^2}{h}k_B \mathcal{T}(1+R), \label{eq:94}\\
S_{12}^q &= S_{34}^q = \frac{1-R}{1+R}S_{14}^q, \label{eq:95}\\ S_{13}^q &= 0, \label{eq:96}\\
S_{24}^{q} &= {S_{24}^{sh}} = -\frac{2e^2}{h}4RT \left(eV\coth \left[\frac{eV}{2k_B \mathcal{T}}\right]-2k_B \mathcal{T}\right). \label{eq:97}
\end{align}
Above Eqs. (\ref{eq:94}), (\ref{eq:95}), (\ref{eq:96}) and (\ref{eq:97}) imply {$S_{14}^{sh} = S_{23}^{sh} = S_{12}^{sh} = S_{34}^{sh}$ = 0 and $S_{24}^{th} = 0$.} {The reason behind shot noise-like contribution vanishing can be understood using the $s$-matrix as in Eq.~(\ref{eq:66}) as vanishing s-matrix elements result in vanishing shot noise. } In this setup {also, one can see that $S_{24}^q$ distinguishes between chiral and helical edge modes since in the chiral case it is zero. This can also be understood using $s$-matrix given in Eq.~(\ref{eq:66})}. The autocorrelation for quantum noise in this setup is only {thermal noise-like}, i.e.,
\begin{equation} \label{eq:98}
\begin{split}
{S_{ii}^q = S_{ii}^{th} = \frac{2e^2}{h}4k_B \mathcal{T}}, \quad i \in \{1,2,3,4\}
\end{split}
\end{equation}
Here the {shot noise-like contributions} to each quantum noise correlation vanish, i.e., {$S_{ii}^{sh} = 0$} for $i \in \{1,2,3,4\}$.
{This can also be explained using the $s$-matrix as in Eq.~(\ref{eq:66}). We can calculate $S_{11}^q$ as follows. The $S_{11}^{\sigma \sigma', th}$ components can be calculated using Eq.~(\ref{eq:62}) and they are given as,}
\begin{equation}
    {S_{11}^{\uparrow \uparrow, th} = S_{11}^{\downarrow \downarrow, th} = \frac{4e^2}{h}\int_0^{\infty} dE f(1-f).}
\end{equation}
{After integrating, we get the thermal noise-like contribution as,}
\begin{equation}
    {S_{11}^{th} = S_{11}^{\uparrow \uparrow, th} + S_{11}^{\downarrow \downarrow, th} = \frac{2e^2}{h}4 k_B \mathcal{T}}.
\end{equation}

{Similarly, the shot noise-like contribution can be calculated. Using Eq.~(\ref{eq:64}), we can calculate the spin-polarized components. $S_{11}^{\uparrow \uparrow, sh}$ is given as,}
\begin{equation}
    \begin{split}
       {S_{11}^{\uparrow \uparrow, sh} = \frac{2e^2}{h}\int_0^{\infty} dE \bigg(f_0(1-f_0) - f(1-f) \bigg) = 0.}
    \end{split}
\end{equation}
{where, we used Eq.~(\ref{eq:55}) in the limit $\mu, \mu_0 \gg k_B \mathcal{T}$. Similarly, $S_{11}^{\downarrow \downarrow, sh}$ is given as,}
\begin{equation}
    {S_{11}^{\downarrow \downarrow, sh} = \frac{2e^2}{h}\int_0^{\infty} dE \bigg(f_0(1-f_0) - f(1-f) \bigg) = 0.}
\end{equation}
{The other spin-polarized components such as $S_{11}^{\uparrow \downarrow, sh}$ and $S_{11}^{\downarrow \uparrow, sh}$ are zero which might become significant in the presence of spin-flip scattering. Therefore, the total shot noise-like contribution $S_{11}^{sh}$ is zero and $S_{11}^q$ is only thermal noise-like. Using the same method used for $S_{11}^q$, we can calculate $S_{22}^q$, $S_{33}^q$ and $S_{44}^q$ and find them all to be thermal noise-like. Here, the autocorrelations are also thermal noise-like and therefore do not provide any distinction between chiral and topological helical edge modes in setup-2.}
\\

\subsubsection{setup 3 [$\mu_1 = \mu_2 = \mu$ and $\mu_3 = \mu_4 = \mu_0$]} \label{section IIIB.3}
For this setup, again the {thermal noise-like contributions} does not change, but the {shot noise-like contributions} to the quantum noise cross-correlation $S_{13}^q$, $S_{24}^q$ are non-zero now. We see distinctive results for chiral and helical edge modes {via $S_{24}^q$} for this setup. The quantum noise cross-correlation between different terminals {can be calculated using Eqs. (\ref{eq: 53}), (\ref{eq:54}), (\ref{eq:57}), (\ref{eq:58}) and they are given as,} 
\begin{equation} \label{eq:99}
\begin{split}
 {S_{14}^q = S_{14}^{th}= S_{23}^q = S_{23}^{th} = -\frac{4e^2}{h}k_B \mathcal{T}(1+R),}\\
{S_{12}^q = S_{12}^{th} = S_{34}^q = S_{34}^{th} = \frac{1-R}{1+R}S_{14}^q,}\\
{S_{13}^{q} = S_{13}^{sh} = S_{24}^q = S_{24}^{sh} = -\frac{2e^2}{h}RT \bigg(eV}{\coth \left[\frac{eV}{2k_B \mathcal{T}}\right]} \\{-2k_B \mathcal{T}\bigg)}.
\end{split}
\end{equation}

Here, Eq. (\ref{eq:99}) implies that
\begin{equation} \label{eq:100}
\begin{split}
{S_{14}^{sh} = S_{23}^{sh} = S_{12}^{sh} = S_{34}^{sh} = 0,}\quad
{S_{13}^{th} = S_{24}^{th} = 0}.
\end{split}
\end{equation}
Similarly, the quantum noise autocorrelations can be derived using Eqs. (\ref{eq:61}), (\ref{eq:62}), (\ref{eq:63}) and (\ref{eq:64}) and they are given as,
\begin{equation} \label{eq:101}
\begin{split}
S_{11}^q = S_{22}^q = \frac{2e^2}{h}\int dE(4f(1-f)+RT(f-f_0)^2),\\
S_{33}^q = S_{44}^q = \frac{2e^2}{h}\int dE(4f_0(1-f_0)+RT(f-f_0)^2).
\end{split}
\end{equation}

which, on integrating gives
\begin{equation} \label{eq:102}
\begin{split}
S_{ii}^q=\frac{2e^2}{h}4k_B \mathcal{T} +\frac{2e^2}{h}RT\bigg(eV\coth\left[\frac{eV}{2k_B \mathcal{T}}\right]-2k_B\mathcal{T}\bigg),\\
i \in \{1,2,3,4\}.
\end{split}
\end{equation}
{Here, the autocorrelations are combinations of equilibrium (thermal noise) and transport (shot noise) fluctuations. The reason behind this behaviour can be  explained in a similar fashion as was done for setup 1 below Eq.~(\ref{eq:86}).}
Here {Here, we have three distinct regimes: $eV \gg k_B \mathcal{T}$, $eV \ll k_B \mathcal{T}$ and $eV = k_B \mathcal{T}$. For $eV \gg k_B \mathcal{T}$, we have,}

\begin{equation} \label{eq:103}
    S_{ii}^q = \frac{2e^2}{h}RT (eV - 2k_B \mathcal{T}), \quad i \in \{1,2,3,4\}.
\end{equation}

for, $eV \ll k_B \mathcal{T}$, we have,
\begin{equation} \label{eq:104}
    S_{ii}^q = \frac{2e^2}{h}4k_B\mathcal{T}, \quad i \in \{1,2,3,4\}.
\end{equation}
and, in the regime $eV = k_B \mathcal{T}$,
\begin{equation} \label{eq:105}
\begin{split}
    S_{ii}^q = \frac{2e^2}{h}4k_B \mathcal{T} + \frac{2e^2}{h}RT \left(\coth \left[\frac{1}{2}\right]-2\right)k_B \mathcal{T}, \\ i \in \{1,2,3,4\}.
    \end{split}
\end{equation}
The thermal noise-like contributions are,
\begin{equation} \label{eq:106}
{S_{ii}^{th} = \frac{2e^2}{h}4k_B\mathcal{T}, \quad i \in \{1,2,3,4\}}.
\end{equation}\\
and the shot noise-like contributions are,
\begin{equation} \label{eq:107}
\begin{split}
S_{ii}^{sh} = \frac{2e^2}{h}RT\left(eV\coth\left[\frac{eV}{2k_B \mathcal{T}}\right]-2k_B\mathcal{T}\right), \\ i \in \{1,2,3,4\}.
\end{split}
\end{equation}\\

In this setup, we also see a difference, i.e., the autocorrelations $S_{22}^q$ and $S_{44}^q$ are {thermal noise-like} for chiral edge modes and for the transport via helical edge modes, they are the sum of {thermal and shot noise-like contributions}.

Due to the existence of helical edge modes, the {shot noise-like contribution} contributes in some cases, such as $S_{24}^{q}$ for {all setups}, and $S_{22}^q$, $S_{44}^q$ in setups 1 and 3, which makes the total quantum noise results for the helical case distinct from the chiral case, as has been highlighted in Fig. \ref{fig:6}(a) and (b) in the regimes $eV \gg k_B \mathcal{T}$ and $eV = k_B \mathcal{T}$ respectively.

\subsection{Results for trivial Helical edge modes} \label{section III C}
\begin{figure}
\centering
\includegraphics[width=1.00\linewidth]{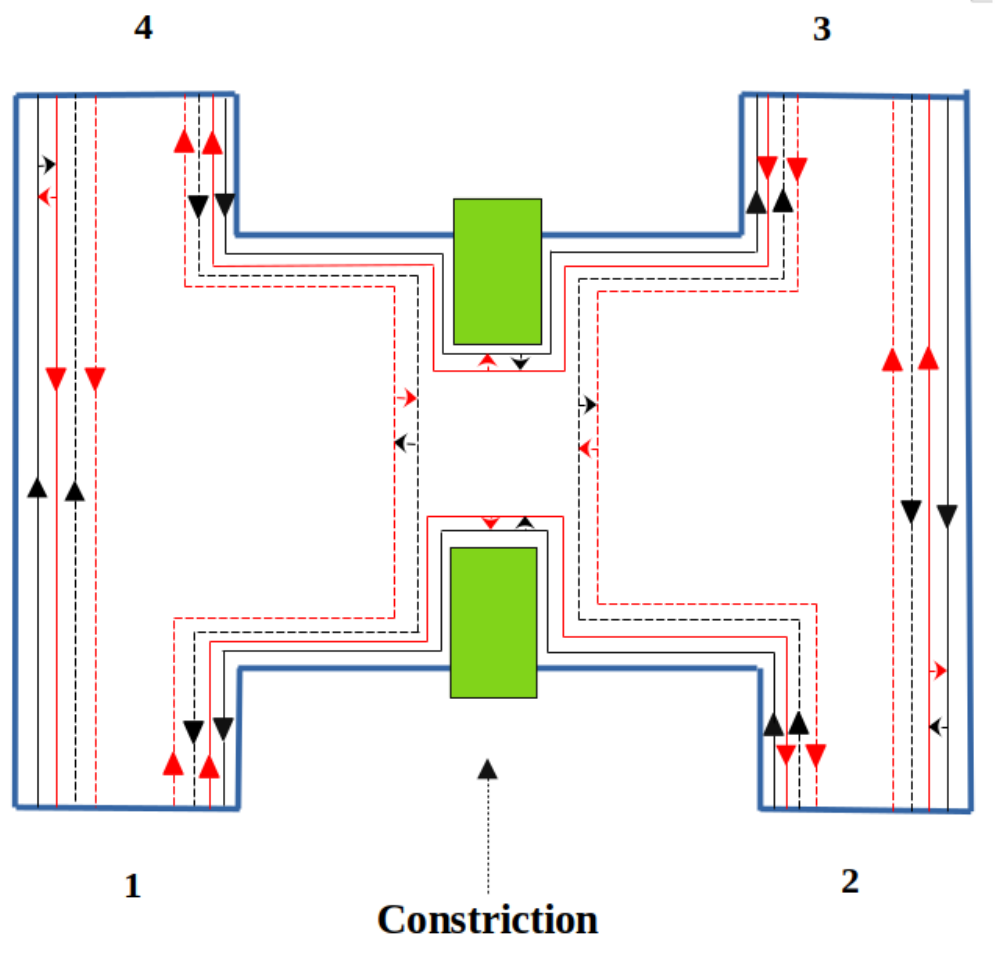}
\caption{Four terminal quantum spin Hall sample having, trivial edge modes with a constriction. The black solid and dashed line depicts the edge mode for a spin-up electron, and the red solid and dashed line depicts the edge mode for a spin-down electron. Solid lines are for the edge modes, transmitting, and dashed lines are for the edge modes reflecting via constriction.}
 \label{fig:7}
\end{figure}

We consider a trivial quantum spin Hall setup shown in Fig. \ref{fig:7}. In this case, up spin electron can get backscattered as a spin-down electron preserving spin-momentum locking \cite{Nichele_2016}. In this situation, the electron has a finite probability $p$ for a change in its direction of motion and spin with the help of intraedge scattering. In our work, we will show how the topological and trivial helical edge modes two different helical edge modes are distinguished from each other via quantum noise cross-correlations as well as autocorrelations.

Similar to chiral and topological helical setups, we can apply different voltages at any of the contacts and measure quantum noise cross and autocorrelation in the trivial QSH setup. Here, we work with three different setups similar to that used for transport via topological helical edge modes. Here, in addition to scattering from the constriction, we have spin-flip scattering, and the general $s$-matrix for this system, as shown in Fig. \ref{fig:7} is\\
\begin{widetext}
\begin{equation} \label{eq:108}
s=
\left(\begin{smallmatrix}
0 & -i\sqrt{p} & te^{i\theta}\sqrt{1-p} & 0 & 0 & 0 & -ire^{i\theta}\sqrt{1-p} & 0\\
-i\sqrt{p} & 0 & 0 & 0 & 0 & 0 & 0 & \sqrt{(1-p)}e^{i\phi_1}\\
0 & 0 & 0 & -i\sqrt{p} & \sqrt{(1-p)}e^{i\phi_2} & 0 & 0 & 0\\
0 & te^{i\theta}\sqrt{1-p} & -i\sqrt{p} & 0 & 0 & -ire^{i\theta}\sqrt{1-p} & 0 & 0\\
0 & 0 & -ire^{i\theta}\sqrt{1-p} & 0 & 0 & -i\sqrt{p} & te^{i\theta}\sqrt{1-p} & 0\\
0 & 0 & 0 & \sqrt{(1-p)}e^{i\phi_2} & -i\sqrt{p} & 0 & 0 & 0\\
\sqrt{(1-p)}e^{i\phi_1} & 0 & 0 & 0 & 0 & 0 & 0 & -i\sqrt{p}\\
0 & -ire^{i\theta}\sqrt{1-p} & 0 & 0 & 0 & te^{i\theta}\sqrt{1-p} & -i\sqrt{p} & 0
\end{smallmatrix} \right),
\end{equation}
\end{widetext}

At $p=0$, we correctly reproduce the s-matrix for topological helical edge modes as shown in Eq. ({\ref{eq:66}}). We herein analyze the thermal noise, shot noise, and quantum noise using the s-matrix given in Eq. (\ref{eq:108}).

\subsubsection{setup 1 [$\mu_1 = \mu_4 = \mu$ and $\mu_2 = \mu_3 = \mu_0$]} \label{section IIIC.1}

From the s-matrix as in Eq. (\ref{eq:108}), we can find {thermal and shot noise-like contributions} to quantum noise cross and autocorrelations. {The quantum noise correlation $S_{14}^q$ is defiined in Eq. (\ref{eq:68}) for a spin-polarized system in general}.
For this setup, the {thermal noise-like contributions} to quantum noise correlation between terminals 1 and 4 is,
\begin{equation} \label{eq:109}
{S_{14}^{th} = S_{14}^{\uparrow \uparrow, th} + S_{14}^{\uparrow \downarrow, th} + S_{14}^{\downarrow \uparrow, th} + S_{14}^{\downarrow \downarrow, th}.}
\end{equation}
{$S_{14}^{\uparrow \uparrow, th}$} from Eq. (\ref{eq:70}) is,
\begin{equation} \label{eq:110}
\begin{split}
{S_{14}^{\uparrow \uparrow, th}} = -\frac{2e^2}{h}\int dE [T_{14}^{\uparrow \uparrow}f_4(1-f_4)+T_{41}^{\uparrow \uparrow}f_1(1-f_1)],
\end{split}
\end{equation}
Making use of integration $\int_0^{\infty}dEf_{\alpha}(1-f_{\alpha}) = k_B \mathcal{T}$ for $\alpha = 1,2,3,4$ and $\mu, \mu_0\gg k_B \mathcal{T}$, we have
\begin{equation} \label{eq:111}
{S_{14}^{\uparrow \uparrow, th}} = -\frac{2e^2}{h}k_B \mathcal{T} (1+R)(1-p).
\end{equation}
Similarly, we can calculate other spin polarised components such as {$S_{14}^{\uparrow \downarrow, th}$, $S_{14}^{\downarrow \uparrow, th}$ and $S_{14}^{\downarrow \downarrow, th}$.} {We get,}
\begin{equation} \label{eq:112}
\begin{split}
    {S_{14}^{\uparrow \downarrow, th} = S_{14}^{\downarrow \uparrow, th} = 0,}\quad
    {S_{14}^{\downarrow \downarrow, th} = S_{14}^{\uparrow \uparrow, th}}
    \end{split}
\end{equation}
The {thermal noise-like contributions} $S_{14}^{th}$ is thus,
\begin{equation} \label{eq:113}
{S_{14}^{th}} = -\frac{4e^2}{h} k_B \mathcal{T}(1+R)(1-p).
\end{equation}

{The thermal noise-like contribution here comes as a result of direct tunneling of electrons from terminal 1 to 4 and viceversa.}

To calculate {shot noise-like contribution} to quantum noise, from, Eq. (\ref{eq:64}), the spin polarised {shot noise-like contribution} to the quantum noise $S_{14}^{q}$ is,
\begin{equation} \label{eq:114}
\begin{split}
{S_{14}^{\sigma \sigma', sh}} = -\frac{2e^2}{h}\int dE \sum_{\gamma, \delta=1,4}\sum_{\rho \rho'=\uparrow,\downarrow}(f_{\gamma}-f_0)(f_{\delta}-f_0)\\\times Tr(s_{1\gamma}^{\sigma \rho^{\dagger}}s_{1\delta}^{\sigma \rho'} s_{4\delta}^{\sigma' \rho'^{\dagger}}s_{4\gamma}^{\sigma' \rho}).
\end{split}
\end{equation}

We have again considered the energy-dependent functions $f_a$ and $f_b$ to be $f_0$, as was shown previously in Sec. \ref{section IIB}.

After calculating the above correlation, we get,
\begin{align}
{S_{14}^{\uparrow \uparrow, sh}} & {= S_{14}^{\downarrow \downarrow, sh} = 0, \label{eq:115}}\\
{S_{14}^{\downarrow \uparrow, sh}} & {= -\frac{2e^2}{h}4p(1-p)\int dE (f-f_0)^2, \label{eq:116}}\\
{S_{14}^{\uparrow \downarrow, sh}} & {= (1-T) S_{14}^{\downarrow \uparrow, sh}}. \label{eq:117}
\end{align}
Using
$\int_0^{\infty}dE(f-f_0)^2 = eV\coth\left[\frac{eV}{2k_BT}\right]-2k_B \mathcal{T}$, we get
the total quantum noise correlation between terminals 1 and 4, $S_{14}^q$ = {$S_{14}^{th} + S_{14}^{sh}$} to be,
\begin{equation} \label{eq:118}
\begin{split}
S_{14}^q = \frac{-2e^2}{h}(1-p)(1+R)\bigg(2 k_B \mathcal{T} + 4p\\\times\bigg(eV \coth \left[\frac{eV}{2k_B \mathcal{T}}\right]-2k_B \mathcal{T}\bigg)\bigg).
\end{split}
\end{equation}

{Here, we can see that as a consequence of spin-flip scattering, shot noise contribution is finite for noise correlation $S_{14}^q$. Here, the spin-flipper responsible for spin-flip scattering, along the edges, acts as a virtual constriction with reflection and transmission probabilities $p$ and $1-p$ respectively. The shot noise vanishes if there is no spin-flipper i.e., $p = 0$ or applied voltage bias is zero.}
\begin{figure}[h!]
\includegraphics[scale=0.50]{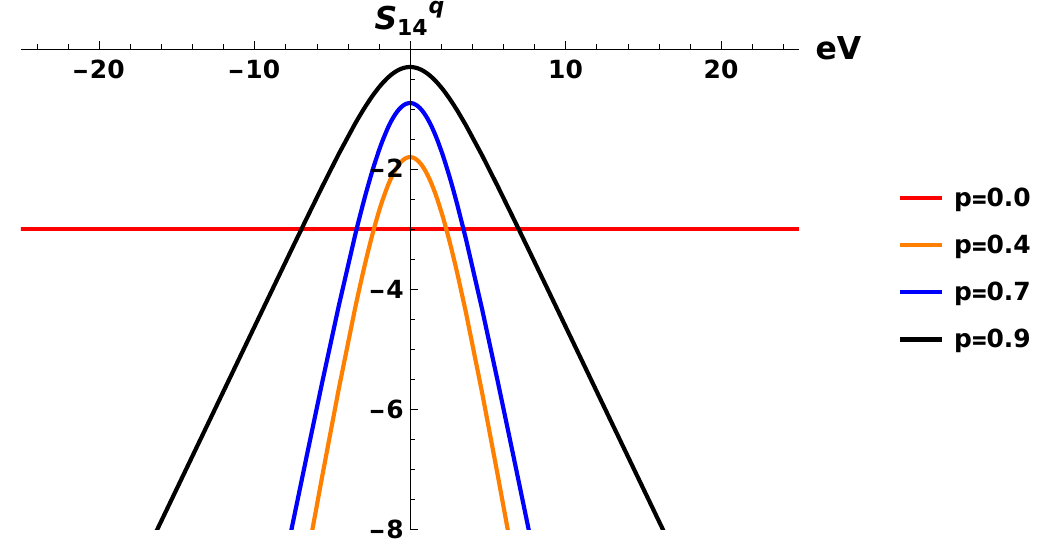}
\caption{Quantum noise $S_{14}^{q}$ in units of $\frac{2e^2}{h} k_B \mathcal{T}$ for setup 1 as a function of applied bias voltage ($eV$) in units of $k_B \mathcal{T}$ at $\mu_0 = 100k_B \mathcal{T}$ and $R=0.5$.}
\label{fig:8}
\end{figure}

From, Eq. {(\ref{eq:118})}, the distinction between the topological and trivial helical edge modes is shown in Fig. \ref{fig:8}. For topological helical edge modes ($p = 0$), the quantum noise is constant. In contrast, quantum noise for, trivial helical edge modes with different spin-flip probabilities changes symmetrically with respect to the applied voltage $eV$.

{There are three different regimes, i.e., $eV \gg k_B \mathcal{T}$, $eV \ll k_B \mathcal{T}$ and $eV = k_B \mathcal{T}$. For, $eV \gg k_B \mathcal{T}$, we get,}
\begin{equation} \label{eq:119}
{S_{14}^q = -\frac{-2e^2}{h}(1-p)(1+R)\bigg(2 k_B \mathcal{T} + 4p (eV - 2k_B \mathcal{T})\bigg).}
\end{equation}

{In this regime, we can easily distinguish between the trivial and topological helical edge modes, where shot noise dominates over thermal noise, which has been highlighted in Fig. (\ref{fig:8}). In the regime $eV \ll k_B \mathcal{T}$, we get,} 
\begin{equation} \label{eq:120}
    S_{14}^q = \frac{-2e^2}{h}2(1-p)(1+R)k_B \mathcal{T}.
\end{equation}
{In this regime as well, we can see a difference in topological and trivial helical edge modes via magnitude, where the thermal noise dominates over shot noise. The quantum noise in this regime is maximum for the transport via topological helical edge modes, whereas it reduces for the transport via trivial helical edge modes. This is a very important result. It explains how the distinction can be made even if there is no applied bias voltage across the setup.}

{For, $eV = k_B \mathcal{T}$, we get,}
\begin{equation} \label{eq:121}
\begin{split}
S_{14}^q = \frac{-2e^2}{h}(1-p)(1+R)\bigg(2  + 0.64 p\bigg)k_B \mathcal{T}.
\end{split}
\end{equation}
Here, in this regime as well, we can see the distinction between topological and trivial helical edge modes via magnitude.

Similarly, other quantum noise cross-correlations are calculated using general definitions as in Eqs. (\ref{eq: 53}), (\ref{eq:54}), (\ref{eq:63}) and (\ref{eq:64}). They are given as,
\begin{equation} \label{eq:122}
\begin{split}
S_{12}^q = S_{34}^q = -\frac{2e^2}{h}(1-R)(1-p)\\\times\bigg(2k_B \mathcal{T}+p\bigg(eV \coth \left[\frac{eV}{2k_B \mathcal{T}}\right]-2k_B \mathcal{T}\bigg)\bigg).
\end{split}
\end{equation}

\begin{figure}[h!]
\includegraphics[scale=0.50]{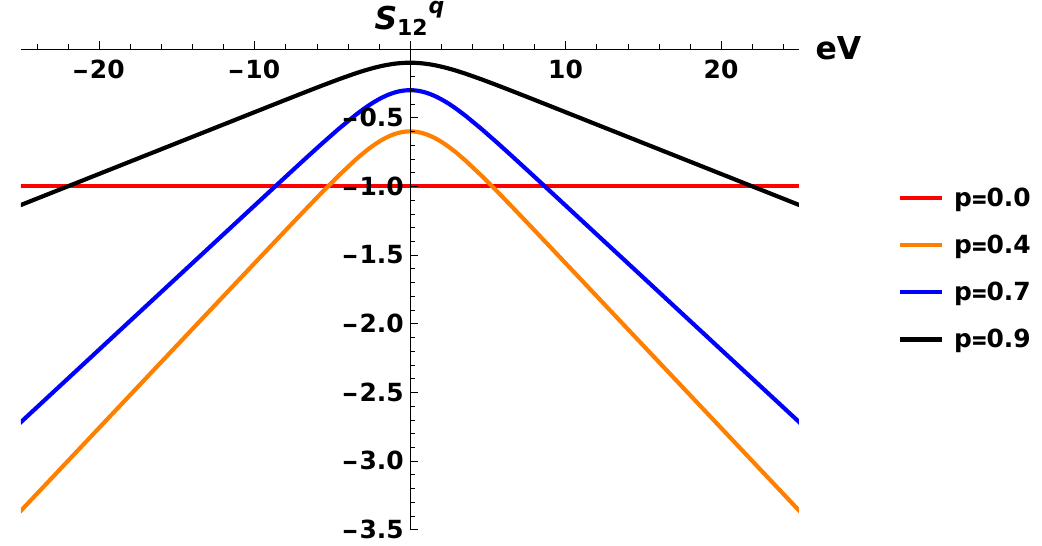}
\caption{Behaviour of quantum noise $S_{12}^{q}$ in units of $\frac{2e^2}{h} k_B \mathcal{T}$ for setup 1 as a function of applied bias voltage ($eV$) in units of $k_B \mathcal{T}$ at $\mu_0 = 100k_B \mathcal{T}$ and $R=0.5$. $S_{34}^{q}$ gives, equally similar results too.}
\label{fig:9}
\end{figure}

{The shot noise appears due to spin-flip scattering. It will be zero if either $p$ is zero or applied voltage bias is zero.}
Eq. (\ref{eq:122}) also distinguishes between topological and trivial helical edge modes, shown in Fig.~\ref{fig:9}. Here as well, we can distinguish between the topological and trivial helical edge modes. The expressions for the quantum noise correlations in these different regimes have been shown via a table in Section~\ref{section IV}.

Similarly, we can calculate the quantum noise correlation between terminals $1, 3$. We had already seen that the total contribution to this correlation comes from {shot noise-like contribution} only. The {thermal noise-like contribution} is zero. So, in the trivial case, the actual contribution comes only from {shot noise-like contribution}. The {shot noise-like contribution} to $S_{13}^q$ reduces the correlation for higher spin-flip scattering and this is given as, 
\begin{equation} \label{eq:123}
\begin{split}
S_{13}^q = {S_{13}^{sh}} = -\frac{2e^2}{h}RT(1-p)^2 \bigg(eV\coth \left[\frac{eV}{2k_B \mathcal{T}}\right]\\-2k_B \mathcal{T}\bigg).
\end{split}
\end{equation}
Similar to $S_{13}^{q}$, we see only the shot noise-like contribution to $S_{24}^q$ is finite, and it is, equal to $S_{13}^q$. However, the quantum noise correlation between terminals $2, 3$ is always thermal noise-like i.e.,
\begin{equation} \label{eq:161}
    S_{23}^q = \frac{-4e^2}{h}k_B \mathcal{T}(1+R)(1-p).
\end{equation}
{Since there is direct tunneling of electrons from terminal 2 to 3 and vice versa with probabilities $T_{23}^{\uparrow \uparrow} = T_{32}^{\downarrow \downarrow} = (1-p)$ and $T_{23}^{\downarrow \downarrow} = T_{32}^{\uparrow \uparrow} = R(1-p)$. Similarly, the shot noise-like contribution can also be calculated. Using Eq. (\ref{eq:108}), we can calculate each spin-polarized component. The general spin-polarized component $S_{23}^{\sigma \sigma'_, sh}$ is given as,}
\begin{equation}
\begin{split}
    {S_{23}^{\sigma \sigma'_, sh} = \frac{-2e^2}{h}\int dE \sum_{\gamma, \delta}\sum_{\rho, \rho' = \uparrow, \downarrow} (f_{\gamma} - f_0)(f_{\delta}-f_0)}\\ {\times Tr (s_{2 \gamma}^{\sigma \rho \dagger}s_{2 \delta}^{\sigma \rho'}s_{3 \delta}^{\sigma' \rho' \dagger}s_{3 \gamma}^{\sigma' \rho})}
    \end{split}
\end{equation}
{For the case $\sigma, \sigma' = \uparrow$, the summation only survives for $\gamma = \delta = 3$, which makes the total quantity zero since $f_3 = f_0$. Similarly, other spin-polarized components will also vanish, which makes the total shot noise-like contribution $S_{23}^{sh}$ to be zero.}
Now, we can see that the shot noise-like contribution is zero for both topological ($p = 0$) and trivial helical edge modes. In this situation, thermal noise can be used as a tool to distinguish trivial from topological helical edge modes via magnitude and this is possible due to the presence of finite temperatures only, since at zero temperature, $S_{23}^q$ vanishes identically for both topological and trivial helical edge modes, which will not help distinguish between these two helical edge modes. It shows that the quantum noise correlation can still distinguish the edge modes even if there is no applied voltage bias across the system via thermal noise-like contribution.

{The quantum noise autocorrelation in each terminal can also be found, using the definition in Eqs. (\ref{eq:61}), (\ref{eq:62}), (\ref{eq:63}), and (\ref{eq:64}). 
The quantum noise autocorrelation in terminal $1$ is $S_{11}^q$ = {$S_{11}^{th} + S_{11}^{sh}$}. Here, we provide the derivation for $S_{11}^q$.}

For setup 1, the thermal noise-like contribution in terminal 1 is,
\begin{equation} \label{eq:124}
\begin{split}
{S_{11}^{th}} = \frac{2e^2}{h}\int_0^{\infty}dE(4(1-p)f(1-f)),
\end{split}
\end{equation}
{where,}
\begin{align}
    {S_{11}^{\uparrow \uparrow, th}} = 
     {\frac{4e^2}{h} \int_0^{\infty}dE f(1-f).}
\end{align}

Similarly, other spin-polarized contributions are 
\begin{equation}
\begin{split}
   {S_{11}^{\uparrow \downarrow, th}} &= {\frac{-4e^2}{h}\int_0^{\infty}dE f(1-f)p},\\{S_{11}^{\downarrow \uparrow, th}} &= {\frac{-4e^2}{h}\int_0^{\infty}dE f(1-f)p,}
    \quad {S_{11}^{\downarrow \downarrow, th}} = 
     {\frac{4e^2}{h} \int_0^{\infty}dE f(1-f).}
    \end{split}
\end{equation}

{and, the shot noise-like contribution is given as,}
\begin{widetext}

\begin{equation} \label{eq:125}
\begin{split}
{S_{11}^{sh}} = \frac{2e^2}{h}\int_{0}^{\infty}dE (T(1-p)(f_0-f)+((1-p^2)-2p(1-p)R-(1-p)^2R^2)f^2-(2p(1-p)T+2(1-p)^2RT)ff_0\\-(1-p)^2T^2f_0^2),
\end{split}
\end{equation}
\end{widetext}

{Using above expressions, one can find quantum noise correlation $S_{11}^q$. For setup 1, the quantum noise correlation $S_{11}^q$ in the limit $\mu, \mu_0 \gg k_B \mathcal{T}$ in terminal 1 (also other autocorrelations) is,}

\begin{widetext}

\begin{equation} \label{eq:126}
\begin{split}
S_{11}^q = S_{22}^q = S_{33}^q = S_{44}^q = \frac{2e^2}{h} &\left(2(1-p)(2-pT^2-RT)+T(1-p)(1-T(1-p))\frac{eV}{k_B \mathcal{T}} \right. \\ &\left. +(2p(1-p)T+2(1-p)^2RT)\frac{eV}{k_B \mathcal{T}\left(e^{\frac{eV}{k_B \mathcal{T}}}-1\right)}\right)k_B \mathcal{T}.
\end{split}
\end{equation}
\end{widetext}

 Here, we can see that all the quantum noise autocorrelations are same in the limit $\mu, \mu_0 \gg k_B \mathcal{T}$. We had already seen that for the topological helical case ($p=0$), the quantum noise autocorrelation changes symmetrically with the applied voltage. In the helical case, both thermal and shot noise are finite. This has been explained below Eq.~(\ref{eq:86}). Similarly, the autocorrelations for trivial helical edge modes are a combination of both thermal and shot noise. They change symmetrically for different values of spin-flip probability ($p$), but they differ in magnitudes in different regimes, which has been highlighted in Table~\ref{Table 4}. Herein the shot noise-like contribution appears due to the spin-flipper being a virtual constriction with spin-flip scattering probability $p$, which means if we put $p = 0$, we will get the topological helical result as shown in Eq.~(\ref{eq:92}).

 {The quantum noise autocorrelation in terminal 1, which is the same to the autocorrelations in other terminals in the limit $\mu, \mu_0\gg k_B \mathcal{T}$. Here, we can have three distinct regimes.}

In the regime, $k_B \mathcal{T}\ll eV$, we get,
\begin{equation}\label{eq:136}
    S_{11}^q = S_{22}^q = S_{33}^q = S_{44}^q = \frac{4e^2}{h} k_B \mathcal{T} (1-p).
\end{equation}
{In this regime thermal noise dominates over shot noise and is maximum for transport due to topological helical edge modes ($p = 0$) and reduces for transport via trivial helical edge modes (for different values of $p$). It shows that the thermal noise comes to the rescue and helps in distinguishing trivial and topological helical edge modes via magnitude when the shot noise is zero. This is only possible at finite temperatures. Similarly, in other regimes such as $eV = k_B \mathcal{T}$ and $eV\gg k_B \mathcal{T}$, we can see the distinction via magnitudes, which has been highlighted in Table~\ref{Table 4}.}

\subsubsection{setup 2 [$\mu_1 = \mu_3 = \mu$ and $\mu_2 = \mu_4 = \mu_0$]} \label{section IIIC.2}
For this setup, we see both quantum noise cross and autocorrelation can distinguish between, trivial and topological helical edge modes, and they are given as,
\begingroup
 \begin{align} \label{eq:127}
\begin{split}
 S_{12}^q = S_{34}^q = \frac{-2e^2}{h}(1-R)(1-p)\\\times\bigg(2 k_B \mathcal{T} + p\bigg(eV \coth \left[\frac{eV}{2k_B \mathcal{T}}\right]-2k_B \mathcal{T}\bigg)\bigg),\\
 S_{14}^{q} = S_{23}^{q} = \frac{-2e^2}{h}(1+R)(1-p)\\\times \bigg(2 k_B \mathcal{T} + p \bigg(eV \coth \left[\frac{eV}{2k_B \mathcal{T}}\right]-2k_B \mathcal{T}\bigg)\bigg), \\
 S_{13}^q = 0,\\S_{24}^q={S_{24}^{sh}}=-\frac{2e^2}{h}4RT(1-p)^2\\ \times\bigg(eV\coth \left[\frac{eV}{2k_B\mathcal{T}}\right]-2k_B\mathcal{T}\bigg).
\end{split}
\end{align}
\endgroup

\begin{figure}[h!]
\includegraphics[scale=0.50]{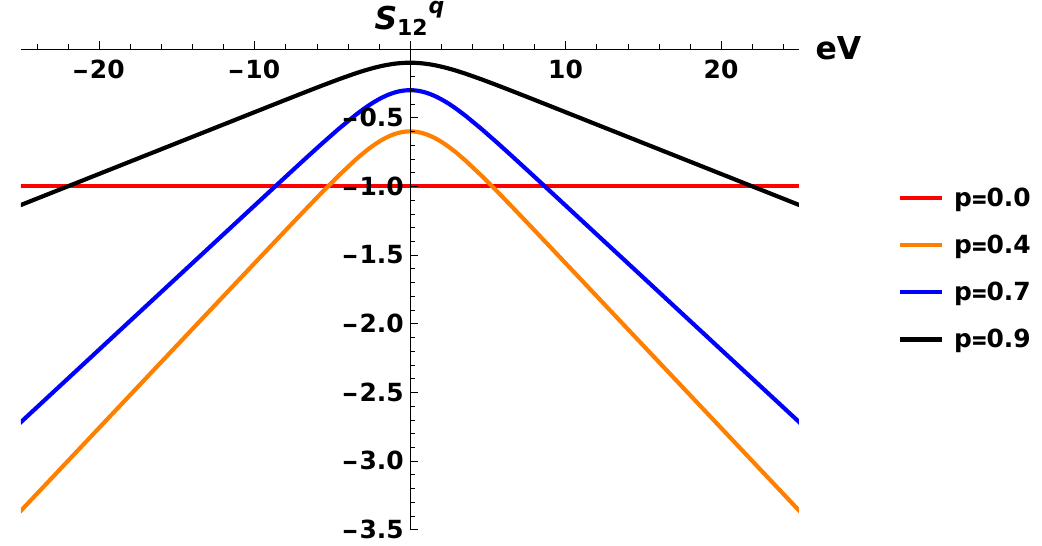}
\caption{Quantum noise $S_{12}^{q}$ in units of $\frac{2e^2}{h} k_B \mathcal{T}$ for setup 2 as a function of applied bias voltage ($eV$) in units of $k_B \mathcal{T}$ at $\mu_0 = 100k_B \mathcal{T}$ and $R=0.5$. $S_{34}^{q}$ gives similar results}
\label{fig:10}
\end{figure}

{We again have three regimes, i.e., $eV \gg k_B \mathcal{T}, eV \ll k_B \mathcal{T}$ and $eV = k_B \mathcal{T}$. For, $eV \gg k_B \mathcal{T}$, we have,}
\begin{equation} \label{eq:128}
    {S_{12}^q=S_{34}^q=\frac{-2e^2}{h}(1-R)(1-p) (2 k_B \mathcal{T}+p(eV-2k_B \mathcal{T}))}.
\end{equation}
{In this regime, shot noise dominates over thermal noise and the distinction between topological and trivial helical edge modes can be easily made, which has been shown in Fig. \ref{fig:10}. }
{For $eV \ll k_B \mathcal{T}$, we have,}
\begin{equation} \label{eq:129}
    {S_{12}^q = S_{34}^q = \frac{-4e^2}{h}(1-R)(1-p)k_B \mathcal{T},}
\end{equation}

{In this regime also, where thermal noise dominates over shot noise, we can see the distinction between topological ($p=0$) and trivial helical edge modes ($p \ne 0$) by magnitude.}
{For $eV = k_B \mathcal{T}$, we have,}
\begin{equation} \label{eq:130}
    {S_{12}^q = S_{34}^q = \frac{-2e^2}{h}(1-R)(1-p)\bigg(2  + 0.16 p\bigg)k_B \mathcal{T}}
\end{equation}

\begin{figure}[h!]
\includegraphics[scale=0.50]{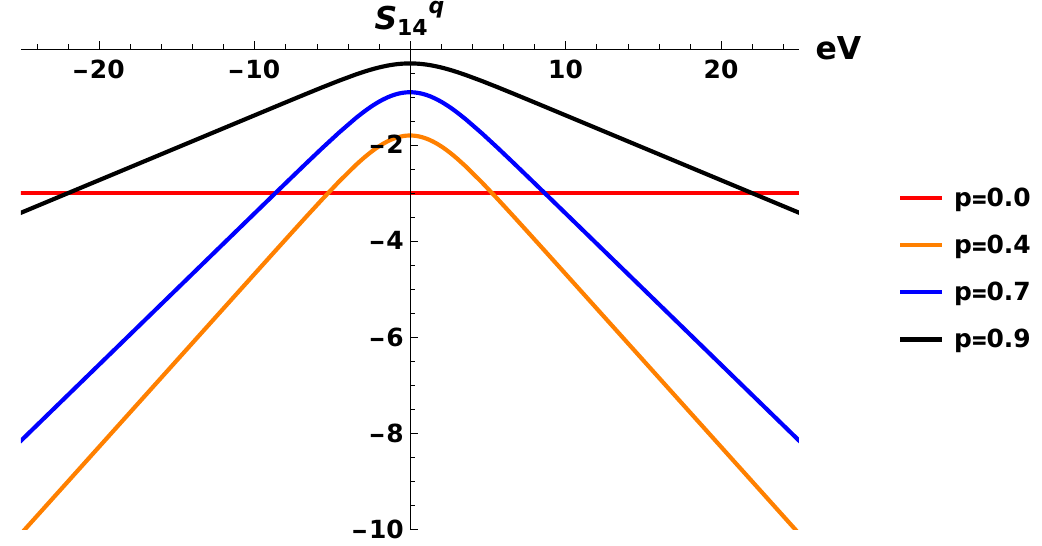}

\caption{Quantum noise $S_{14}^{q}$ in units of $\frac{2e^2}{h} k_B \mathcal{T}$ for setup 2 as a function of applied bias voltage ($eV$) in units of $k_B \mathcal{T}$ for $\mu_0 = 100k_B \mathcal{T}$ and $R=0.5$. $S_{23}^{q}$ gives similar results.}
\label{fig:11}
\end{figure}

{In this regime as well, we can see a clear difference between topological and trivial helical edge modes by magnitude.}
{Eq. (\ref{eq:127})  implies that the thermal noise-like contribution to $S_{12}^q, S_{34}^q, S_{14}^q, S_{23}^q$ are given as,}
\begin{equation} \label{eq:131}
\begin{split}
    {S_{12}^{th} = S_{34}^{th} = -\frac{2e^2}{h}(1-R)(1-p)2k_B \mathcal{T},}\\
    {S_{14}^{th} = S_{23}^{th} = \frac{(1+R)}{(1-R)}S_{12}^{th}.}
    \end{split}
\end{equation}
{and the shot noise-like contributions are,}
\begin{equation} \label{eq:132}
\begin{split}
    {S_{12}^{sh}} &= S_{34}^{sh} = -\frac{2e^2}{h}(1-R)p(1-p)\\&\times\bigg(eV\coth\left[\frac{eV}{2k_B \mathcal{T}}\right]-2k_B \mathcal{T}\bigg),\\ 
    {S_{14}^{sh}} &= S_{23}^{sh} = \frac{(1+R)}{(1-R)}S_{12}^{sh}.
    \end{split}
\end{equation}
In Fig. \ref{fig:10}, we see that for topological helical edge modes, $S_{12}^q$ is constant since it is contributed by {thermal noise} only. In contrast, for, trivial helical edge modes, i.e., for non-zero spin-flip scattering values, it changes symmetrically with the applied voltage ($eV$). It effectively distinguishes between topological and trivial helical edge modes. Similar results are obtained for other quantum noise cross-correlations such as $S_{14}^q$ and $S_{23}^q$ shown in Fig. \ref{fig:11}. {Similar to the quantum noise correlations $S_{14}^q$ and $S_{12}^q$ of the setup 1, here also we can see the distinction between the topological and trivial helical edge modes in three distinct regimes, which has been shown via tables in Section \ref{section IV}.}

{In setup 2, the thermal noise-like contribution in terminal 1 is given as,}
\begin{equation} \label{eq:133}
{S_{11}^{th} = \frac{2e^2}{h}\int_0^{\infty}dE(4(1-p)f(1-f)),}
\end{equation}

{The thermal noise-like contribution is same as that of setup as this is independent of the setup as explained in Eq.~(\ref{eq:65}).}
{The shot noise-like contribution is given as,}

\begin{equation} \label{eq:134}
\begin{split}
S_{11}^{sh} = \frac{2e^2}{h}\int_{0}^{\infty}dE (2(1-p)(f_0-f)+2(1-p^2)f^2\\-4p(1-p)ff_0-((1-p)^2T^2+2(1-p)^2RT\\+(1-p)^2(R^2+1))f_0^2).
\end{split}
\end{equation}

{ We can find the quantum noise autocorrelation using Eqs.~(\ref{eq:133}) and (\ref{eq:134}) and in the limit $\mu$, $\mu_0 \gg k_B \mathcal{T}$, this is given as,}

\begin{equation} \label{eq:135}
\begin{split}
S_{11}^q &= S_{22}^q = S_{33}^q = S_{44}^q = \\ &\left(4(1-p)^2 + 2 p(1-p) \frac{eV}{k_B \mathcal{T}}+\frac{4p(1-p) eV}{k_B \mathcal{T}(e^{\frac{eV}{k_B \mathcal{T}}}-1)}\right)k_B \mathcal{T}.
\end{split}
\end{equation}

\begin{figure}[h!]
\includegraphics[scale=0.50]{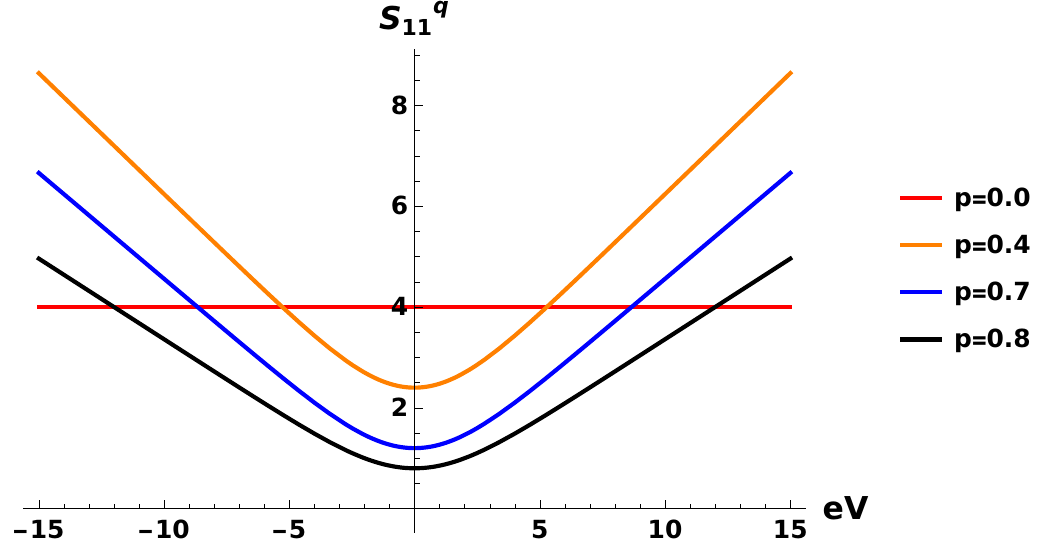}
\caption{Behaviour of autocorelation quantum noise $S_{11}^q = S_{22}^q$ = $S_{33}^q = S_{44}^q$ in units of $\frac{2e^2}{h} k_B \mathcal{T}$ for setup 2 as a function of applied voltage ($eV$) in units of $k_B \mathcal{T}$ with $R=0.5$ and $\mu_0 = 100 k_B \mathcal{T}$.}
\label{fig:12}
\end{figure}

{As we know in this setup for topological helical edge modes ($p = 0$), the quantum noise auto-correlations are only thermal noise-like, See Eq.~(\ref{eq:98}). Here, the shot noise appears as a result of spin-flip scattering due to the spin-flipper being a virtual constriction, which means if either $p = 0$ or $eV \rightarrow 0$, it vanishes. We can see the quantum noise autocorrelations helps in distinguishing trivial from topological helical edge modes as has been shown in Fig.~\ref{fig:12}.}

{The quantum noise autocorrelation in terminal 1, which is the same as the autocorrelations in other terminals in the limit $\mu, \mu_0\gg k_B \mathcal{T}$. Here, we can have three distinct regimes.}

In the regime, $k_B \mathcal{T}\ll eV$, we get,
\begin{equation}\label{eq:136}
    S_{11}^q = S_{33}^q = 4(1-p) k_B \mathcal{T}.
\end{equation}

{Here also, we see that thermal noise can help distinguish trivial and topological helical edge modes via magnitudes in the regime $eV \ll  k_B \mathcal{T}$. Similarly, in the regime, $eV \gg k_B \mathcal{T}$, shot noise dominates over thermal noise and the distinction has been shown via Fig. \ref{fig:12} and highlighted in Table \ref{Table 5}.}
{In the regime $eV = k_B \mathcal{T}$, also, we can see the distinction between the topological ($p =0$) and trivial ($p \ne 0$) helical edge modes by magnitude.}

\subsubsection{setup 3 [$\mu_1 = \mu_2 = \mu$ and $\mu_3 = \mu_4 = \mu_0$]} \label{section IIIC.3}
In this setup, the difference between the topological and trivial helical edge modes can be shown via quantum noise cross-correlations: $S_{12}^q$, $S_{14}^q$ and $S_{23}^q$. The quantum noise correlation between different terminals is given as,

\begin{equation} \label{eq:137}
\begin{split}
S_{12}^{q} = -\frac{2e^2}{h}(1-R)(1-p)\bigg(2k_B \mathcal{T}+4p\\\times\bigg[eV\coth \left[\frac{eV}{2k_B \mathcal{T}}\right]-2k_B \mathcal{T}\bigg]\bigg),\\
S_{14}^{q} = S_{23}^{q} = -\frac{2e^2}{h}(1+R)(1-p)\bigg(2k_B \mathcal{T}+p\\\times\bigg[eV\coth \left[\frac{eV}{2k_B \mathcal{T}}\right]-2k_B \mathcal{T}\bigg]\bigg),\\
S_{34}^q = -\frac{4e^2}{h}(1-R)(1-p)k_B \mathcal{T},\\
S_{13}^q = S_{24}^q = {S_{13}^{sh} = S_{24}^{sh}} = -\frac{2e^2}{h}R(1-R)(1-p)^2\\\times\bigg(eV\coth \left[\frac{eV}{2k_B \mathcal{T}}\right]-2k_B \mathcal{T}\bigg).
\end{split}
\end{equation}

{Here, thermal noise-like contributions are independent of  setups and same as that of other setups. The shot noise-like contribution comes due to spin-flip scattering meaning if either $p$ or $eV$ are zero, it vanishes.}
{In this setup, we can have three regimes, i.e., $eV \gg k_B \mathcal{T}, eV \ll k_B \mathcal{T}$ and $eV = k_B \mathcal{T}$. For, $eV \gg k_B \mathcal{T}$, we have,}
\begin{equation} \label{eq:138}
    {S_{12}^q = \frac{-2e^2}{h}(1-R)(1-p)(2k_B \mathcal{T} + 4p(eV -2k_B \mathcal{T})) }.
\end{equation}

\begin{figure}
\includegraphics[scale=0.50]{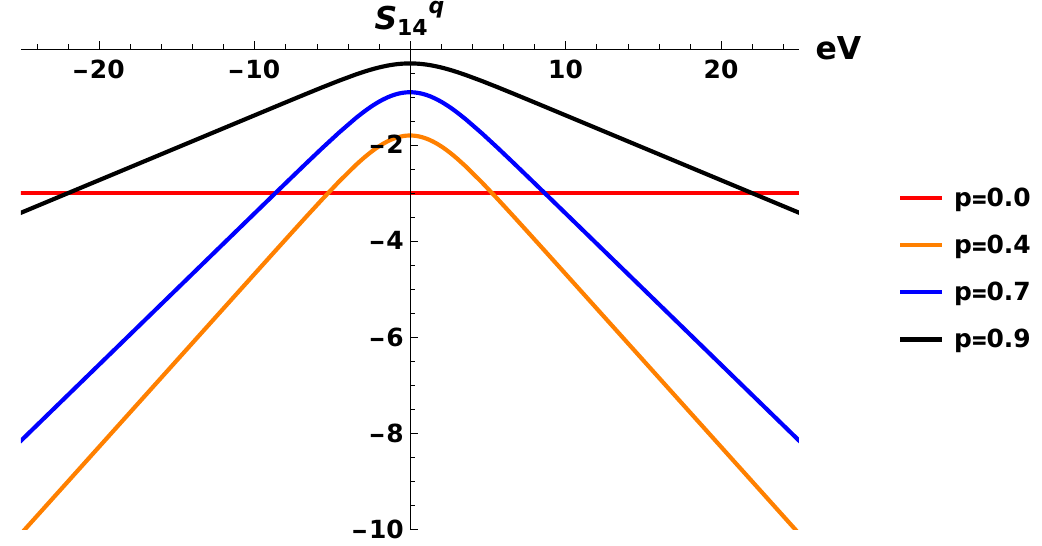}
\caption{Quantum noise $S_{14}^{q} = S_{23}^q$ in units of $\frac{2e^2}{h} k_B \mathcal{T}$ for setup 3 as a function of applied voltage bias ($eV$) in units of $k_B \mathcal{T}$ at $\mu_0 = 100k_B \mathcal{T}$ and $R=0.5$.}
\label{fig:13}
\end{figure}

{In this regime, shot noise dominates over thermal noise and the distinction between topological and trivial helical edge modes can be easily made, which has been shown in Fig.~\ref{fig:13}. }
{For $eV \ll k_B \mathcal{T}$, we have,}
\begin{equation} \label{eq:139}
    {S_{12}^q =  \frac{-4e^2}{h}(1-R)(1-p)k_B \mathcal{T},}
\end{equation}

{In this regime also, where thermal noise dominates over shot noise, we can see the distinction between topological ($p=0$) and trivial helical edge modes ($p \ne 0$) by magnitude.}
{For $eV = k_B \mathcal{T}$, we have,}
\begin{equation} \label{eq:140}
    {S_{12}^q =  \frac{-2e^2}{h}(1-R)(1-p)\bigg(2  + 0.64 p\bigg)k_B \mathcal{T}}
\end{equation}

{In this regime as well, we can see a clear difference between topological and trivial helical edge modes by magnitude. Similarly, we can see the distinction between these two helical edge modes via other cross-correlations such as $S_{14}^q, S_{23}^q$.}

Eq. (\ref{eq:137}) implies that the {thermal noise-like contributionss} to $S_{12}^q, S_{14}^q, S_{23}^q, S_{13}^q, S_{24}^q$ are given as,

\begin{equation} \label{eq:141}
\begin{split}
{S_{12}^{th} = \frac{2e^2}{h}(1-R)(1-p)2k_B \mathcal{T},} \\{S_{14}^{th} = S_{23}^{th} = \frac{(1+R)}{(1-R)}S_{12}^{th}}, 
{S_{13}^{th} = S_{24}^{th} = 0.}
\end{split}
\end{equation}
and the {shot noise-like contributions are},
\begin{equation} \label{eq:142}
\begin{split}
S_{12}^{sh} &= -\frac{8e^2}{h}(1-R)p(1-p) \bigg(eV \coth \left[\frac{eV}{2k_B \mathcal{T}}\right]-2k_B \mathcal{T}\bigg),\\
S_{14}^{sh} &= S_{23}^{sh} = \frac{(1+R)}{4(1-R)}S_{12}^{sh}.
\end{split}
\end{equation}

In this setup as well, the {quantum noise auto-correlations can also be derived. In this setup, $S_{11}^q$ and $S_{22}^q$ are same, while $S_{33}^q$ and $S_{44}^q$ are also same, which can be found by interchanging $\mu$ and $\mu_0$ in $S_{11}^q$.}

{The thermal noise-like contributions in terminal 1 in setup-3 is,}
\begin{equation} \label{eq:143} 
\begin{split}
{S_{11}^{th} = S_{22}^{th} = \frac{2e^2}{h}\int_0^{\infty}dE 4(1-p)f(1-f),}
\end{split}
\end{equation}

{and, the shot noise-like contribution is,}
\begin{widetext}

\begin{equation} \label{eq:144}
\begin{split}
{S_{11}^{sh} = \frac{2e^2}{h}\int_0^{\infty}dE ((R+1)(1-p)(f_0-f)}{+(2(1-p^2)}{-2p(1-p)T} {-T^2(1-p)^2)f^2}{-(2pR(1-p)} \\{+2RT(1-p)^2+2p(1-p))ff_0}{-(R^2+1)(1-p)^2f_0^2).}
\end{split}
\end{equation}
\end{widetext}

The total quantum noise correlation $S_{11}^q$ can be found out using Eqs. (\ref{eq:143}) and (\ref{eq:144}) and is given as,

 \begin{widetext}
\begin{equation} \label{eq:145}
\begin{split}
S_{11}^q = S_{22}^q = S_{33}^q = S_{44}^q = &\left(2(1-p)(2-p-R+(1-p)R^2)+(1-p)(p+R-(1-p)R^2)\frac{eV}{k_B \mathcal{T}} \right. \\ &\left.+ (2pR(1-p)+2RT(1-p)^2 + 2p(1-p))\frac{eV}{k_B \mathcal{T}(e^{\frac{eV}{k_B \mathcal{T}}}-1)}\right)k_B \mathcal{T}.
\end{split}
\end{equation}
\end{widetext}

\begin{figure}
\includegraphics[scale=0.50]{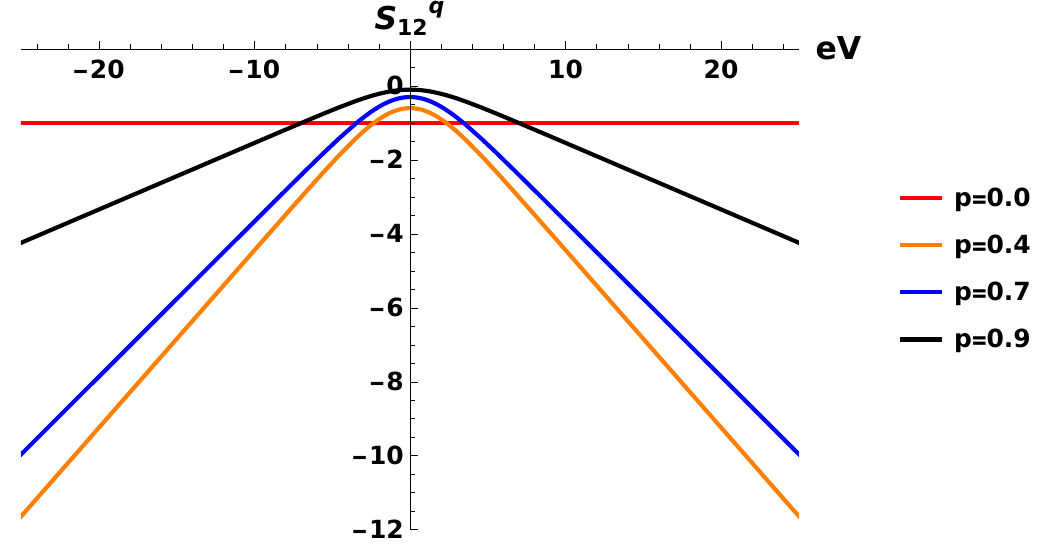}
\caption{Quantum noise $S_{12}^{q}$ in units of $\frac{2e^2}{h} k_B \mathcal{T}$ for setup 3 as a function of applied voltage bias ($eV$) in units of $k_B \mathcal{T}$ at $\mu_0 = 100k_B \mathcal{T}$ and $R=0.5$.}
\label{fig:14}
\end{figure}
Similar to setup-1, we can see that all the quantum noise autocorrelations are same in the limit $\mu, \mu_0 \gg k_B \mathcal{T}$. We had already seen that for the topological helical case ($p=0$), the quantum noise autocorrelation changes symmetrically with the applied voltage. {In this case, both thermal and shot noise appears. This has been explained below Eq.~(\ref{eq:100}).} Similarly, the autocorrelations for trivial helical edge modes change symmetrically for different values of spin-flip probability ($p$), but they differ in magnitudes in different regimes, which has been highlighted in Table~\ref{Table 6}. {The extra shot noise contribution appears as a result of spin-flip scattering $p$, which means at $p = 0$, we will get the topological helical result as in Eq.~(\ref{eq:102}). Similar to setup 1, the quantum noise autocorrelations change symmetrically with applied voltage $eV$ for both the topological ($p=0$) and trivial ($p \ne 0$) helical edge modes, which does not distinguish these two helical edge modes via behaviour, but they distinguish via magnitude as shown in Table~\ref{Table 6} in different regimes. }

In setup 3, we observe that the topological and trivial helical edge modes are distinguished via the quantum noise correlations $S_{12}^q, S_{14}^q,  S_{23}^q$ as shown in Figs. \ref{fig:13} and \ref{fig:14}. Like setup 2, we don't see any instance where the quantum noise autocorrelation can discriminate between topological and trivial helical edge states.

\section{Analysis} \label{section IV}
This section analyses all the possible cross and auto-quantum noise correlations for different setups via Tables {in the regimes $eV\gg k_B \mathcal{T}$, $eV = k_B \mathcal{T}$ and $eV\ll k_B \mathcal{T}$.}

In TABLE~\ref{Table 1}, we have summarised the behaviour of all the possible cross and autocorrelations as a function of the voltage bias ($eV = \mu - \mu_0$) for setup 1 with the condition $\mu_1 = \mu_4 = \mu$ and $\mu_2 = \mu_3 = \mu_0$. We observe that the quantum noise cross-correlations behave very similarly in the presence of chiral and helical edge modes except $S_{24}^q$, shown in Fig.~\ref{fig:3}. For chiral edge modes, it is zero. In contrast, in the presence of topological helical edge modes transport contribution is finite, which gives nonvanishing quantum noise correlation $S_{24}^q$. Similarly, for setup 1, we see the distinction between, trivial helical and topological helical edge modes via quantum noise correlation such as $S_{12}^q, S_{14}^q, S_{34}^q$, which are all constant for topological helical edge modes, while change while being symmetric with respect to the sign of voltage bias ($eV$) for trivial helical edge modes. It is shown in Figs.~\ref{fig:8} and \ref{fig:9}. Similarly, we have also looked into autocorrelations and summarised them in TABLE~\ref{Table 1}. Chiral is distinguished from topological helical edge mode transport via the quantum noise autocorrelations $S_{22}^q$ and $S_{44}^q$ as in Fig. \ref{fig:6} (a). {We have analyzed the results via magnitudes in the regimes such as $eV\gg k_B \mathcal{T}$, $eV = k_B \mathcal{T}$ and $eV\ll k_B \mathcal{T}$} in TABLE~\ref{Table 4}.

Similarly, quantum noise cross and autocorrelations for setup 2 with the condition $\mu_1 = \mu_3 = \mu$ and $\mu_2 = \mu_4 = \mu_0$ are summarised in TABLE \ref{Table 2}. We did not observe any examples of chiral and topological helical edge modes distinguished via quantum noise cross-correlation in this setup. Similar to setup 1, most of the examples clearly distinguish between the helical edge modes, trivial and topological, via the quantum noise cross-correlations $S_{12}^q, S_{14}^q, S_{23}^q, S_{34}^q$ as shown in Figs. \ref{fig:10} and \ref{fig:11}. Here, autocorrelations do not discriminate between chiral and topological helical edge modes. However, they distinguish between topological and trivial helical edge modes, which can be seen in Fig. \ref{eq:12}. The autocorrelations are constant due to topological helical edge modes and change symmetrically with respect to the change in sign of voltage bias for, trivial helical. {We have also analyzed the results via magnitudes in the regimes such as $eV \gg k_B \mathcal{T}$, $eV = k_B \mathcal{T}$ and $eV \ll k_B \mathcal{T}$} in TABLE \ref{Table 5}.

For the third setup with the condition $\mu_1 = \mu_2 = \mu$ and $\mu_2 = \mu_3 = \mu_0$, the results are summarised in TABLE III. The distinction between chiral and topological helical is not seen for the cross-correlations of quantum noise. However, it is seen between topological and trivial helical edge modes for the quantum noise correlations $S_{12}^q, S_{14}^q, S_{23}^q$ as shown in Figs. \ref{fig:13} and \ref{fig:14}. Similar to setup 1, the quantum noise autocorrelations $S_{22}^q$ and $S_{44}^q$ clearly distinguish between the chiral and topological helical edge modes as shown in Fig. \ref{fig:6} (a). {Herein too, We have analyzed the results via magnitudes in the regimes such as $eV \gg k_B \mathcal{T}$, $eV = k_B \mathcal{T}$ and $eV \ll k_B \mathcal{T}$} in TABLE~\ref{Table 6}.

\begin{widetext}

\begin{table}[h!] 
\centering
\caption{Nature of quantum noise cross and autocorrelation as a function of voltage bias for setup 1.}
\scalebox{0.92}{\begin{tabular}{|l|l|l|l|l|l|l|l|}
\hline
Edge Modes & $S_{12}^q=S_{34}^q$ &  $S_{14}^q$ & $S_{24}^q$ & $S_{22}^q=S_{44}^q$ \\ \hline
Chiral & Constant (Eq. \ref{eq: 28}) & Constant (Eq. \ref{eq: 27}) & Zero (Eq. \ref{eq: 29}, Fig. \ref{fig:3}(a)) & Constant (Eq. \ref{eq:41}, Fig. \ref{eq:6}(a)) \\ \hline
Helical(Topological) & Constant (Eq. \ref{eq:75}, Fig. \ref{fig:9}) & Constant (Eq.~\ref{eq:74}, Fig.~\ref{fig:8}) & Symmetric (Eq.~\ref{eq:82}, Fig. \ref{fig:3}(a)) & Symmetric (Eq.~\ref{eq:93}, Fig.~\ref{fig:6}(a)) \\ \hline
Helical(Trivial) & Symmetric (Eq.~\ref{eq:119}, Fig.~\ref{fig:9}) & Symmetric (Eq.~\ref{eq:122}, Fig. \ref{fig:8}) & Symmetric (Eq.~\ref{eq:123}) & Symmetric (Eq.~\ref{eq:126}) \\ \hline
\end{tabular}}
\label{Table 1}
\end{table}

\begin{table}[h!] 
\centering
\caption{Nature of quantum noise cross and autocorrelation as a function of voltage bias for setup 2. }
\scalebox{0.92}{\begin{tabular}{|l|l|l|l|l|l|}
\hline
Edge Modes & $S_{12}^q=S_{34}^q$  & $S_{14}^q=S_{23}^q$ & $S_{24}^q$ & $S_{11}^q=S_{22}^q=S_{33}^q=S_{44}^q$ \\ \hline
Chiral & Constant (Eq.~\ref{eq:43}) & Constant (Eq.~\ref{eq:42}) & Zero (Eq. \ref{eq:44}, Fig. \ref{eq:3}(b)) & Constant (Eq. \ref{eq:45}) \\ \hline
Helical(Topological) & Constant (Eq. \ref{eq:95}, Fig. \ref{fig:10}) & Constant (Eq. \ref{eq:94}, Fig. \ref{fig:11}) & Symmetric (Eq. \ref{eq:97}, Fig. \ref{eq:3}(b)) & Constant (Eq. \ref{eq:98}, Fig. \ref{fig:12})  \\ \hline
Helical(Trivial) & Symmetric (Eq. \ref{eq:127}, Fig. \ref{fig:10}) & Symmetric (\ref{eq:127}, Fig. \ref{fig:11}) & Symmetric (Eq. \ref{eq:127}) & Symmetric (Eq. \ref{eq:135}, Fig. \ref{fig:12}) \\ \hline
\end{tabular}}
\label{Table 2}
\end{table}

\begin{table}[h!] 
\centering
\caption{Nature of quantum noise cross and autocorrelation as a function of voltage bias for setup 3.}
\scalebox{0.92}{\begin{tabular}{|l|l|l|l|l|l|l|l|}
\hline
Edge Modes & $S_{12}^q$ & $S_{14}^q=S_{23}^q$ & $S_{24}^q$ &  $S_{22}^q=S_{44}^q$ \\ \hline
Chiral & Constant (Eq. \ref{eq:46}) & Constant (Eq. \ref{eq:46}) & Zero (Eq. \ref{eq:46}, Fig. \ref{fig:3}(a)) & Constant (Eq. \ref{eq:47}, Fig. \ref{fig:6}(a)) \\ \hline
Helical(Topological) & Constant (Eq. \ref{eq:99}, Fig. \ref{fig:14}) &  Constant (Eq. \ref{eq:99}, Fig. \ref{fig:13}) & Symmetric (Eq. \ref{eq:99}, Fig. \ref{fig:3}(a)) & Symmetric (Eq. \ref{eq:102}, Fig. \ref{fig:6}(a)) \\ \hline
Helical(Trivial) & Symmetric (Eq. \ref{eq:137}, Fig. \ref{fig:14}) & Symmetric (Eq. \ref{eq:137}, Fig. \ref{fig:13}) & Symmetric (Eq. \ref{eq:137}) & Symmetric (Eq. \ref{eq:145}) \\ \hline
\end{tabular}}
\label{Table 3}
\end{table}

\begin{table}[h!]
\centering
\caption{Quantum noise cross and autocorrelation for setup 1 (magnitudes are normalized by $\frac{2e^2}{h}$).}
\renewcommand{\arraystretch}{4}
\scalebox{0.34}{ \begin{tabular}{|l|lll|lll|lll|lll|}
\hline
\scalebox{1.50}{Edge Modes} & \multicolumn{3}{c|}{\scalebox{1.50}{$S_{12}^q = S_{34}^q$}} & \multicolumn{3}{c|}{\scalebox{1.50}{$S_{14}^q$}} & \multicolumn{3}{c|}{\scalebox{1.50}{$S_{24}^q$}} & \multicolumn{3}{c|}{\scalebox{1.50}{$S_{22}^q = S_{44}^q$}} \\ \hline
& \multicolumn{1}{l|}{\scalebox{1.50}{$eV\gg k_B \mathcal{T}$}} & \multicolumn{1}{l|}{\scalebox{1.50}{$eV = k_B \mathcal{T}$}} &
\scalebox{1.50}{$eV\ll k_B\mathcal{T}$} & \multicolumn{1}{l|}{\scalebox{1.50}{$eV\gg k_B \mathcal{T}$}} & \multicolumn{1}{l|}{\scalebox{1.50}{$eV= k_B \mathcal{T}$}} & {\scalebox{1.50}{$eV\ll k_B \mathcal{T}$}} & \multicolumn{1}{l|}{\scalebox{1.50}{$eV\gg k_B \mathcal{T}$}} & \multicolumn{1}{l|}{\scalebox{1.50}{$eV= k_B\mathcal{T}$}} & {\scalebox{1.50}{$eV\ll k_B\mathcal{T}$}} & \multicolumn{1}{l|}{\scalebox{1.50}{$eV\gg k_B\mathcal{T}$}} & \multicolumn{1}{l|}{\scalebox{1.50}{$eV= k_B\mathcal{T}$}} & {\scalebox{1.50}{$eV\ll k_B\mathcal{T}$}}\\ \hline
\scalebox{1.50}{Chiral} & \multicolumn{1}{l|}{\scalebox{1.50}{$-2T k_B \mathcal{T}$}} & \multicolumn{1}{l|}{\scalebox{1.50}{$-2T k_B \mathcal{T}$}} & \multicolumn{1}{l|}{\scalebox{1.50}{$-2T k_B \mathcal{T}$}} & \multicolumn{1}{l|}{\scalebox{1.50}{$\frac{1+R}{T}S_{12}^q$}} & \multicolumn{1}{l|}{\scalebox{1.50}{$\frac{1+R}{T}S_{12}^q$}} & \multicolumn{1}{l|}{\scalebox{1.50}{$\frac{1+R}{T}S_{12}^q$}} & \multicolumn{1}{l|}{\scalebox{1.50}{Zero}} & \multicolumn{1}{l|}{\scalebox{1.50}{Zero}} & \multicolumn{1}{l|}{\scalebox{1.50}{Zero}} & \multicolumn{1}{l|}{\scalebox{1.50}{$4 k_B \mathcal{T}$}} & \multicolumn{1}{l|}{\scalebox{1.50}{$ 4 k_B \mathcal{T}$}} & \multicolumn{1}{l|}{\scalebox{1.50}{$ 4 k_B \mathcal{T}$}} \\ \hline
\multicolumn{1}{|l|}{\begin{tabular}[c]{@{}l@{}}\scalebox{1.50}{Helical}\\ \scalebox{1.50}{(Topological)}\end{tabular}} & \multicolumn{1}{l|}{\scalebox{1.50}{$-2T k_B \mathcal{T}$}} & \multicolumn{1}{l|}{\scalebox{1.50}{$-2T k_B \mathcal{T}$}} &
\multicolumn{1}{l|}{\scalebox{1.50}{$-2T k_B \mathcal{T}$}} & \multicolumn{1}{l|}{\scalebox{1.50}{$\frac{1+R}{T}S_{12}^q$}} & \multicolumn{1}{l|}{\scalebox{1.50}{$\frac{1+R}{T}S_{12}^q$}} &
\multicolumn{1}{l|}{\scalebox{1.50}{$\frac{1+R}{T}S_{12}^q$}} & \multicolumn{1}{l|} {\begin{tabular}[c]{@{}l@{}}\scalebox{1.50}{-$RT$}\\ \scalebox{1.50}{$\times\left(eV -2k_B \mathcal{T}\right)$}\end{tabular}} & \multicolumn{1}{l|}{\scalebox{1.50}{$-0.16RT k_B \mathcal{T}$}} & \multicolumn{1}{l|}{\scalebox{1.50}{Zero}} & \multicolumn{1}{l|}{\begin{tabular}[c]{@{}l@{}}\scalebox{1.50}{$(4-2 R T)k_B \mathcal{T} $}\\ \scalebox{1.50}{$+RT eV $}\end{tabular}} & \multicolumn{1}{l|}{\begin{tabular}[c]{@{}l@{}}\scalebox{1.50}{$4k_B \mathcal{T}$}\\ \scalebox{1.50}{$+0.16RTk_B \mathcal{T}$}\end{tabular}} &
\multicolumn{1}{l|}{\scalebox{1.50}{$4k_B \mathcal{T}$}} \\ \hline
\multicolumn{1}{|l|}{\begin{tabular}[c]{@{}l@{}}\scalebox{1.50}{Helical}\\ \scalebox{1.50}{(Trivial)}\end{tabular}} & \multicolumn{1}{l|}{\begin{tabular}[c]{@{}l@{}}\scalebox{1.50}{$-T(1-p)\bigg(2k_B \mathcal{T}$}\\ \scalebox{1.50}{$+p\left(eV -2k_B \mathcal{T}\right)\bigg)$}\end{tabular}} & \multicolumn{1}{l|}{\begin{tabular}[c]{@{}l@{}}\scalebox{1.50}{$-T k_B \mathcal{T}(1-p)$}\\ \scalebox{1.50}{$\times(2 + 0.16p)$}\end{tabular}} & \multicolumn{1}{l|}{\begin{tabular}[c]{@{}l@{}}\scalebox{1.50}{$-2T k_B \mathcal{T}$}\\ \scalebox{1.50}{$\times(1-p)$}\end{tabular}} &
\multicolumn{1}{l|} {\begin{tabular}[c]{@{}l@{}}\scalebox{1.50}{$-(1+R)(1-p)\bigg(2k_B \mathcal{T}$}\\ \scalebox{1.50}{$+4p\left(eV -2k_B \mathcal{T}\right)\bigg)$}\end{tabular}} & \multicolumn{1}{l|}{\begin{tabular}[c]{@{}l@{}}\scalebox{1.50}{$-(1+R) k_B \mathcal{T}$}\\ \scalebox{1.50}{$(1-p)(2 + 0.64p)$}\end{tabular}} & \multicolumn{1}{l|}{\scalebox{1.50}{$\frac{1+R}{T}S_{12}^q$}} &
\multicolumn{1}{l|} {\begin{tabular}[c]{@{}l@{}}\scalebox{1.50}{$-RT (1-p)^2 $}\\ \scalebox{1.50}{$ \times\left(eV-2k_B \mathcal{T}\right)$}\end{tabular}} & \multicolumn{1}{l|}{\begin{tabular}[c]{@{}l@{}}\scalebox{1.50}{$-(1-p)^2$}\\ \scalebox{1.50}{$ \times 0.16 RT k_B \mathcal{T}$}\end{tabular}} & \multicolumn{1}{l|}{\scalebox{1.50}{Zero}} & \multicolumn{1}{l|}{\scalebox{1.50}{$A_{1}$}} & \multicolumn{1}{l|}{\scalebox{1.50}{$A_2$}} & \multicolumn{1}{l|}{\begin{tabular}[c]{@{}l@{}}\scalebox{1.50}{$4 k_B \mathcal{T}$}\\ \scalebox{1.50}{$\times(1-p)$}\end{tabular}} \\ \hline
\end{tabular}}
\label{Table 4}
\end{table}

\begin{table}[h!]
\centering
\caption{Quantum noise cross and autocorrelation for setup 2 (magnitudes are normalized by $\frac{2e^2}{h}$).}
\renewcommand{\arraystretch}{4}
\scalebox{0.35}{ \begin{tabular}{|l|lll|lll|lll|lll|}
\hline
\scalebox{1.50}{Edge Modes} & \multicolumn{3}{c|}{\scalebox{1.50}{$S_{12}^q = S_{34}^q$}} & \multicolumn{3}{c|}{\scalebox{1.50}{$S_{14}^q = S_{23}^q$}} & \multicolumn{3}{c|}{\scalebox{1.50}{$S_{24}^q$}} & \multicolumn{3}{c|}{\scalebox{1.50}{$S_{11}^q = S_{22}^q = S_{33}^q = S_{44}^q$}} \\ \hline
& \multicolumn{1}{l|}{\scalebox{1.50}{$eV\gg k_B \mathcal{T}$}} & \multicolumn{1}{l|}{\scalebox{1.50}{$eV = k_B \mathcal{T}$}} &
\scalebox{1.50}{$eV\ll k_B\mathcal{T}$} & \multicolumn{1}{l|}{\scalebox{1.50}{$eV\gg k_B \mathcal{T}$}} & \multicolumn{1}{l|}{\scalebox{1.50}{$eV = k_B \mathcal{T}$}} & {\scalebox{1.50}{$eV\ll k_B \mathcal{T}$}} & \multicolumn{1}{l|}{\scalebox{1.50}{$eV\gg k_B \mathcal{T}$}} & \multicolumn{1}{l|}{\scalebox{1.50}{$eV = k_B\mathcal{T}$}} & {\scalebox{1.50}{$eV\ll k_B\mathcal{T}$}} & \multicolumn{1}{l|}{\scalebox{1.50}{$eV\gg k_B\mathcal{T}$}} & \multicolumn{1}{l|}{\scalebox{1.50}{$eV = k_B\mathcal{T}$}} & {\scalebox{1.50}{$eV\ll k_B\mathcal{T}$}}\\ \hline
\scalebox{1.50}{Chiral} & \multicolumn{1}{l|}{\scalebox{1.50}{$-2T k_B \mathcal{T}$}} & \multicolumn{1}{l|}{\scalebox{1.50}{$-2T k_B \mathcal{T}$}} & \multicolumn{1}{l|}{\scalebox{1.50}{$-2T k_B \mathcal{T}$}} & \multicolumn{1}{l|}{\scalebox{1.50}{$\frac{1+R}{T}S_{12}^q$}} & \multicolumn{1}{l|}{\scalebox{1.50}{$\frac{1+R}{T}S_{12}^q$}} & \multicolumn{1}{l|}{\scalebox{1.50}{$\frac{1+R}{T}S_{12}^q$}} & \multicolumn{1}{l|}{\scalebox{1.50}{Zero}} & \multicolumn{1}{l|}{\scalebox{1.50}{Zero}} & \multicolumn{1}{l|}{\scalebox{1.50}{Zero}} & \multicolumn{1}{l|}{\scalebox{1.50}{$4k_B \mathcal{T}$}} & \multicolumn{1}{l|}{\scalebox{1.50}{$4k_B \mathcal{T}$}} & \multicolumn{1}{l|}{\scalebox{1.50}{$4k_B \mathcal{T}$}} \\ \hline
\multicolumn{1}{|l|}{\begin{tabular}[c]{@{}l@{}}\scalebox{1.50}{Helical}\\ \scalebox{1.50}{(Topological)}\end{tabular}} & \multicolumn{1}{l|}{\scalebox{1.50}{$-2T k_B \mathcal{T}$}} & \multicolumn{1}{l|}{\scalebox{1.50}{$-2T k_B \mathcal{T}$}} & \multicolumn{1}{l|}{\scalebox{1.50}{$-2T k_B \mathcal{T}$}} &
\multicolumn{1}{l|}{\scalebox{1.50}{$\frac{1+R}{T}S_{12}^q$}} & \multicolumn{1}{l|}{\scalebox{1.50}{$\frac{1+R}{T}S_{12}^q$}} & \multicolumn{1}{l|}{\scalebox{1.50}{$\frac{1+R}{T}S_{12}^q$}} & \multicolumn{1}{l|}{\begin{tabular}[c]{@{}l@{}}\scalebox{1.50}{-$4RT$}\\ \scalebox{1.50}{$\left(eV -2k_B \mathcal{T}\right)$}\end{tabular}} & \multicolumn{1}{l|}{\scalebox{1.50}{$-0.64RTk_B \mathcal{T}$}} & \multicolumn{1}{l|}{\scalebox{1.50}{Zero}} & \multicolumn{1}{l|}{\scalebox{1.50}{$4k_B \mathcal{T}$}} & \multicolumn{1}{l|}{\scalebox{1.50}{$4k_B \mathcal{T}$}} & \multicolumn{1}{l|}{\scalebox{1.50}{$4k_B \mathcal{T}$}} \\ \hline
\multicolumn{1}{|l|}{\begin{tabular}[c]{@{}l@{}}\scalebox{1.50}{Helical}\\ \scalebox{1.50}{(Trivial)}\end{tabular}} & \multicolumn{1}{l|}{\begin{tabular}[c]{@{}l@{}}\scalebox{1.50}{$-T(1-p)\bigg(2k_B \mathcal{T}$}\\ \scalebox{1.50}{$+p\left(eV -2k_B \mathcal{T}\right)\bigg)$}\end{tabular}} & \multicolumn{1}{l|}{\begin{tabular}[c]{@{}l@{}}\scalebox{1.50}{$-T k_B \mathcal{T}(1-p)$}\\ \scalebox{1.50}{$\times(2 + 0.16p)$}\end{tabular}} & \multicolumn{1}{l|}{\begin{tabular}[c]{@{}l@{}}\scalebox{1.50}{$-2T k_B \mathcal{T}$}\\ \scalebox{1.50}{$\times(1-p)$}\end{tabular}} & \multicolumn{1}{l|}{\begin{tabular}[c]{@{}l@{}}\scalebox{1.50}{$-(1+R)(1-p)\bigg(2k_B \mathcal{T}$}\\ \scalebox{1.50}{$+p\left(eV -2k_B \mathcal{T}\right)\bigg)$}\end{tabular}} & \multicolumn{1}{l|}{\scalebox{1.50}{$\frac{1+R}{T}S_{12}^q$}} & \multicolumn{1}{l|}{\scalebox{1.50}{$\frac{1+R}{T}S_{12}^q$}} & \multicolumn{1}{l|}{\begin{tabular}[c]{@{}l@{}}\scalebox{1.50}{$-4RT(1-p)^2$}\\ \scalebox{1.50}{$\times \left(eV-2k_B \mathcal{T}\right)$}\end{tabular}} & \multicolumn{1}{l|}{\begin{tabular}[c]{@{}l@{}}\scalebox{1.50}{$-(1-p)^2$}\\ \scalebox{1.50}{$\times 0.64RTk_B \mathcal{T}$}\end{tabular}} & \multicolumn{1}{l|}{\scalebox{1.50}{Zero}} & \multicolumn{1}{l|}{\scalebox{1.50}{$B_{1}$}} & \multicolumn{1}{l|}{\scalebox{1.50}{$B_2$}} & \multicolumn{1}{l|}{\begin{tabular}[c]{@{}l@{}}\scalebox{1.50}{$4k_B\mathcal{T}$}\\ \scalebox{1.50}{$\times(1-p)$}\end{tabular}} \\ \hline
\end{tabular}}
\label{Table 5}
\end{table}

\begin{table}[h!]
\centering
\caption{Quantum noise cross and autocorrelation for setup 3 (magnitudes are normalized by $\frac{2e^2}{h}$).}
\renewcommand{\arraystretch}{4}
\scalebox{0.34}{ \begin{tabular}{|l|lll|lll|lll|lll|}
\hline
\scalebox{1.50}{Edge Modes} & \multicolumn{3}{c|}{\scalebox{1.50}{$S_{12}^q$}} & \multicolumn{3}{c|}{\scalebox{1.50}{$S_{14}^q = S_{23}^q$}} & \multicolumn{3}{c|}{\scalebox{1.50}{$S_{24}^q$}} & \multicolumn{3}{c|}{\scalebox{1.50}{$S_{22}^q = S_{44}^q$}} \\ \hline
& \multicolumn{1}{l|}{\scalebox{1.50}{$eV\gg k_B \mathcal{T}$}} & \multicolumn{1}{l|}{\scalebox{1.50}{$eV = k_B \mathcal{T}$}} &
\scalebox{1.50}{$eV\ll k_B\mathcal{T}$} & \multicolumn{1}{l|}{\scalebox{1.50}{$eV\gg k_B \mathcal{T}$}} & \multicolumn{1}{l|}{\scalebox{1.50}{$eV = k_B \mathcal{T}$}} & {\scalebox{1.50}{$eV\ll k_B \mathcal{T}$}} & \multicolumn{1}{l|}{\scalebox{1.50}{$eV\gg k_B \mathcal{T}$}} & \multicolumn{1}{l|}{\scalebox{1.50}{$eV = k_B\mathcal{T}$}} & {\scalebox{1.50}{$eV\ll k_B\mathcal{T}$}} & \multicolumn{1}{l|}{\scalebox{1.50}{$eV\gg k_B\mathcal{T}$}} & \multicolumn{1}{l|}{\scalebox{1.50}{$eV = k_B\mathcal{T}$}} & {\scalebox{1.50}{$eV\ll k_B\mathcal{T}$}}\\ \hline
\scalebox{1.50}{Chiral} & \multicolumn{1}{l|}{\scalebox{1.50}{$-2Tk_B \mathcal{T}$}} & \multicolumn{1}{l|}{\scalebox{1.50}{$-2Tk_B \mathcal{T}$}} & \multicolumn{1}{l|}{\scalebox{1.50}{$-2Tk_B \mathcal{T}$}} & \multicolumn{1}{l|}{\scalebox{1.50}{$\frac{1+R}{T}S_{12}^q$}} & \multicolumn{1}{l|}{\scalebox{1.50}{$\frac{1+R}{T}S_{12}^q$}} & \multicolumn{1}{l|}{\scalebox{1.50}{$\frac{1+R}{T}S_{12}^q$}} & \multicolumn{1}{l|}{\scalebox{1.50}{Zero}} & \multicolumn{1}{l|}{\scalebox{1.50}{Zero}} & \multicolumn{1}{l|}{\scalebox{1.50}{Zero}} & \multicolumn{1}{l|}{\scalebox{1.50}{$4 k_B \mathcal{T}$}} & \multicolumn{1}{l|}{\scalebox{1.50}{$4 k_B \mathcal{T}$}} & \multicolumn{1}{l|}{\scalebox{1.50}{$4 k_B \mathcal{T}$}} \\ \hline
\multicolumn{1}{|l|}{\begin{tabular}[c]{@{}l@{}}\scalebox{1.50}{Helical}\\ \scalebox{1.50}{(Topological)}\end{tabular}} & \multicolumn{1}{l|}{\scalebox{1.50}{$-2Tk_B \mathcal{T}$}} & \multicolumn{1}{l|}{\scalebox{1.50}{$-2Tk_B \mathcal{T}$}} & \multicolumn{1}{l|}{\scalebox{1.50}{$-2Tk_B \mathcal{T}$}} & \multicolumn{1}{l|}{\scalebox{1.50}{$\frac{1+R}{T}S_{12}^q$}} & \multicolumn{1}{l|}{\scalebox{1.50}{$\frac{1+R}{T}S_{12}^q$}} & \multicolumn{1}{l|}{\scalebox{1.50}{$\frac{1+R}{T}S_{12}^q$}} & \multicolumn{1}{l|}{\begin{tabular}[c]{@{}l@{}}\scalebox{1.50}{$-RT $}\\ \scalebox{1.50}{$\left(eV-2k_B \mathcal{T}\right)$}\end{tabular}} & \multicolumn{1}{l|}{\scalebox{1.50}{$-0.16RTk_B \mathcal{T}$}} & \multicolumn{1}{l|}{\scalebox{1.50}{Zero}} & \multicolumn{1}{l|}{\begin{tabular}[c]{@{}l@{}}\scalebox{1.50}{$(4-2RT)k_B \mathcal{T} $}\\ \scalebox{1.50}{$+ RT eV$}\end{tabular}} & \multicolumn{1}{l|} {\begin{tabular}[c]{@{}l@{}}\scalebox{1.50}{$4k_B \mathcal{T} $}\\ \scalebox{1.50}{$+ 0.16 RT k_B \mathcal{T})$}\end{tabular}} & \multicolumn{1}{l|}{\scalebox{1.50}{$4 k_B \mathcal{T}$}} \\ \hline
\multicolumn{1}{|l|}{\begin{tabular}[c]{@{}l@{}}\scalebox{1.50}{Helical}\\ \scalebox{1.50}{(Trivial)}\end{tabular}} & \multicolumn{1}{l|}{\begin{tabular}[c]{@{}l@{}}\scalebox{1.50}{$-T(1-p)\bigg(2k_B \mathcal{T}$}\\ \scalebox{1.50}{$+4p\left(eV -2k_B \mathcal{T}\right)\bigg)$}\end{tabular}} & \multicolumn{1}{l|}{\begin{tabular}[c]{@{}l@{}}\scalebox{1.50}{$-Tk_B \mathcal{T}(1-p)$}\\ \scalebox{1.50}{$\times(2+0.64p)$}\end{tabular}} & \multicolumn{1}{l|}{\begin{tabular}[c]{@{}l@{}}\scalebox{1.50}{$-2T k_B \mathcal{T}$}\\ \scalebox{1.50}{$\times(1-p)$}\end{tabular}} & \multicolumn{1}{l|}{\begin{tabular}[c]{@{}l@{}}\scalebox{1.50}{$-(1+R)(1-p)\bigg(2k_B \mathcal{T}$}\\ \scalebox{1.50}{$+p\left(eV -2k_B \mathcal{T}\right)\bigg)$}\end{tabular}} & \multicolumn{1}{l|}{\begin{tabular}[c]{@{}l@{}}\scalebox{1.50}{$-(1+R)k_B \mathcal{T}(1-p)$}\\ \scalebox{1.50}{$\times(2+0.16p)$}\end{tabular}} & \multicolumn{1}{l|}{\scalebox{1.50}{$\frac{1+R}{T}S_{12}^q$}} & \multicolumn{1}{l|} {\begin{tabular}[c]{@{}l@{}}\scalebox{1.50}{$-RT(1-p)^2$}\\ \scalebox{1.50}{$\times\left(eV -2k_B \mathcal{T}\right)$}\end{tabular}} & \multicolumn{1}{l|}{\begin{tabular}[c]{@{}l@{}}\scalebox{1.50}{$-(1-p)^2$}\\ \scalebox{1.50}{$\times 0.16RTk_B \mathcal{T}$}\end{tabular}} & \multicolumn{1}{l|}{\scalebox{1.50}{Zero}} & \multicolumn{1}{l|}{\scalebox{1.50}{$C_{1}$}} & \multicolumn{1}{l|}{\scalebox{1.50}{$ C_2$}} & \multicolumn{1}{l|}{\begin{tabular}[c]{@{}l@{}}\scalebox{1.50}{$4k_B \mathcal{T}$}\\ \scalebox{1.50}{$\times(1-p)$}\end{tabular}} \\ \hline
\end{tabular}}
\label{Table 6}
\end{table}

\begingroup
\begin{align} \label{eq:146}
    A_{1} = &\left(2(1-p)(2-pT^2-RT)+T(1-p)(1-T(1-p))\frac{eV}{k_B \mathcal{T}}\right)k_B \mathcal{T},\nonumber\\
    A_{2} = &(4 + 0.16RT + p(-3.84+(1.84-2R)R-2.33RT)+p^2(-0.16+(-0.84+R)R+1.16RT))k_B \mathcal{T},\nonumber\\
    B_{1} = &\left(4(1-p)^2 + 2 p(1-p) \frac{eV}{k_B \mathcal{T}}\right)k_B \mathcal{T}, \quad
    B_{2} = \left(4 - p(3.68+0.33 p )\right)k_B \mathcal{T},\nonumber\\
  C_{1} = &\bigg(2(1-p)(2-p-R+(1-p)R^2)+(1-p)(p+R-(1-p)R^2)\frac{eV}{k_B \mathcal{T}}\bigg)k_B \mathcal{T},\nonumber\\
  C_{2} = &\bigg(2(1- p)(2-p-R+(1-p)R^2)+2.16((1-p)p+R-pR-(1-p)^2R^2)\bigg)k_B\mathcal{T}.
    \end{align}
\endgroup

\end{widetext}

\subsection{{Question of finite frequency quantum noise correlations}}

{In the previous sections, we have concentrated only on zero-frequency quantum noise correlations. However, one may take finite frequency too. In that case, as we show below, quantum noise successfully distinguishes between different edge modes at finite temperatures. The calculation of quantum noise correlations can be extended to finite frequency as well by using the general expression given in Eq. (\ref{eq:7}) for chiral edge modes, and in Eq. (\ref{eq:51}) for both topological helical and trivial helical edge modes. In our setups, we don't have energy dependence in the s-matrix. The only frequency dependence comes from the Fermi-Dirac distribution. First, we will explain the effect of finite frequency for setup 1 (see Figs. \ref{fig:1} and \ref{fig:2}) in the case of both chiral and topological helical edge modes, where $\mu_1 = \mu_4$ and $\mu_2 = \mu_3$. Next, we will discuss the distinction between topological and trivial helical.}

{In setup 1, see Fig. \ref{fig:6}(a), chiral and topological helical edge modes can be easily distinguished via the quantum noise autocorrelation $S_{22}^q$ at $\omega = 0$. For chiral, the shot noise-like contribution vanishes but for the topological helical case, both thermal as well as shot noise contribute.  }

\begin{widetext}
    
\begin{figure}[h!] 
\centering
\includegraphics[width=1.00\linewidth]{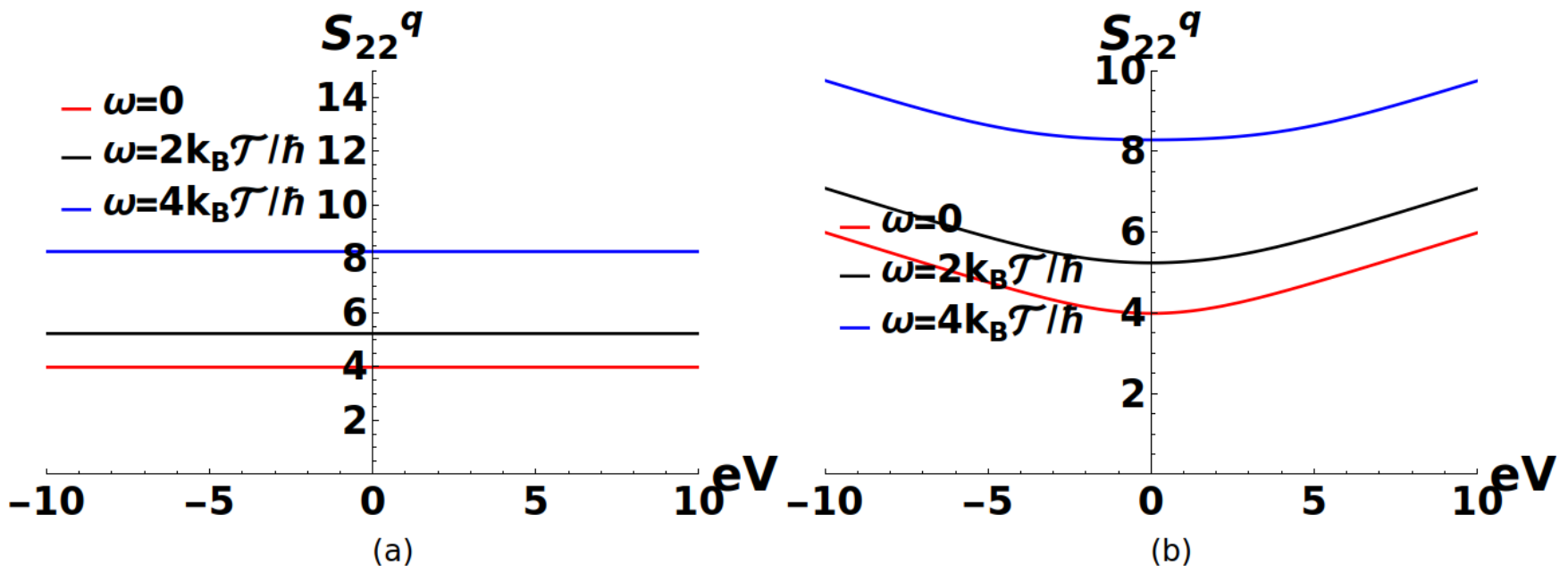}
\caption{{(a) Quantum noise correlation $S_{22}^q$ for chiral edge modes, (b) Quantum noise autocorrelation $S_{22}^q$ in units of $\frac{2e^2}{h}k_B \mathcal{T}$ versus the voltage bias $eV$ in units of $k_B \mathcal{T}$ with $R = 0.5$ and $\mu_0 = 100k_B \mathcal{T}$ at zero frequency as well as finite frequencies in setup 1 and 3.}}
\label{fig:15}
\end{figure}
\end{widetext}

{The quantum noise correlation given in Eq. (\ref{eq:7}) are at finite frequency. Now, the quantum noise autocorrelation $S_{22}^q$ can be written for chiral case as,}
\begin{equation} \label{eq:147}
\begin{split}
S_{22}^q(\omega) =\frac{2e^2}{h}\sum_{\gamma,\delta} \int dE  [A_{\gamma \delta}(2;E,E+\hbar \omega)
\\\times A_{\delta \gamma}(2; E +\hbar \omega,E)(f_{\gamma}(E)[1-f_{\delta}(E+\hbar \omega)]\\+[1-f_{\gamma}(E)]f_{\delta}(E+\hbar \omega)).
\end{split}
\end{equation}

{where, $A_{\gamma \delta}(2;E,E+\hbar \omega) = \delta_{2 \gamma} \delta_{2 \delta} - s_{2 \gamma}^{\dagger}(E) s_{2 \delta}(E + \hbar \omega)$. The s-matrix elements are energy-independent, thus $S_{22}^q$ reduces to,}
\begin{equation} \label{eq:148}
\begin{split}
S_{22}^q(\omega) =\frac{2e^2}{h}\sum_{\gamma,\delta} \int dE  [A_{\gamma \delta}(2)
A_{\delta \gamma}(2)\\\times(f_{\gamma}(E)[1-f_{\delta}(E+\hbar \omega)]+[1-f_{\gamma}(E)]f_{\delta}(E+\hbar \omega)).
\end{split}
\end{equation}

{Using the s-matrix for the chiral case as given in Eq. (\ref{eq:12}), we get,}
\begin{equation} \label{eq:149}
\begin{split}
    {S_{22}^q = \frac{4e^2}{h}\int_0^{\infty}dE (f_0(E)(1-f_0(E+\hbar \omega)}\\ {+ (1-f_0)(f_0(E + \hbar \omega))}.
    \end{split}
\end{equation}
where,
\begin{equation*} 
\begin{split}
    {f_0(E) = \frac{1}{1 + e^{\frac{E-\mu_0}{k_B \mathcal{T}}}}, }
    {\quad f_0(E + \hbar \omega) = \frac{1}{1 + e^{\frac{E-\mu_0 + \hbar \omega}{k_B \mathcal{T}}}}}.
    \end{split}
\end{equation*}

{At $\omega = 0$, Eq. (\ref{eq:149}) becomes,}

\begin{equation} \label{eq:150}
    S_{22}^q = \frac{8e^2}{h}\int_0^{\infty} dE f_0(1-f_0),
\end{equation}
which is same as Eq. (\ref{eq:39}) for chiral edge mode transport in setup 1.

{Eq. (\ref{eq:149}) gives us the effect of finite frequency as seen in Fig. \ref{fig:15}}. {For $\omega = 0$, we recover the zero frequency autocorrelation, which is only thermal-noise-like for chiral case, as also shown in Fig. \ref{fig:15}(a), it is constant with respect to the applied voltage $eV$. At finite frequencies also, it is constant with a different magnitude and with only thermal noise-like contribution, see Eq. (\ref{eq:150}).}

{Similarly, we can analyze the quantum noise correlation at finite temperature as well as finite frequency in the case of topological helical edge modes in setup 1 using Eq. (\ref{eq:51}) and the $s$-matrix given in Eq. (\ref{eq:66}). The general expression for quantum noise autocorrelation $S_{22}^q$ is}
\begin{equation} \label{eq:151}
    S_{22}^q =  \sum_{\sigma, \sigma' = \uparrow/\downarrow} S_{22}^{\sigma \sigma'_, q} .
\end{equation}
where,
\begin{equation} \label{eq:152}
\begin{split}
S_{22}^{\sigma \sigma'_, q} =\frac{e^2}{h}\sum_{\rho, \rho' = \uparrow, \downarrow}\sum_{\gamma,\delta}\int dE Tr[A_{\gamma \delta}^{\rho \rho'}(2,\sigma)A_{\delta \gamma}^{\rho' \rho}(2, \sigma')]\\ \times
(f_{\gamma}(E)[1-f_{\delta}(E + \hbar \omega)]+[1-f_{\gamma}(E + \hbar \omega)]f_{\delta}(E)),
\end{split}
\end{equation}

 with, $A_{\gamma \delta}^{\rho \rho'}(2,\sigma) = \delta_{2 \gamma} \delta_{2 \delta} \delta_{\sigma \rho}\delta_{\sigma \rho'} - s_{2 \gamma}^{\sigma \rho \dagger} s_{2 \delta}^{\sigma \rho'}$. For the helical case, we do not have energy dependence in $s$-matrix elements, so the only frequency dependence in $S_{22}^q$ comes from the Fermi-Dirac distribution.
{Taking care of all the spin-polarised components, the quantum noise autocorrelation $S_{22}^q$ for non-zero frequencies is given as, }
\begin{widetext}

\begin{equation} \label{eq:153}
    \begin{split}
        S_{22}^q = \frac{e^2}{h}\int_0^{\infty}dE (2(f_0(E)(1-f_0(E+\hbar \omega))+(1-f_0(E))f_0(E + \hbar \omega)) + T^2(f(E)(1-f(E + \hbar \omega))\\+f(E + \hbar \omega)(1-f(E)))RT(f_0(E)(1-f(E + \hbar \omega))+(1-f_0(E))f(E + \hbar \omega)+f(E)(1-f_0(E + \hbar \omega))\\+f_0(E + \hbar \omega)(1-f(E)))+R^2(f_0(E)(1-f_0(E + \hbar \omega))+f_0(E + \hbar \omega)(1-f_0(E)))).
    \end{split}
\end{equation}
\end{widetext}

\begin{widetext}
    
 \begin{figure}[h!] 
\centering
\includegraphics[width=1.00\linewidth]{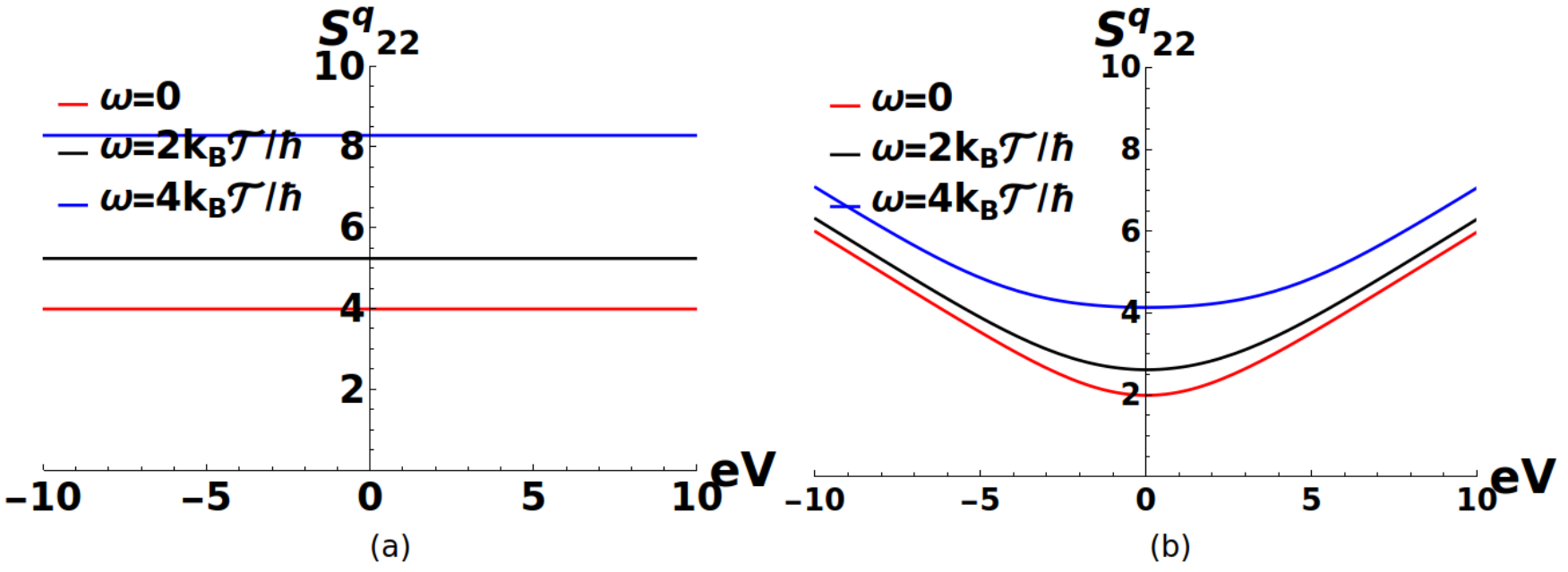}
\caption{{(a) Quantum noise correlation $S_{22}^q$ for topological helical edge modes, (b) Quantum noise autocorrelation $S_{22}^q$ in units of $\frac{2e^2}{h}k_B \mathcal{T}$ versus the voltage bias $eV$ in units of $k_B \mathcal{T}$ with $R = 0.5, p = 0.5$ and $\mu_0 = 100k_B \mathcal{T}$ at zero frequency as well as finite frequencies in setup 2.}}
\label{fig:16}
\end{figure}
\end{widetext}

{For $\omega = 0$, we get back Eq. (\ref{eq:93}) for the topological helical case. The behaviour is shown in the plot Fig. \ref{fig:17}(b). We can see that for finite frequencies also the quantum noise autocorrelation $S_{22}^q$ changes with applied bias voltage $eV$. We see there is a clear difference between chiral and topological helical edge modes at zero frequency as well as at finite frequencies. The ability for quantum noise to distinguish between chiral and topological at zero frequency and finite temperature is retained at finite frequencies and finite temperatures too.}

{Now, to distinguish trivial helical from topological helical, we look at $S_{22}^q$ in setup 2, because in setup 1, $S_{22}^q$ is unable to distinguish between trivial and topological helical edge modes, because in setup 1 $S_{22}^q$ consists of both thermal noise as well as shot noise-like contributions for both trivial and topological helical edge modes, which makes the distinction between them impossible. To calculate $S_{22}^q$ for trivial helical, we use Eq. (\ref{eq:152}) and the $s$-matrix given in Eq. (\ref{eq:108}). Taking care of all the spin-polarised components, we get $S_{22}^q$ to be,}

\begin{widetext}
    \begin{equation} \label{eq:154}
    \begin{split}
        S_{22}^q = \frac{e^2}{h}\int_0^{\infty}dE (2(1-p)^2(f(E)(1-f(E + \hbar \omega))+(1-f(E))f(E + \hbar \omega))+2(1-p)p(f_0(E)(1-f(E + \hbar \omega))\\+f(E + \hbar \omega)(1-f_0(E)))+2(1+(1-p)^2)(f_0(E)(1-f_0(E + \hbar \omega))+f_0(E + \hbar \omega)(1-f_0(E)))\\+2(1-p)p(f(E)(1-f_0(E + \hbar \omega))+f_0(E + \hbar \omega)(1-f(E)))).
        \end{split}
    \end{equation}
\end{widetext}

{At zero frequency in Eq. (\ref{eq:154}), in case of topological helical edge modes ($p=0$), we get back Eq. (\ref{eq:98}), and for trivial helical edge modes, we get back Eq. (\ref{eq:135}) for setup 2. One can distinguish between trivial and topological helical edge modes as shown in Fig. \ref{fig:16}. In Figs. \ref{fig:16}(a),(b), we have plotted the quantum noise autocorrelation $S_{22}^q$ for topological (a) and trivial helical edge modes (b) in setup 2 with $\mu_1 = \mu_3$ and $\mu_2 = \mu_4$. At zero frequency, for topological helical edge mode, $S_{22}^q$ is constant as a function of applied voltage bias ($eV$), but for trivial helical edge mode, it changes symmetrically, which distinguishes trivial from topological helical edge modes at zero frequency. Similarly, at finite frequencies too, the autocorrelation $S_{22}^q$ for topological helical is constant, whereas for trivial helical it changes with applied voltage bias symmetrically. Therefore, we conclude that in the presence of finite frequencies also, one can see the distinction between different edge modes of quantum transport.}

\section{Experimental realization and Conclusion} \label{section V}

\begin{figure}
\centering
\includegraphics[width=1.00\linewidth]{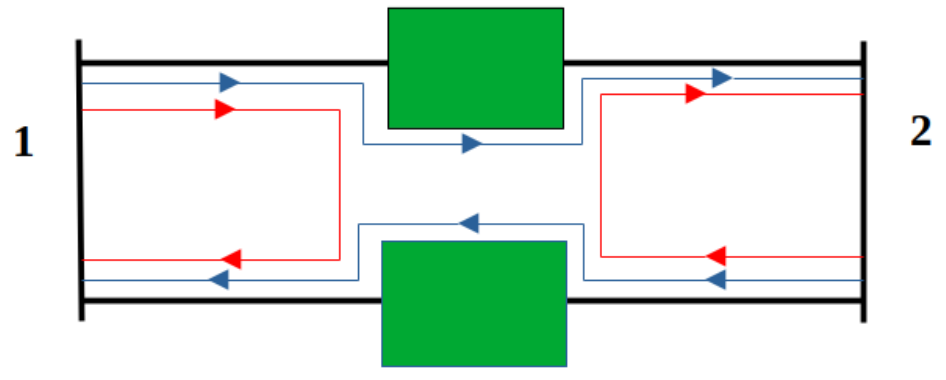}
\caption{{Two terminal quantum Hall setup with chiral edge modes with one constriction. Blue solid lines are for the edge modes transmitting and red solid lines are for the edge modes reflecting via constriction.}}
 \label{fig:17}
\end{figure}

{First, we will look at how to distinguish between thermal and shot noise experimentally in our setups, by looking at a simple 2-terminal setup.  The general expression for both quantum noise autocorrelation and cross-correlation in the case of a two-terminal quantum Hall sample in the presence of a constriction as shown in Fig. \ref{fig:17} is,}
\begin{equation} \label{eq:155}
\begin{split}
        {S_{11}^q = S_{22}^q = -S_{12}^q = -S_{21}^q}  {= \frac{2e^2}{h}T 4 k_B \mathcal{T}}\\ {+ \frac{4e^2}{h}|eV|T(1-T)\left(\coth\left[\frac{eV}{2k_B \mathcal{T}}\right]-\frac{2k_B \mathcal{T}}{eV}\right).}
         \end{split}
    \end{equation}
    {with, $T$ = the transmission probability of the constriction, \\$V$ = the voltage bias applied between terminals 1 and 2. \\$\mathcal{T}$ is Temperature of the setup, and $e$ is the electronic charge.}

{The derivation of Eq. (\ref{eq:155}) is shown in Appendix~\ref{Appendix A.1} using wave-packet approach and Appendix~\ref{Appendix A.2} using Buttiker's approach. The first term in Eq.~(\ref{eq:155}), which is proportional to $k_B \mathcal{T}$ is thermal noise-like contribution ($S_{12}^{th}$). The second term is the shot noise-like contribution ($S_{12}^{sh}$), which is proportional to both the applied voltage $|e V|$ and $T(1-T)$ and arises due to the quantum nature of charge particles. We can distinguish the thermal noise and shot noise experimentally in the following way.}
     \begin{enumerate}
        \item  {Having a finite temperature is a necessary and sufficient condition to see thermal noise. So, it will be significant if the applied voltage bias goes to zero or in the limit $eV \ll k_B \mathcal{T}$. The shot noise like contribution ($S_{12}^{sh}$) according to Eq. (\ref{eq:155}) is,}
        
        \begin{equation} \label{eq:156}
            {S_{12}^{sh} = -\frac{4e^2}{h}|eV|T(1-T)\bigg(\coth \left[\frac{eV}{2k_B \mathcal{T}}\right] - \frac{2k_B \mathcal{T}}{eV}\bigg)}
        \end{equation}
        {In the regime, $eV\ll k_B \mathcal{T}$, $\coth \left[\frac{eV}{2k_B \mathcal{T}}\right] = \frac{2k_B \mathcal{T}}{eV}$, which gives,}
        \begin{equation} \label{eq:157}
            {S_{12}^{sh} = 0.}
        \end{equation}
        {Thus, the total quantum noise is given as,}
        \begin{equation} \label{eq:158}
            {S_{12}^q = S_{12}^{th} = -\frac{8e^2}{h}Tk_B \mathcal{T}}
        \end{equation}
        
        \item  {Experimentally, we can distinguish the shot noise in the regime $eV \gg k_B \mathcal{T}$ or at zero kelvin. At zero temperature, the thermal noise-like contribution $S_{12}^{th}$ vanishes, and there will be only a shot noise-like contribution. At $\mathcal{T} \longrightarrow 0$, the shot noise contribution ($S_{12}^{sh}$) becomes,}
        \begin{equation} \label{eq:159}
            {S_{12}^{sh} = -\frac{4e^2}{h}|eV| T(1-T) }
        \end{equation}
        
    \end{enumerate}

To conclude, this work has made an essential contribution to distinguishing between topological and trivial helical edge modes. By analyzing the finite
temperature quantum noise cross-correlation and autocorrelation, we have been able to distinguish between different edge modes, such as chiral, topological, and trivial
helical edge modes. The results, presented in Tables~\ref{Table 1}, \ref{Table 2}, \ref{Table 3}, \ref{Table 4}, \ref{Table 5}, \ref{Table 6} provide valuable insights into the behaviour of
these edge modes under different conditions and demonstrate the usefulness of noise correlations in distinguishing between topological and trivial helical edge modes. Our
work has considered finite temperatures and a diversity of
setups, which are critical factors in experimental studies. The findings of our study have important implications
for future research and may help guide the development
of new experimental techniques.

This work provides a significant result, which is of experimental relevance since we have taken care of finite temperatures. {This work can be further extended to systems with a temperature gradient instead of a voltage gradient, so-called $\Delta_T$ noise, where the average current in the system is zero. This field is exciting and the experimental work \cite{Lumbroso} has shown the $\Delta_T$ noise in a two-terminal ballistic sample. This $\Delta_T$ noise has been studied theoretically in \cite{PhysRevLett.125.086801},  \cite{PhysRevResearch.4.043191} in the context of fractional quantum Hall systems.} {By studying the $\Delta_T$ noise, one can also have a more in-depth study of the nature of edge modes as well, which is one of our future aims. The nature of edge modes can still be studied in the presence of both temperature gradient as well as voltage gradient by imposing the average current to be zero.} We also plan to extend our study to thermoelectric properties, which would also
be an exciting area of research that could shed further
light on the nature of topological and trivial edge modes.

\acknowledgments {The grants supported this work ``Josephson junctions with strained Dirac materials and their application in quantum information processing'' from the Science and Engineering Research Board (SERB), New Delhi, Government of India, under Grant No. CRG/2019/006258.}

\section{Appendix}{In Appendix \ref{Appendix A}, we have given a historical perspective regarding the calculation of quantum noise correlations via the wavepacket approach in Appendix \ref{Appendix A.1} and directly by Buttiker's approach using Eq. (\ref{eq:8}) in Appendix \ref{Appendix A.2} using a two-terminal QH sample and showed the inconsistency in using Buttiker's result, while calculating quantum noise cross-correlation. Although the thermal noise-like contributions were similar to that of the wavepacket approach, but shot noise-like correlations were not matching with that of the wavepacket approach, which has been resolved by introducing two energy-dependent functions $f_a$ and $f_b$. Next, we derived thermal noise-like in Appendix \ref{Appendix B.1} as well-shot noise-like contributions in Appendix \ref{Appendix B 2} for chiral edge mode transport in a multiterminal QH sample. Similarly, the derivations for both thermal noise-like and shot noise-like contributions for helical edge mode transport are given in Appendices \ref{Appendix C 1} and \ref{Appendix C 2} for a multiterminal QSH sample. In the main text, we have used these expressions to distinguish different modes of transport such as chiral, helical, and trivial edge modes.  }
\appendix
\section{{Historical Perspective on Quantum noise}} \label{Appendix A}
{Here, we give a historical perspective on the need for two energy-dependent functions $f_a$ and $f_b$ in a more detailed way via an example.} {Consider a QH setup with two reservoirs with Fermi-Dirac distributions $f_1$ and $f_2$ with chemical potentials $\mu_1$ and $\mu_2$ respectively at finite temperature $\mathcal{T}$ as shown in Fig. (\ref{fig:17}). Suppose, we apply a voltage bias across the conductor i.e. $\mu_1 - \mu_2 = eV$. Inside the setup, there is a scatterer, with an s-matrix given by}
\begin{equation} \label{eq:A1}
    {s = \begin{pmatrix}
        r & t\\
        t & r
    \end{pmatrix}} ,
\end{equation}
{where, $r$ and $t$ are the reflection and, transmission amplitude of the scatterer, and $R = |r|^2, T = |t|^2$ are the reflection and, transmission probabilities of the scatterer, respectively. In the following subsections, we will derive the quantum noise cross-correlation $S_{12}^q$ or $S_{21}^q$ via two approaches, one is by wave-packet approach as given by \cite{LANDAUER1991167, PhysRevB.45.1742,imry} in and another is by using the formulae derived above using the approach used in \cite{PhysRevB.46.12485, BUTTIKER1991199}.}
\subsection{{Wave-packet approach or Martin and Landaeur's approach \cite{LANDAUER1991167, PhysRevB.45.1742,imry}}} \label{Appendix A.1}
{Here, we derive the two terminal quantum noise in a chiral quantum Hall setup using the wavepacket approach. The current-current correlation between any terminal $\alpha$ and $\beta$ is given as,}
\begin{equation} \label{eq:A2}
    {S_{\alpha \beta}(t) = \frac{1}{\bar{T}}\int dt' \langle \Delta I_{\alpha}(t') \Delta I_{\beta}(t+t') \rangle}.
\end{equation}
{Where, $\Delta I_{\alpha}(t')$ is the deviation of current from its average, i.e., $\Delta I_{\alpha}(t') = I_{\alpha}(t') - \langle I_{\alpha}(t') \rangle$}
{where $\bar{T}$ is the time period. The noise power (quantum noise) can be found by taking the Fourier transform of current $I(t')$. The noise power is given as,}
\begin{equation} \label{eq:A3}
   {S_{11}(\omega)} {=}  \
    {\lim_{\bar{T}\to\infty} \frac{1}{\bar{T}} \langle |\Delta I_1(\omega)|^2 \rangle},
\end{equation}
{where, $\langle |\Delta I(\omega)^2| \rangle = \langle I_1(\omega)^2 \rangle - \langle I_1(\omega) \rangle^2$. According to the wavepacket approach, the current in terminal 1 through a sample is given as,}
\begin{equation} \label{eq:A4}
    {I_1(t) = \sum_n j_n(t - n\tau)g_n}.
\end{equation}

{Here, the current is seen as a superposition of successive wavepackets, with $j_n(t-n\tau)$ being the current 
associated with $n$th wavepacket, and $\tau$ is the difference in time between two successive wavepackets. These wavepackets, which represent the electrons, are separated in time and can in principle overlap in space. Considering the energy states in the energy interval $[E - \Delta E/2]$ and $[E + \Delta E/2]$, means within the small energy range $\Delta E$, which is very small, $\tau$ becomes $\hbar / \Delta E$, which can ensure that the successive wavepackets can be orthogonal to each other. The wavepackets can be viewed as pulses, with $g_n$ as an occupation factor associated with the particular pulse as shown in Fig. 2 of \cite{LANDAUER1991167}. It will take the value +1 if the electron, travels from the left terminal to the right terminal of the sample, -1 if the electron travels in the opposite direction, and 0 if no electron is transmitted at all. Now, using the Fourier component of current to be,}
\begin{equation} \label{eq:A5}
    {I_1(\omega) = \sum_n g_n j(\omega)e^{i \omega n \tau}}.
\end{equation}
{For an electric charge, $\langle |I_1(\omega)|^2 \rangle$ is given as,}
\begin{equation} \label{eq:A6}
    {\langle |I_1(\omega)|^2 \rangle = \int_0^{\infty} dt |I_1 (\omega)|^2 = e^2 \sum_{m,n} \langle g_m g_n \rangle e^{i \omega \tau(m-n)}.}
\end{equation}
{where,}
\begin{equation} \label{eq:A7}
    {\int_{0}^{\infty} dt j(\omega)^2 = e^2}.
\end{equation}
{Now, since the $m$th and $n$th wavepacket do not overlap and are orthogonal to each other, then only the term $m = n$ will survive to the summation. So, the above equation becomes,}
\begin{equation} \label{eq:A8}
    {\langle |I_1(\omega)|^2 \rangle = e^2 \sum_n \langle g_n^2 \rangle}.
\end{equation}
{Simlarly, the square of the average of $|\langle I_1(\omega) \rangle|^2 = e^2 \sum_n \langle g_n \rangle^2$. So, $|\Delta I_1 (\omega)|^2$ is give as,}
\begin{equation} \label{eq:A9}
    \langle|\Delta I_1 (\omega)|^2\rangle = \langle |I_1 (\omega)|^2 \rangle - \langle |I_1(\omega)| \rangle^2 = e^2 \sum_n (\langle g_n^2 \rangle - \langle g_n \rangle^2)
\end{equation}
{All pulses contribute to the noise in the same fashion, so the summation over $n$ in Eq. (\ref{eq:A9}) is $\bar{T}/\tau$.  Considering states in the wavepacket with energy states within the energy interval $\Delta E$ and choosing $\tau = \hbar/\Delta E$ and $\omega = 0$, Eq. (\ref{eq:A9}) becomes}
\begin{equation} \label{eq:A10}
    \langle|\Delta I_1 (\omega)|^2\rangle = \frac{e^2 \bar{T} \Delta E}{\pi \hbar} (\langle g^2 \rangle - \langle g \rangle^2)
\end{equation}

{Here, we dropped the index $n$ and used $g_n = g$ for any $n$. The quantum noise autocorrelation using Eq. (\ref{eq:A3}) at $\omega = 0$ is,}
\begin{equation} \label{eq:A11}
    S_{11}(\omega = 0) = \frac{4 \Delta E e^2}{h} (\langle g^2 \rangle - \langle g \rangle^2)
\end{equation}

{Now, doing the summation over all energy intervals, we get the quantum noise autocorrelation in terminal 1 at zero frequency to be,}
\begin{equation} \label{eq:A12}
    {S_{11}(\omega = 0) = \frac{4e^2}{h}\int dE(\langle g^2 \rangle - \langle g \rangle^2)}.
\end{equation}

{In Eq. (\ref{eq:A12}), the extra factor of 2 has been taken to take care of spin degeneracy. We calculate $\langle g^2 \rangle - \langle g \rangle^2$ as follows. The reservoirs have the occupation probabilities, given by Fermi-Dirac distributions $f_1$ and $f_2$. Below, we describe all possibilities by considering the trajectories of two electrons from terminal 1 and terminal 2.}


\begin{enumerate}
\item {When an electron, transmits from the left terminal to the right terminal, the probability $P_1$ (weight factor) is $f_1(1-f_2)T$. Here, we can understand this process in the following way. The electron incident from the left with probability $f_1$ then transmits via the barrier with probability $T$ and then is incident onto the right terminal of an unoccupied electron state with probability ($1-f_2$). We can thus find the total probability for this process. The $g$ value is +1, since the electron has, transmitted from terminal 1 to terminal 2.}

\item {The probability $P_2$ (weight factor) for an electron to transmit from terminal 2 to 1 is given as $f_2(1-f_1)T$ and the value of $g$ recorded is -1.}

\item {The probability (weight factor) for the electron to reflect terminal 1 after it is incident from terminal 1 is $f_1(1-f_2)(1-T)$. Here, after the reflection the electron is incident on the unoccupied state with probability ($1-f_2$). This situation is quite similar to the situation (1) but with reflection. Here, no current results, with the pulse $g = 0.$}

\item {The probability (weight factor) for the electron to reflect to terminal 2 after it is incident from terminal 2 is $f_2(1-f_1)(1-T)$, where again no current results, with the pulse $g=0$.}
\end{enumerate}
{$\langle g \rangle$ is the average value of the pulse. Now, the deviation of $g$ value from its average $\langle g \rangle$ in each of the above trajectory are given as,}

{(1) $1 - \langle g \rangle$,}\quad {(2) $-1 - \langle g \rangle$,}\quad {(3) -$\langle g \rangle$,} \quad {(4) -$\langle g \rangle$}

{Now, the expression $\langle g^2 \rangle - \langle g \rangle^2$ is found by squaring the above deviations and multiplying by its weight factor. This can be written as,}
\begin{equation} \label{eq:A13}
\begin{split}
    {\langle g^2 \rangle - \langle g \rangle^2 = (1 - \langle g \rangle)^2 P_1 + (1 + \langle g \rangle)^2 P_2} \\{+ \langle g \rangle^2 (1 - P_1 - P_2)}
    \end{split}
\end{equation}
{Eq. (\ref{eq:A13}) is rewritten as,}
\begin{equation} \label{eq:A14}
    {\langle g^2 \rangle - \langle g \rangle^2 = P_1 + P_2 - 2\langle g \rangle (P_1 - P_2) + \langle g \rangle^2}.
\end{equation}

{The average net pulse $\langle g \rangle$ can be written as $P_1 - P_2$. Using $\langle g \rangle$ in Eq. (\ref{eq:A14}), we get,}
\begin{equation} \label{eq:A15}
    {\langle g^2 \rangle - \langle g \rangle^2 = P_1 + P_2 - (P_1 - P_2)^2}.
\end{equation}

{Substituting the value of $P_1$ and $P_2$, we get,}
\begin{equation} \label{eq:A16}
\begin{split}
    {\langle g^2 \rangle - \langle g \rangle^2 = f_1(1-f_1)T + f_2(1-f_2)T}\\
    {+ T(1-T)(f_1 - f_2)^2}.
    \end{split}
\end{equation}

{So, according to the wavepacket approach, the quantum noise correlation is given as,}
\begin{equation} \label{eq:A17}
\begin{split}
    {S_{11}^q = \frac{4e^2}{h}\int_0^{\infty} dE(T f_1(1-f_1) + T f_2(1-f_2)} \\{+ T(1-T) (f_1 - f_2)^2)}.
    \end{split}
\end{equation}
{According to the current conservation, we require $\Delta I_1 = - \Delta I_2$, which gives rise to $\langle (\Delta I_1)^2 \rangle = - \langle \Delta I_1 \Delta I_2 \rangle$}
{Then the quantum noise cross-correlation $S_{12}^q$ is given as,}
\begin{equation} \label{eq:A18}
    \begin{split}
    {S_{12}^q = -\frac{4e^2}{h}\int_0^{\infty} dE(T f_1(1-f_1) + T f_2(1-f_2)} \\{+ T(1-T) (f_1 - f_2)^2)}.
    \end{split}
\end{equation}
{This is the appropriate form for the quantum noise cross correlation between terminals 1 and 2. The first two terms are thermal noise-like contribution, which deals with the direct tunneling of electrons from one terminal to another, and always thermal noise-like. The second term is transport contribution, which is a non-equilibrium phenomenon. Now, the integrals involving the Fermi function can be written as,}
\begin{equation} \label{eq:A19}
\begin{split}
    {\int_0^{\infty}dE f_{1}(1-f_1)} = {k_B \mathcal{T} \left(1 - \frac{1}{1 + e^{\frac{\mu_1}{k_B \mathcal{T}}}}\right),}\\
    {\int_0^{\infty}dE f_{2}(1-f_2)} = {k_B \mathcal{T} \left(1 - \frac{1}{1 + e^{\frac{\mu_2}{k_B \mathcal{T}}}}\right),}\\
    {\int_0^{\infty}dE (f_1 - f_2)^2 = k_B \mathcal{T}\bigg[-2 + \frac{1}{1 + e^{\frac{\mu_1}{k_B \mathcal{T}}}} + \frac{1}{1 + e^{\frac{\mu_1}{k_B \mathcal{T}}}}} \\ {+ \coth \left(\frac{\mu - \mu_0}{k_B \mathcal{T}}\right)\left(\log \left[1 + e^{\frac{\mu_1}{k_B \mathcal{T}}}\right] - \log \left[1 + e^{\frac{\mu_2}{k_B \mathcal{T}}}\right]\right)\bigg].}
    \end{split}
\end{equation}

For $\mu_1, \mu_2 \gg k_B \mathcal{T}$,
\begin{equation} \label{eq:A20}
\begin{split}
    {\int_0^{\infty}dE f_1(1-f_1) = \int_0^{\infty}dE f_2(1-f_2) = k_B \mathcal{T},}\\
    {\int_0^{\infty}dE (f_1 - f_2)^2 = (\mu_1 - \mu_2)\coth\left[\frac{(\mu_1 - \mu_2)}{2k_B \mathcal{T}} - 2k_B \mathcal{T}\right]}.
    \end{split}
\end{equation}
{We get the final quantum noise cross-correlation to be,}
    \begin{equation} \label{eq:A21}
\begin{split}
    {S_{12}^q}  {= -\frac{8e^2}{h}k_B  \mathcal{T}T  - \frac{4e^2}{h} RT\left(eV \coth \left[\frac{eV}{2k_B \mathcal{T}}\right]-2k_B \mathcal{T}\right).}
    \end{split}
\end{equation}

{where, $\mu_1 - \mu_2 = eV$. At zero temperature and for $|eV| = 0$, quantum noise vanishes, which can be confirmed by Eq. (\ref{eq:A21}). The first term in Eq. (\ref{eq:A21}) is the thermal noise-like contribution, whereas the second term is a shot noise-like contribution.}
\subsection{{Buttiker's approach\cite{PhysRevB.46.12485},\cite{BUTTIKER1991199}}} \label{Appendix A.2}

{In a two-terminal QH setup, current conservation demands $\Delta I_1 = - \Delta I_2$, where $\Delta I_i$ is fluctuation in the current from its average value, i.e., $\Delta I_i = I_i - \langle I_i \rangle$ for $i = 1, 2$. We can also find quantum noise cross-correlation ($S_{12}^q$) for a two-terminal QH setup with chiral edge modes directly using Eq. (\ref{eq:8}), which is Buttiker's approach. According to Eq. (\ref{eq:8}), $S_{12}^q$ is given as,}
    \begin{equation} \label{eq:A22}
\begin{split}
S_{12}^q =\frac{2e^2}{h}\sum_{\gamma,\delta = 1,2}\int dE [A_{\gamma \delta}(1)
A_{\delta \gamma}(2)]\\\times (f_{\gamma}(E)[1-f_{\delta}(E)]+[1-f_{\gamma}(E)]f_{\delta}(E)).
\end{split}
\end{equation}
{where, $A_{\gamma \delta}(i) = \delta_{i \gamma}\delta_{i \delta} - s_{i \gamma}^{\dagger}s_{i \delta}$ for $i = 1,2$. Using the $s$-matrix given in Eq. (\ref{eq:A1}), we get $S_{12}^q$ as,}
\begin{equation} \label{eq:A23}
    \begin{split}
    {S_{12}^q = -\frac{4e^2}{h}\int_0^{\infty} dE(T f_1(1-f_1) + T f_2(1-f_2)} \\{+ T(1-T) (f_1 + f_2)^2)}.
    \end{split}
\end{equation}


{The first two terms are thermal noise-like contributions, which is same as that of Eq. (\ref{eq:A18}), but the third term, which is expected to be shot noise-like is not same as that of Eq. (\ref{eq:A18}). Further, doing the integration in the limit $\mu_1, \mu_2 \gg k_B \mathcal{T}$ for shot noise-like contribution, we get,}
\begin{equation} \label{eq:A24}
\begin{split}
    {S_{12}^{sh} =  -\frac{4e^2}{h} RT \bigg( \mu_1 + 3\mu_2} {- 2\left(\frac{eV}{(e^{\frac{eV}{k_B \mathcal{T}}}-1)}\right)-2k_B \mathcal{T}\bigg)}.
    \end{split}
\end{equation}
{As we can see, this expression is not the same as that in Eq. (\ref{eq:A21}). Secondly at $eV = 0$, i.e., $\mu_1 = \mu_2$, we get,} 
\begin{equation} \label{eq:A25}
    {S_{12}^{sh} = -\frac{4e^2}{h}RT 4\mu_2}.
\end{equation}

{From Eq. (\ref{eq:A25}), we see that for $eV = 0$, the shot noise-like contribution $S_{12}^{sh}$ is nonzero as $\mu_2$ is nonzero in the limit $\mu_1, \mu_2 \gg k_B \mathcal{T}$, which contradicts the results obtained from the wavepacket approach in Eq. (\ref{eq:A21}) and this is an unphysical result, since for zero applied bias voltage, there will be no current. Therefore the shot noise should vanish in such a situation. This problem can be resolved in the following way. We first decompose the quantum noise expression $S_{\alpha \beta}^q$ into thermal noise-like and shot noise-like contributions and further modify shot noise-like contribution as done in \cite{PhysRevB.46.12485, BUTTIKER1991199} to match with that of the wave packet approach.}

If both the terminals are at the same chemical potential means the same Fermi-Dirac distribution $f_1$, then Eq. (\ref{eq:A22}) becomes,
\begin{equation} \label{eq:A26}
\begin{split}
    S_{12}^q = \frac{4e^2}{h}\int dE \left(f_1(1-f_1) \sum_{\gamma, \delta = 1,2} [A_{\gamma \delta}(1)A_{\delta \gamma}(2)]\right).
    \end{split}
\end{equation}
Now, 
\begin{equation} \label{eq:A27}
\begin{split}
{\sum_{\gamma, \delta = 1,2}Tr[A_{\gamma \delta}(1)A_{\delta \gamma}(2)] = \sum_{\gamma, \delta = 1,2}Tr(I_{1}\delta_{1 \gamma}\delta_{1 \delta}\delta_{2 \gamma}\delta_{2 \delta}}\\{-\delta_{2 \gamma}\delta_{2 \delta}s_{1 \gamma}^{\dagger}s_{1 \delta}-\delta_{1 \gamma}\delta_{1 \delta}s_{2 \delta}^{\dagger}s_{2 \gamma}+s_{1 \gamma}^{\dagger}s_{1 \delta}s_{2 \delta}^{\dagger}s_{2 \gamma})}.
\end{split}
\end{equation}

{Doing the summation over $\gamma$ and $\delta$ and using unitarity of s-matrix $\sum_{\delta} s_{\delta \beta}^{\dagger}s_{\delta \gamma} = I_{\beta}\delta_{\beta \gamma}$, we get,}
\begin{equation} \label{eq:A28}
\begin{split}
{\sum_{\gamma, \delta}Tr[A_{\gamma \delta}(1)A_{\delta \gamma}(2)] =Tr[-s_{1 2}^{\dagger}s_{1 2}-s_{2 1}^{\dagger}s_{2 1}]},\\
= -T_{21} - T_{12},
\end{split}
\end{equation}
$T_{\alpha \beta}$ is the transmission probability for an electron to scatter from terminal $\beta$ to terminal $\alpha$ for $\alpha, \beta = 1,2$.
So, the thermal noise-like contribution when all the terminals have same Fermi-Dirac distribution is,
\begin{equation} \label{eq:A29}
    S_{12}^{th} = \frac{-4e^2}{h}\int_0^{\infty} dE f_1(1-f_1)[T_{12}+ T_{21}].
\end{equation}
Since terminals 1 and 2 are maintained at different chemical potentials i.e. $f_1$ and $f_2$ respectively, then the thermal noise-like contribution in this case is,
\begin{equation}\label{eq:A30}
     S_{12}^{th} = \frac{-4e^2}{h}\int_0^{\infty} dE [f_2(1-f_2)T_{12}+ f_1(1-f_1)T_{21}].
\end{equation}
As we know $T_{12} = T_{21} = T$ and after the integration in the limit $\mu_1, \mu_2 \gg k_B \mathcal{T}$, we get,
\begin{equation} \label{eq:A31}
    S_{12}^{th} = \frac{-8e^2}{h}T k_B \mathcal{T}.
\end{equation}
Here, we recover the correct thermal noise-like contribution ($S_{12}^{th}$) to $S_{12}^q$, which is the same as both Wavepacket-approach as in Eq. (\ref{eq:A21}).

{For shot noise-like contribution $S_{12}^{sh}$, we have to subtract thermal noise-like contribution $S_{12}^{th}$ as given in Eq. (\ref{eq:A30}) from the quantum noise cross-correlation $S_{12}^q$ as given in Eq. (\ref{eq:A22}), i.e. $S_{12}^{sh} = S_{12}^q - S_{12}^{th}$.}
Using the relation 
 \begin{equation} \label{eq:A32}
\begin{split}
\sum_{\delta = 1,2}f_{\delta}\left(\sum_{\gamma = 1,2} Tr[A_{\gamma \delta}(1)A_{\delta \gamma}(2)]\right) \\= \sum_{\gamma = 1,2}f_{\gamma}\left(\sum_{\delta = 1,2} Tr[A_{\gamma \delta}(1)A_{\delta \gamma}(2)]\right) = -T_{12}f_{2} - T_{21}f_{1},
\end{split}
\end{equation}

This gives the shot noise-like contribution to be,
\begin{equation} \label{eq:A33}
{S_{12}^{sh}} = -4\frac{e^2}{h}\int dE \sum_{\gamma, \delta = 1,2}f_{\gamma}f_{\delta}Tr(s_{1 \gamma}^{\dagger}s_{1 \delta}s_{2 \delta}^{\dagger}s_{2 \gamma}).
\end{equation}

Summing over $\gamma$ and $\delta$, we get,
\begin{equation}  \label{eq:A34}
    {S_{12}^{sh}} = -4\frac{e^2}{h}\int dE RT (f_1 + f_2)^2.
\end{equation}
As we can see the shot noise-like expression is same as that of in Eq. (\ref{eq:A23}) and we have already discussed its disadvantage i.e. at zero applied voltage bias, it does not vanish, which is an unphysical result. This can be tackled by introducing two energy-dependent functions $f_a$ and $f_b$ in Eq. (\ref{eq:A33}). So, the modified shot noise-like cross-correlation is given as \cite{PhysRevB.46.12485},

\begin{equation} \label{eq:A35}
{S_{1 2}^{sh}} = -4\frac{e^2}{h}\int dE \sum_{\gamma, \delta}(f_{\gamma}-f_a)(f_{\delta}-f_b)Tr(s_{1 \gamma}^{\dagger}s_{1 \delta}s_{2 \delta}^{\dagger}s_{2 \gamma}).
\end{equation}
Now, using $f_a$ = $f_b = f_2$, we get $S_{12}^{sh}$ to be,
\begin{equation} \label{eq:A36}
    S_{12}^{sh} = \frac{-4e^2}{h}\int_0^{\infty}dE RT (f_1 - f_2)^2.
\end{equation}
Now, after the integration in the limit $\mu_1, \mu_2 \gg k_B \mathcal{T}$, we get,
\begin{equation} \label{eq:A37}
    S_{12}^{sh} = \frac{-4e^2}{h}\left(eV \coth \left[\frac{eV}{2k_B \mathcal{T}}\right]-2k_B \mathcal{T}\right).
\end{equation}

Here, $f_a$ and $f_b$ is choosen to be $f_2$. There is no unique choice for $f_a$ and $f_b$. In general, it can be any energy-dependent function. The proper reason behind choosing $f_a$ and $f_b$ to be $f_2$ is it allows us to correctly reproduce the result of shot noise-like contribution as derived via the wave packet approach as mentioned in \ref{Appendix A.1}. Secondly with this choice, the shot noise-like contribution vanishes for $eV = 0$. 

So, the total quantum noise cross-correlation $S_{12}^q$ in a two-terminal QH sample using the approach of Buttiker is given as,

\begin{equation} \label{eq:A38}
\begin{split}
    {S_{12}^q}  {= -\frac{8e^2}{h}Tk_B \mathcal{T}  - \frac{4e^2}{h} RT\left(eV \coth \left[\frac{eV}{2k_B \mathcal{T}}\right]-2k_B \mathcal{T}\right).}
    \end{split}
\end{equation}
This expression is same as that of the wavepacket approach as in Eq. (\ref{eq:A21}).

\section{Derivation of {thermal noise-like 
 and shot noise-like contributions} for chiral edge mode transport for a multiterminal QH sample} \label{Appendix B}

In this section, we separately derive the thermal noise-like and shot noise-like contributions for the transport via chiral edge modes.
 \subsection{Derivation of thermal noise-like 
 contribution} \label{Appendix B.1}
The quantum noise correlation for chiral edge modes as derived in Eq. (\ref{eq:8}) is
\begin{equation} \label{eq:B1}
\begin{split}
S_{\alpha \beta}^q =\frac{2e^2}{h}\sum_{\gamma,\delta}\int dE \quad Tr[A_{\gamma \delta}(\alpha)A_{\delta \gamma}(\beta)]\\\times
 (f_{\gamma}(E)[1-f_{\delta}(E)]+[1-f_{\gamma}(E)]f_{\delta}(E)),
\end{split}
\end{equation}
If the system is in thermal equilibrium at temperature $\mathcal{T}$, the Fermi-Dirac distribution in all the terminals is, that (assuming the chemical potential in each terminal are same), and we denote it by $f(E)$. Now, the above expression reduces to
\begin{equation} \label{eq:B2}
{S_{\alpha \beta}^{th}} = \frac{4e^2}{h}\int dE f(E)[1-f(E)] \sum_{\gamma, \delta}Tr[A_{\gamma\delta}(\alpha)A_{\delta \gamma}(\beta)].
\end{equation}
Here, {$A_{\gamma \delta}(\alpha) = I_{\alpha}\delta_{\alpha \gamma}\delta_{\alpha \delta} - s_{\alpha \gamma}^{\dagger}s_{\alpha \delta}$ and similarly, $A_{\delta \gamma}(\beta) = I_{\beta}\delta_{\beta \delta}\delta_{\beta \gamma} - s_{\beta \delta}^{\dagger}s_{\beta \gamma}$.}
{Next, we evaluate $S_{\alpha \beta}^{th}$ in Eq. (\ref{eq:B2}) by first doing a summation of $Tr[A_{\gamma \delta}(\alpha)A_{\delta \gamma}(\beta)]$ over $\gamma$ and $\delta$.} 
\begin{equation} \label{eq:B3}
\begin{split}
{\sum_{\gamma, \delta}Tr[A_{\gamma \delta}(\alpha)A_{\delta \gamma}(\beta)] = \sum_{\gamma \delta}Tr(I_{\alpha}\delta_{\alpha \gamma}\delta_{\alpha \delta}\delta_{\beta \gamma}\delta_{\beta \delta}}\\{-\delta_{\beta \gamma}\delta_{\beta \delta}s_{\alpha \gamma}^{\dagger}s_{\alpha \delta}-\delta_{\alpha \gamma}\delta_{\alpha \delta}s_{\beta \delta}^{\dagger}s_{\beta \gamma}+s_{\alpha \gamma}^{\dagger}s_{\alpha \delta}s_{\beta \delta}^{\dagger}s_{\beta \gamma})},
\end{split}
\end{equation}

{After the summation over $\gamma$ and $\delta$ and using unitarity of s-matrix $\sum_{\alpha} s_{\alpha \beta}^{\dagger}s_{\alpha \gamma} = I_{\beta}\delta_{\beta \gamma}$, we get,}
\begin{equation} \label{eq:B4}
\begin{split}
{\sum_{\gamma, \delta}Tr[A_{\gamma \delta}(\alpha)A_{\delta \gamma}(\beta)] =Tr[2I_{\alpha}\delta_{\beta \alpha}-s_{\alpha \beta}^{\dagger}s_{\alpha \beta}-s_{\beta \alpha}^{\dagger}s_{\beta \alpha}]}.
\end{split}
\end{equation}

{Thus, Eq.~(\ref{eq:B2}) can be written as,}

\begin{equation} \label{eq:B5}
\begin{split}
{S_{\alpha \beta}^{th}} = \frac{4e^2}{h}\int dE f(E)[1-f(E)] [2N_{\alpha}\delta_{\alpha \beta}-Tr(s_{\alpha \beta}^{\dagger}s_{\alpha \beta}\\-s_{\beta \alpha}^{\dagger}s_{\beta \alpha})].
\end{split}
\end{equation}
where, $Tr[I_{\alpha}] = N_{\alpha} $ is the number of edge modes in terminal $\alpha$. This equation is the \textit{Nyquist-Johnson} noise. It is a result of thermal fluctuations alone.

{For the autocorrelation($\alpha=\beta$), we see the thermal noise-like contributions to be,} 
\begin{equation}  \label{eq:B6}
{S_{\alpha \alpha}^{th}} = 8\frac{e^2}{h}\int dE (f(E)(1-f(E))) [N_{\alpha}-R_{\alpha \alpha}],
\end{equation}
$R_{\alpha \alpha}$ is the reflection probability for an electron to reflect from terminal $\alpha$.

For $\alpha \ne \beta$, {the thermal noise-like fluctuation is} 
\begin{equation} \label{eq:B7}
{S_{\alpha \beta}^{th}} = -\frac{4e^2}{h}\int dE (f(E)(1-f(E)))(T_{\alpha \beta} + T_{\beta \alpha}),
\end{equation}
Here, we have assumed the contacts were connected to the reservoirs with the same chemical potential. For the general case, wherein the distribution function at terminals $\alpha$ and $\beta$ are different, which is given as,
\begin{equation} \label{eq:B8}
{S_{\alpha \beta}^{th}}= -\frac{4e^2}{h}\int dE[T_{\alpha \beta}f_{\beta}(1-f_{\beta})+T_{\beta \alpha}f_{\alpha}(1-f_{\alpha})].
\end{equation}
where $f_{\beta}$ and $f_{\alpha}$ are Fermi-Dirac distribution function of terminals $\beta$ and $\alpha$. 
\\
\subsection{Derivation of {shot noise-like contribution}} \label{Appendix B 2}
To derive the expression for {transport like correlation}, we have to subtract the {thermal noise-like fluctuation} from the {quantum noise correlation,} {and for autocorrelation ($\alpha = \beta$), it is given as,} 
\begin{equation} \label{eq:B9}
{S_{\alpha \alpha}^{sh} = S_{\alpha \alpha}^q - S_{\alpha \alpha}^{th}},
\end{equation}
{where,}
\begin{equation} \label{eq:B10}
    \begin{split}
        {S_{\alpha \alpha}^q =\frac{2e^2}{h}\sum_{\gamma,\delta}\int dE \quad Tr[A_{\gamma \delta}(\alpha)}A_{\delta \gamma}(\alpha)]\\\times
{(f_{\gamma}(E)[1-f_{\delta}(E)]+[1-f_{\gamma}(E)]f_{\delta}(E)),}
    \end{split}
\end{equation}
{and,}
\begin{equation} \label{eq:B11}
   {S_{\alpha \alpha}^{th} = 8\frac{e^2}{h}\int dE (f(E)(1-f(E)))[N_{\alpha}-R_{\alpha \alpha}]}.
\end{equation}
Using the relation given in Eq. (\ref{eq:B4}), we can simplify Eq. (\ref{eq:B10}) as,
\begin{equation} \label{eq:B12}
\begin{split}
\sum_{\delta}f_{\delta}\left(\sum_\gamma Tr[A_{\gamma \delta}(\alpha)A_{\delta \gamma}(\beta)]\right)\\ =Tr[I_{\alpha}-2s_{\alpha \alpha}^{\dagger} s_{\alpha \alpha}]f_{\alpha}+\sum_{\alpha}Tr(s_{\alpha \delta}s_{\alpha \delta}^{\dagger})f_{\delta}.
\end{split}
\end{equation}
After doing the algebra, we find {$S_{\alpha \alpha}^{sh}$} to be,
\begin{equation} \label{eq:B13}
\begin{split}
{S_{\alpha \alpha}^{sh}} = \frac{4e^2}{h}\int dE \bigg(\sum_{\gamma}T_{\alpha \gamma}(f_{\gamma} - f_{\alpha})+M_{\alpha}f_{\alpha}^2-\\\sum_{\gamma \delta}Tr(s_{\alpha \gamma}s_{\alpha \delta}s_{\alpha \delta}s_{\alpha \gamma})\bigg).
\end{split}
\end{equation}

For the cross-correlation $\alpha \ne \beta$, the {shot noise-like contribution is,}
\begin{equation} \label{eq:B14}
{S_{\alpha \beta}^{sh} = S_{\alpha \beta}^q - S_{\alpha \beta}^{th}},
\end{equation}
where,
\begin{equation} \label{eq:B15}
\begin{split}
S_{\alpha \beta}^q =\frac{2e^2}{h}\sum_{\gamma,\delta}\int dE \quad Tr[A_{\gamma \delta}(\alpha)A_{\delta \gamma}(\beta)]\\\times
 (f_{\gamma}(E)[1-f_{\delta}(E)]+[1-f_{\gamma}(E)]f_{\delta}(E)),
\end{split}
\end{equation}
{and from Eq. (\ref{eq:B8})}
\begin{equation} \label{eq:B16}
{S_{\alpha \beta}^{th}}= -\frac{4e^2}{h}\int dE[T_{\alpha \beta}f_{\beta}(1-f_{\beta})+T_{\beta \alpha}f_{\alpha}(1-f_{\alpha})].
\end{equation}
Again using Eq. (\ref{eq:B4}), we get, 
\begin{equation} \label{eq:B17}
\begin{split}
\sum_{\delta}f_{\delta}\left(\sum_\gamma Tr[A_{\gamma \delta}(\alpha)A_{\delta \gamma}(\beta)]\right) \\= \sum_{\gamma}f_{\gamma}\left(\sum_\delta Tr[A_{\gamma \delta}(\alpha)A_{\delta \gamma}(\beta)]\right) = -T_{\alpha \beta}f_{\beta} - T_{\beta \alpha}f_{\alpha},
\end{split}
\end{equation}

We get the {shot noise-like contribution} to be,
\begin{equation} \label{eq:B18}
{S_{\alpha \beta}^{sh}} = -4\frac{e^2}{h}\int dE \sum_{\gamma, \delta}f_{\gamma}f_{\delta}Tr(s_{\alpha \gamma}^{\dagger}s_{\alpha \delta}s_{\beta \delta}^{\dagger}s_{\beta \gamma}).
\end{equation}

Usage of Eq. (\ref{eq:B18}), we have certain disadvantages, which can be understood by studying a simple two-terminal QH sample as has been done in Appendix \ref{Appendix A.2}, which motivated Buttiker to use two energy-dependent functions $f_a$ and $f_b$ in Eq. (\ref{eq:B18}).
So, the final correct expression for the {shot noise-like contribution \cite{PhysRevB.46.12485}} is,
\begin{equation} \label{eq:B19}
{S_{\alpha \beta}^{sh}} = -4\frac{e^2}{h}\int dE \sum_{\gamma, \delta}(f_{\gamma}-f_a)(f_{\delta}-f_b)Tr(s_{\alpha \gamma}^{\dagger}s_{\alpha \delta}s_{\beta \delta}^{\dagger}s_{\beta \gamma}).
\end{equation}\\

\section{Derivation of {thermal noise-like contributions} for helical edge mode transport for a multiterminal QSH sample} \label{Appendix C}
In this section, we derive the thermal and shot noise-like contributions separately for helical edge modes. The quantum noise correlation $S_{\alpha \beta}^q$ in setup with helical edge mode transport is given as,
\begin{equation} \label{eq:C1}
    S_{\alpha \beta}^q = \sum_{\sigma, \sigma' = \uparrow/\downarrow} S_{\alpha \beta}^{\sigma \sigma',q}.
\end{equation}
where $S_{\alpha \beta}^{\sigma \sigma',q}$ is a spin-polarised component to $S_{\alpha \beta}^q$.\\

\subsection{Derivation of thermal noise-like contributions} \label{Appendix C 1}

The spin-polarised components of the quantum noise in the presence of helical edge modes are
\begin{equation} \label{eq:C2}
\begin{split}
S_{\alpha \beta}^{\sigma \sigma'_, q} =\frac{e^2}{h}\sum_{\rho, \rho' = \uparrow, \downarrow}\sum_{\gamma,\delta}\int dE \quad Tr[A_{\gamma \delta}^{\rho \rho'}(\alpha,\sigma)
A_{\delta \gamma}^{\rho' \rho}(\beta, \sigma')]\\\times(f_{\gamma}(E)[1-f_{\delta}(E)]+[1-f_{\gamma}(E)]f_{\delta}(E)),
\end{split}
\end{equation}
The quantum noise expression if the Fermi-Dirac distribution in all the terminals are same, {is thermal noise-like}, which is given by
\begin{equation} \label{eq:C3}
\begin{split}
{S_{\alpha \beta}^{\sigma \sigma', th}} = 2\frac{e^2}{h}\int dE f(E)[1-f(E)] \\\sum_{\gamma, \delta}\sum_{\rho, \rho'}Tr[A_{\gamma\delta}^{\rho \rho'}(\alpha, \sigma)A_{\delta \gamma}^{\rho' \rho}(\beta, \sigma')].
\end{split}
\end{equation}
We proceed further using the expression for $A_{\gamma \delta}^{\rho \rho'}(\alpha, \sigma)$ and we get,
\begin{equation} \label{eq:C4}
\begin{split}
{S_{\alpha \beta}^{\sigma \sigma', th}} = \frac{2e^2}{h}\int dE f(E)[1-f(E)]\\ \sum_{\gamma,\delta}Tr[(\delta_{\alpha \gamma}\delta_{\alpha \delta}-s_{\alpha \gamma}^{\sigma \rho \dagger}s_{\alpha \delta}^{\sigma \rho'})(\delta_{\beta \delta}\delta_{\beta \gamma}-s_{\beta \delta}^{\sigma' \rho' \dagger}s_{\beta \gamma}^{\sigma' \rho})].
\end{split}
\end{equation}
Further, by simplifying the algebra as we did for the chiral case, we get
\begin{equation} \label{eq:C5}
\begin{split}
Tr[A_{\gamma \delta}^{\rho \rho'}(\alpha, \sigma)A_{ \delta \gamma}^{\rho' \rho}(\beta, \sigma')] = [2N_{\alpha}\delta_{\beta \alpha}\delta_{\sigma \sigma'}\\-Tr(s_{\alpha \beta}^{\sigma \sigma' \dagger}s_{\alpha \beta}^{\sigma \sigma'})-Tr(s_{\beta \alpha}^{\sigma \sigma' \dagger}s_{\beta \alpha}^{\sigma \sigma'})].
\end{split}
\end{equation}
{So, the final expression for thermal noise-like contributions is given as,}
\begin{equation} \label{eq:C6}
{S_{\alpha \beta}^{\sigma \sigma', th}} = \frac{2e^2}{h}\int dE (f(E)(1-f(E))\times -(T_{\alpha \beta}^{\sigma \sigma'} + T_{\beta \alpha}^{\sigma \sigma'}).
\end{equation}
Here, we have assumed the contacts were connected to the reservoirs with the same chemical potential. For a general case, where the distribution function at terminals $\alpha$ and $\beta$ is different, we can write,
\begin{equation} \label{eq:C7}
{S_{\alpha \beta}^{\sigma \sigma', th}= -\frac{2e^2}{h}\int dE[T_{\alpha \beta}^{\sigma \sigma'}f_{\beta}(1-f_{\beta})+T_{\beta \alpha}^{\sigma \sigma'}f_{\alpha}(1-f_{\alpha})]}.
\end{equation}
Where $f_{\beta}$ and $f_{\alpha}$ are Fermi-Dirac distribution of contacts $\beta$ and $\alpha$. {The thermal noise-like contributions $S_{\alpha \beta}^{th}$ is thermal noise-like only, which deals with direct tunneling of electrons from one terminal to another, which can be written as,}
\begin{equation} \label{eq:C8}
    {S_{\alpha \beta}^{th} = \sum_{\sigma, \sigma' = \uparrow/\downarrow} S_{\alpha \beta}^{\sigma \sigma', th}}.
\end{equation}
Similarly, for $\alpha = \beta$,
the {thermal noise-like contributions} is,
\begin{equation} \label{eq:C9}
S_{\alpha \alpha}^{\sigma \sigma', th} = \frac{4e^2}{h}\int dE f_{\alpha}(1-f_{\alpha})(N_{\alpha}\delta_{\sigma \sigma'} - R_{\alpha \alpha}^{\sigma \sigma'}).
\end{equation}\\
\subsection{Derivation of {shot noise-like contributions}} \label{Appendix C 2}
{To derive the shot noise-like contribution, we} need to subtract the {thermal noise-like fluctuation} from the quantum noise {correlation}. For the autocorrelation $\alpha = \beta$, we have to subtract Eq. (\ref{eq:C9}) from Eq. (\ref{eq:C2}). After doing some algebra, we get the {shot noise-like contribution} to be,
\begin{equation} \label{eq:C10}
\begin{split}
{S_{\alpha \alpha}^{\sigma \sigma', sh}} = \frac{2e^2}{h}\int dE \bigg(\sum_{\gamma}T_{\alpha \gamma}^{\sigma \sigma'}(f_{\gamma} - f_{\alpha})\\+M_{\alpha}\delta_{\sigma \sigma'}f_{\alpha}^2-\sum_{\gamma \delta}\sum_{\rho \rho' = \uparrow, \downarrow}Tr(s_{\alpha \gamma}^{\sigma \rho^{\dagger}}s_{\alpha \delta}^{\sigma \rho'}s_{\alpha \delta}^{\sigma' \rho'^{\dagger}}s_{\alpha \gamma}^{\sigma' \rho})\bigg).
\end{split}
\end{equation}

For the cross-correlation case $\alpha \ne \beta$, we find
\begin{equation} \label{eq:C11}
{S_{\alpha \beta}^{\sigma \sigma', sh}} = S_{\alpha \beta}^{\sigma \sigma' q}-{S_{\alpha \beta}^{\sigma \sigma', th}}.
\end{equation}
We use the formula,
\begin{equation} \label{eq:C12}
\begin{split}
\sum_{\delta}f_{\delta}\left(\sum_{\gamma}Tr[A_{\gamma \delta}^{\rho \rho'}(\alpha, \sigma)A_{\delta \gamma}^{\rho' \rho}(\alpha, \sigma)]\right)=\\\sum_{\gamma}f_{\gamma}\left(\sum_{\delta}Tr[A_{\gamma \delta}^{\rho \rho'}(\alpha, \sigma)A_{\delta \gamma}^{\rho' \rho}(\alpha, \sigma)]\right) = -T_{\alpha \beta}^{\sigma \sigma'}f_{\beta}-T_{\beta \alpha}^{\sigma \sigma'}f_{\alpha}.
\end{split}
\end{equation}\\

Using, Eq. (\ref{eq:C12}) and (\ref{eq:C13}) in, Eq. (\ref{eq:C2}), we find that
\begin{equation} \label{eq:C13}
\begin{split}
{S_{\alpha \beta}^{\sigma \sigma', sh}} = -2\frac{e^2}{h}\int dE \sum_{\gamma, \delta}f_{\gamma}f_{\delta}Tr(s_{\alpha \gamma}^{\sigma \rho \dagger}s_{\alpha \delta}^{\sigma \rho'}s_{\beta \delta}^{\sigma' \rho \dagger}s_{\beta \gamma}^{\sigma' \rho'}).
\end{split}
\end{equation}
{Similar to the chiral case, One can study the quantum noise cross-correlation via wavepacket approach, and by Buttiker's approach in a two-terminal QSH sample with a constriction, again will find inconsistency in Buttiker's result, which will require the introduction of two energy-dependent functions $f_a$ and $f_b$ in, Eq. (\ref{eq:C13}). }

{So, the final expression for the shot noise-like contribution for the cross-correlation is given as,}
\begin{equation} \label{eq:C14}
    \begin{split}
        {S_{\alpha \beta}^{\sigma \sigma', sh}} = -2\frac{e^2}{h}\int dE \sum_{\gamma, \delta}(f_{\gamma}-f_a)(f_{\delta}-f_b)\\Tr(s_{\alpha \gamma}^{\sigma \rho \dagger}s_{\alpha \delta}^{\sigma \rho'}s_{\beta \delta}^{\sigma' \rho \dagger}s_{\beta \gamma}^{\sigma' \rho'}).
    \end{split}
\end{equation}\\

\nocite{*}
\bibliography{apssamp.bib}

\providecommand{\noopsort}[1]{}\providecommand{\singleletter}[1]{#1}%
\begin{thebibliography}{23}%
\makeatletter
\providecommand \@ifxundefined [1]{%
 \@ifx{#1\undefined}
}%
\providecommand \@ifnum [1]{%
 \ifnum #1\expandafter \@firstoftwo
 \else \expandafter \@secondoftwo
 \fi
}%
\providecommand \@ifx [1]{%
 \ifx #1\expandafter \@firstoftwo
 \else \expandafter \@secondoftwo
 \fi
}%
\providecommand \natexlab [1]{#1}%
\providecommand \enquote  [1]{``#1''}%
\providecommand \bibnamefont  [1]{#1}%
\providecommand \bibfnamefont [1]{#1}%
\providecommand \citenamefont [1]{#1}%
\providecommand \href@noop [0]{\@secondoftwo}%
\providecommand \href [0]{\begingroup \@sanitize@url \@href}%
\providecommand \@href[1]{\@@startlink{#1}\@@href}%
\providecommand \@@href[1]{\endgroup#1\@@endlink}%
\providecommand \@sanitize@url [0]{\catcode `\\12\catcode `\$12\catcode
  `\&12\catcode `\#12\catcode `\^12\catcode `\_12\catcode `\%12\relax}%
\providecommand \@@startlink[1]{}%
\providecommand \@@endlink[0]{}%
\providecommand \url  [0]{\begingroup\@sanitize@url \@url }%
\providecommand \@url [1]{\endgroup\@href {#1}{\urlprefix }}%
\providecommand \urlprefix  [0]{URL }%
\providecommand \Eprint [0]{\href }%
\providecommand \doibase [0]{https://doi.org/}%
\providecommand \selectlanguage [0]{\@gobble}%
\providecommand \bibinfo  [0]{\@secondoftwo}%
\providecommand \bibfield  [0]{\@secondoftwo}%
\providecommand \translation [1]{[#1]}%
\providecommand \BibitemOpen [0]{}%
\providecommand \bibitemStop [0]{}%
\providecommand \bibitemNoStop [0]{.\EOS\space}%
\providecommand \EOS [0]{\spacefactor3000\relax}%
\providecommand \BibitemShut  [1]{\csname bibitem#1\endcsname}%
\let\auto@bib@innerbib\@empty
\bibitem [{\citenamefont {Klitzing}\ \emph {et~al.}(1980)\citenamefont
  {Klitzing}, \citenamefont {Dorda},\ and\ \citenamefont
  {Pepper}}]{PhysRevLett.45.494}%
  \BibitemOpen
  \bibfield  {author} {\bibinfo {author} {\bibfnamefont {K.~v.}\ \bibnamefont
  {Klitzing}}, \bibinfo {author} {\bibfnamefont {G.}~\bibnamefont {Dorda}},\
  and\ \bibinfo {author} {\bibfnamefont {M.}~\bibnamefont {Pepper}},\ }\href
  {https://doi.org/10.1103/PhysRevLett.45.494} {\bibfield  {journal} {\bibinfo
  {journal} {Phys. Rev. Lett.}\ }\textbf {\bibinfo {volume} {45}},\ \bibinfo
  {pages} {494} (\bibinfo {year} {1980})}\BibitemShut {NoStop}%
\bibitem [{\citenamefont {B\"uttiker}(1988)}]{PhysRevB.38.9375}%
  \BibitemOpen
  \bibfield  {author} {\bibinfo {author} {\bibfnamefont {M.}~\bibnamefont
  {B\"uttiker}},\ }\href {https://doi.org/10.1103/PhysRevB.38.9375} {\bibfield
  {journal} {\bibinfo  {journal} {Phys. Rev. B}\ }\textbf {\bibinfo {volume}
  {38}},\ \bibinfo {pages} {9375} (\bibinfo {year} {1988})}\BibitemShut
  {NoStop}%
\bibitem [{\citenamefont {Datta}(1995)}]{datta_1995}%
  \BibitemOpen
  \bibfield  {author} {\bibinfo {author} {\bibfnamefont {S.}~\bibnamefont
  {Datta}},\ }\href {https://doi.org/10.1017/CBO9780511805776} {\emph {\bibinfo
  {title} {Electronic Transport in Mesoscopic Systems}}},\ Cambridge Studies in
  Semiconductor Physics and Microelectronic Engineering\ (\bibinfo  {publisher}
  {Cambridge University Press},\ \bibinfo {year} {1995})\BibitemShut {NoStop}%
\bibitem [{\citenamefont {Bernevig}\ \emph {et~al.}(2006)\citenamefont
  {Bernevig}, \citenamefont {Hughes},\ and\ \citenamefont {Zhang}}]{zhang}%
  \BibitemOpen
  \bibfield  {author} {\bibinfo {author} {\bibfnamefont {B.~A.}\ \bibnamefont
  {Bernevig}}, \bibinfo {author} {\bibfnamefont {T.~L.}\ \bibnamefont
  {Hughes}},\ and\ \bibinfo {author} {\bibfnamefont {S.-C.}\ \bibnamefont
  {Zhang}},\ }\href {https://doi.org/10.1126/science.1133734} {\bibfield
  {journal} {\bibinfo  {journal} {Science}\ }\textbf {\bibinfo {volume}
  {314}},\ \bibinfo {pages} {1757} (\bibinfo {year} {2006})},\ \Eprint
  {https://arxiv.org/abs/https://www.science.org/doi/pdf/10.1126/science.1133734}
  {https://www.science.org/doi/pdf/10.1126/science.1133734} \BibitemShut
  {NoStop}%
\bibitem [{\citenamefont {König}\ \emph {et~al.}(2007)\citenamefont {König},
  \citenamefont {Wiedmann}, \citenamefont {Brüne}, \citenamefont {Roth},
  \citenamefont {Buhmann}, \citenamefont {Molenkamp}, \citenamefont {Qi},\ and\
  \citenamefont {Zhang}}]{Konig}%
  \BibitemOpen
  \bibfield  {author} {\bibinfo {author} {\bibfnamefont {M.}~\bibnamefont
  {König}}, \bibinfo {author} {\bibfnamefont {S.}~\bibnamefont {Wiedmann}},
  \bibinfo {author} {\bibfnamefont {C.}~\bibnamefont {Brüne}}, \bibinfo
  {author} {\bibfnamefont {A.}~\bibnamefont {Roth}}, \bibinfo {author}
  {\bibfnamefont {H.}~\bibnamefont {Buhmann}}, \bibinfo {author} {\bibfnamefont
  {L.}~\bibnamefont {Molenkamp}}, \bibinfo {author} {\bibfnamefont
  {X.}~\bibnamefont {Qi}},\ and\ \bibinfo {author} {\bibfnamefont
  {S.}~\bibnamefont {Zhang}},\ }\href {https://doi.org/10.1126/science.1148047}
  {\bibfield  {journal} {\bibinfo  {journal} {Science}\ }\textbf {\bibinfo
  {volume} {318}},\ \bibinfo {pages} {776770} (\bibinfo {year}
  {2007})}\BibitemShut {NoStop}%
\bibitem [{\citenamefont {Roth}\ \emph {et~al.}(2009)\citenamefont {Roth},
  \citenamefont {Brüne}, \citenamefont {Buhmann}, \citenamefont {Molenkamp},
  \citenamefont {Maciejko}, \citenamefont {Qi},\ and\ \citenamefont
  {Zhang}}]{Roth_2009}%
  \BibitemOpen
  \bibfield  {author} {\bibinfo {author} {\bibfnamefont {A.}~\bibnamefont
  {Roth}}, \bibinfo {author} {\bibfnamefont {C.}~\bibnamefont {Brüne}},
  \bibinfo {author} {\bibfnamefont {H.}~\bibnamefont {Buhmann}}, \bibinfo
  {author} {\bibfnamefont {L.~W.}\ \bibnamefont {Molenkamp}}, \bibinfo {author}
  {\bibfnamefont {J.}~\bibnamefont {Maciejko}}, \bibinfo {author}
  {\bibfnamefont {X.-L.}\ \bibnamefont {Qi}},\ and\ \bibinfo {author}
  {\bibfnamefont {S.-C.}\ \bibnamefont {Zhang}},\ }\href
  {https://doi.org/10.1126/science.1174736} {\bibfield  {journal} {\bibinfo
  {journal} {Science}\ }\textbf {\bibinfo {volume} {325}},\ \bibinfo {pages}
  {294} (\bibinfo {year} {2009})}\BibitemShut {NoStop}%
\bibitem [{\citenamefont {Shen}(2017)}]{Shen_2017}%
  \BibitemOpen
  \bibfield  {author} {\bibinfo {author} {\bibfnamefont {S.-Q.}\ \bibnamefont
  {Shen}},\ }\href {https://doi.org/10.1007/978-981-10-4606-3} {\emph {\bibinfo
  {title} {Topological Insulators}}}\ (\bibinfo  {publisher} {Springer
  Singapore},\ \bibinfo {year} {2017})\BibitemShut {NoStop}%
\bibitem [{\citenamefont {Nichele}\ \emph {et~al.}(2016)\citenamefont
  {Nichele}, \citenamefont {Suominen}, \citenamefont {Kjaergaard},
  \citenamefont {Marcus}, \citenamefont {Sajadi}, \citenamefont {Folk},
  \citenamefont {Qu}, \citenamefont {Beukman}, \citenamefont {de~Vries},
  \citenamefont {van Veen}, \citenamefont {Nadj-Perge}, \citenamefont
  {Kouwenhoven}, \citenamefont {Nguyen}, \citenamefont {Kiselev}, \citenamefont
  {Yi}, \citenamefont {Sokolich}, \citenamefont {Manfra}, \citenamefont
  {Spanton},\ and\ \citenamefont {Moler}}]{Nichele_2016}%
  \BibitemOpen
  \bibfield  {author} {\bibinfo {author} {\bibfnamefont {F.}~\bibnamefont
  {Nichele}}, \bibinfo {author} {\bibfnamefont {H.~J.}\ \bibnamefont
  {Suominen}}, \bibinfo {author} {\bibfnamefont {M.}~\bibnamefont
  {Kjaergaard}}, \bibinfo {author} {\bibfnamefont {C.~M.}\ \bibnamefont
  {Marcus}}, \bibinfo {author} {\bibfnamefont {E.}~\bibnamefont {Sajadi}},
  \bibinfo {author} {\bibfnamefont {J.~A.}\ \bibnamefont {Folk}}, \bibinfo
  {author} {\bibfnamefont {F.}~\bibnamefont {Qu}}, \bibinfo {author}
  {\bibfnamefont {A.~J.~A.}\ \bibnamefont {Beukman}}, \bibinfo {author}
  {\bibfnamefont {F.~K.}\ \bibnamefont {de~Vries}}, \bibinfo {author}
  {\bibfnamefont {J.}~\bibnamefont {van Veen}}, \bibinfo {author}
  {\bibfnamefont {S.}~\bibnamefont {Nadj-Perge}}, \bibinfo {author}
  {\bibfnamefont {L.~P.}\ \bibnamefont {Kouwenhoven}}, \bibinfo {author}
  {\bibfnamefont {B.-M.}\ \bibnamefont {Nguyen}}, \bibinfo {author}
  {\bibfnamefont {A.~A.}\ \bibnamefont {Kiselev}}, \bibinfo {author}
  {\bibfnamefont {W.}~\bibnamefont {Yi}}, \bibinfo {author} {\bibfnamefont
  {M.}~\bibnamefont {Sokolich}}, \bibinfo {author} {\bibfnamefont {M.~J.}\
  \bibnamefont {Manfra}}, \bibinfo {author} {\bibfnamefont {E.~M.}\
  \bibnamefont {Spanton}},\ and\ \bibinfo {author} {\bibfnamefont {K.~A.}\
  \bibnamefont {Moler}},\ }\href
  {https://doi.org/10.1088/1367-2630/18/8/083005} {\bibfield  {journal}
  {\bibinfo  {journal} {New Journal of Physics}\ }\textbf {\bibinfo {volume}
  {18}},\ \bibinfo {pages} {083005} (\bibinfo {year} {2016})}\BibitemShut
  {NoStop}%
\bibitem [{\citenamefont {Mani}\ and\ \citenamefont
  {Benjamin}(2017)}]{Mani_2017}%
  \BibitemOpen
  \bibfield  {author} {\bibinfo {author} {\bibfnamefont {A.}~\bibnamefont
  {Mani}}\ and\ \bibinfo {author} {\bibfnamefont {C.}~\bibnamefont
  {Benjamin}},\ }\bibfield  {journal} {\bibinfo  {journal} {Scientific
  Reports}\ }\textbf {\bibinfo {volume} {7}},\ \href
  {https://doi.org/10.1038/s41598-017-06820-w} {10.1038/s41598-017-06820-w}
  (\bibinfo {year} {2017})\BibitemShut {NoStop}%
\bibitem [{\citenamefont {Blanter}\ and\ \citenamefont
  {B\"uttiker}(2000)}]{BLANTER20001}%
  \BibitemOpen
  \bibfield  {author} {\bibinfo {author} {\bibfnamefont {Y.~M.}\ \bibnamefont
  {Blanter}}\ and\ \bibinfo {author} {\bibfnamefont {M.}~\bibnamefont
  {B\"uttiker}},\ }\href
  {https://doi.org/https://doi.org/10.1016/S0370-1573(99)00123-4} {\bibfield
  {journal} {\bibinfo  {journal} {Physics Reports}\ }\textbf {\bibinfo {volume}
  {336}},\ \bibinfo {pages} {1} (\bibinfo {year} {2000})}\BibitemShut {NoStop}%
\bibitem [{\citenamefont {Stevens}\ \emph {et~al.}(2019)\citenamefont
  {Stevens}, \citenamefont {Li}, \citenamefont {Du},\ and\ \citenamefont
  {Natelson}}]{10.1063/1.5111626}%
  \BibitemOpen
  \bibfield  {author} {\bibinfo {author} {\bibfnamefont {L.~A.}\ \bibnamefont
  {Stevens}}, \bibinfo {author} {\bibfnamefont {T.}~\bibnamefont {Li}},
  \bibinfo {author} {\bibfnamefont {R.-R.}\ \bibnamefont {Du}},\ and\ \bibinfo
  {author} {\bibfnamefont {D.}~\bibnamefont {Natelson}},\ }\bibfield  {journal}
  {\bibinfo  {journal} {Applied Physics Letters}\ }\textbf {\bibinfo {volume}
  {115}},\ \href {https://doi.org/10.1063/1.5111626} {10.1063/1.5111626}
  (\bibinfo {year} {2019}),\ \bibinfo {note} {052107},\ \Eprint
  {https://arxiv.org/abs/https://pubs.aip.org/aip/apl/article-pdf/doi/10.1063/1.5111626/14526545/052107\_1\_online.pdf}
  {https://pubs.aip.org/aip/apl/article-pdf/doi/10.1063/1.5111626/14526545/052107\_1\_online.pdf}
  \BibitemShut {NoStop}%
\bibitem [{\citenamefont {Qu}\ \emph {et~al.}(2015)\citenamefont {Qu},
  \citenamefont {Beukman}, \citenamefont {Nadj-Perge}, \citenamefont {Wimmer},
  \citenamefont {Nguyen}, \citenamefont {Yi}, \citenamefont {Thorp},
  \citenamefont {Sokolich}, \citenamefont {Kiselev}, \citenamefont {Manfra},
  \citenamefont {Marcus},\ and\ \citenamefont
  {Kouwenhoven}}]{PhysRevLett.115.036803}%
  \BibitemOpen
  \bibfield  {author} {\bibinfo {author} {\bibfnamefont {F.}~\bibnamefont
  {Qu}}, \bibinfo {author} {\bibfnamefont {A.~J.~A.}\ \bibnamefont {Beukman}},
  \bibinfo {author} {\bibfnamefont {S.}~\bibnamefont {Nadj-Perge}}, \bibinfo
  {author} {\bibfnamefont {M.}~\bibnamefont {Wimmer}}, \bibinfo {author}
  {\bibfnamefont {B.-M.}\ \bibnamefont {Nguyen}}, \bibinfo {author}
  {\bibfnamefont {W.}~\bibnamefont {Yi}}, \bibinfo {author} {\bibfnamefont
  {J.}~\bibnamefont {Thorp}}, \bibinfo {author} {\bibfnamefont
  {M.}~\bibnamefont {Sokolich}}, \bibinfo {author} {\bibfnamefont {A.~A.}\
  \bibnamefont {Kiselev}}, \bibinfo {author} {\bibfnamefont {M.~J.}\
  \bibnamefont {Manfra}}, \bibinfo {author} {\bibfnamefont {C.~M.}\
  \bibnamefont {Marcus}},\ and\ \bibinfo {author} {\bibfnamefont {L.~P.}\
  \bibnamefont {Kouwenhoven}},\ }\href
  {https://doi.org/10.1103/PhysRevLett.115.036803} {\bibfield  {journal}
  {\bibinfo  {journal} {Phys. Rev. Lett.}\ }\textbf {\bibinfo {volume} {115}},\
  \bibinfo {pages} {036803} (\bibinfo {year} {2015})}\BibitemShut {NoStop}%
\bibitem [{\citenamefont {Zhuang}\ \emph {et~al.}(2021)\citenamefont {Zhuang},
  \citenamefont {Mitrovi\ifmmode~\acute{c}\else \'{c}\fi{}},\ and\
  \citenamefont {Marston}}]{PhysRevB.104.045144}%
  \BibitemOpen
  \bibfield  {author} {\bibinfo {author} {\bibfnamefont {Z.}~\bibnamefont
  {Zhuang}}, \bibinfo {author} {\bibfnamefont {V.~F.}\ \bibnamefont
  {Mitrovi\ifmmode~\acute{c}\else \'{c}\fi{}}},\ and\ \bibinfo {author}
  {\bibfnamefont {J.~B.}\ \bibnamefont {Marston}},\ }\href
  {https://doi.org/10.1103/PhysRevB.104.045144} {\bibfield  {journal} {\bibinfo
   {journal} {Phys. Rev. B}\ }\textbf {\bibinfo {volume} {104}},\ \bibinfo
  {pages} {045144} (\bibinfo {year} {2021})}\BibitemShut {NoStop}%
\bibitem [{\citenamefont {de~Medeiros}\ \emph {et~al.}(2021)\citenamefont
  {de~Medeiros}, \citenamefont {Teixeira}, \citenamefont {Sipahi},\ and\
  \citenamefont {Dias~da Silva}}]{PhysRevB.104.195307}%
  \BibitemOpen
  \bibfield  {author} {\bibinfo {author} {\bibfnamefont {M.~H.~L.}\
  \bibnamefont {de~Medeiros}}, \bibinfo {author} {\bibfnamefont {R.~L. R.~C.}\
  \bibnamefont {Teixeira}}, \bibinfo {author} {\bibfnamefont {G.~M.}\
  \bibnamefont {Sipahi}},\ and\ \bibinfo {author} {\bibfnamefont {L.~G. G.~V.}\
  \bibnamefont {Dias~da Silva}},\ }\href
  {https://doi.org/10.1103/PhysRevB.104.195307} {\bibfield  {journal} {\bibinfo
   {journal} {Phys. Rev. B}\ }\textbf {\bibinfo {volume} {104}},\ \bibinfo
  {pages} {195307} (\bibinfo {year} {2021})}\BibitemShut {NoStop}%
\bibitem [{\citenamefont {B\"uttiker}(1992)}]{PhysRevB.46.12485}%
  \BibitemOpen
  \bibfield  {author} {\bibinfo {author} {\bibfnamefont {M.}~\bibnamefont
  {B\"uttiker}},\ }\href {https://doi.org/10.1103/PhysRevB.46.12485} {\bibfield
   {journal} {\bibinfo  {journal} {Phys. Rev. B}\ }\textbf {\bibinfo {volume}
  {46}},\ \bibinfo {pages} {12485} (\bibinfo {year} {1992})}\BibitemShut
  {NoStop}%
\bibitem [{\citenamefont {Landauer}\ and\ \citenamefont
  {Martin}(1991)}]{LANDAUER1991167}%
  \BibitemOpen
  \bibfield  {author} {\bibinfo {author} {\bibfnamefont {R.}~\bibnamefont
  {Landauer}}\ and\ \bibinfo {author} {\bibfnamefont {T.}~\bibnamefont
  {Martin}},\ }\href
  {https://doi.org/https://doi.org/10.1016/0921-4526(91)90710-V} {\bibfield
  {journal} {\bibinfo  {journal} {Physica B: Condensed Matter}\ }\textbf
  {\bibinfo {volume} {175}},\ \bibinfo {pages} {167} (\bibinfo {year}
  {1991})},\ \bibinfo {note} {analogies in Optics and
  Micro-Electronics}\BibitemShut {NoStop}%
\bibitem [{\citenamefont {Martin}\ and\ \citenamefont
  {Landauer}(1992)}]{PhysRevB.45.1742}%
  \BibitemOpen
  \bibfield  {author} {\bibinfo {author} {\bibfnamefont {T.}~\bibnamefont
  {Martin}}\ and\ \bibinfo {author} {\bibfnamefont {R.}~\bibnamefont
  {Landauer}},\ }\href {https://doi.org/10.1103/PhysRevB.45.1742} {\bibfield
  {journal} {\bibinfo  {journal} {Phys. Rev. B}\ }\textbf {\bibinfo {volume}
  {45}},\ \bibinfo {pages} {1742} (\bibinfo {year} {1992})}\BibitemShut
  {NoStop}%
\bibitem [{\citenamefont {Imry}(1997)}]{imry}%
  \BibitemOpen
  \bibfield  {author} {\bibinfo {author} {\bibfnamefont {Y.}~\bibnamefont
  {Imry}},\ }\href {https://doi.org/10.1017/CBO9780511805776} {\emph {\bibinfo
  {title} {Introduction to Mesoscopic Physics}}},\ Oxford University Press\
  (\bibinfo  {publisher} {Cambridge University Press},\ \bibinfo {year}
  {1997})\BibitemShut {NoStop}%
\bibitem [{\citenamefont {Büttiker}(1991)}]{BUTTIKER1991199}%
  \BibitemOpen
  \bibfield  {author} {\bibinfo {author} {\bibfnamefont {M.}~\bibnamefont
  {Büttiker}},\ }\href
  {https://doi.org/https://doi.org/10.1016/0921-4526(91)90713-O} {\bibfield
  {journal} {\bibinfo  {journal} {Physica B: Condensed Matter}\ }\textbf
  {\bibinfo {volume} {175}},\ \bibinfo {pages} {199} (\bibinfo {year}
  {1991})},\ \bibinfo {note} {analogies in Optics and
  Micro-Electronics}\BibitemShut {NoStop}%
\bibitem [{\citenamefont {Dragomirova}\ and\ \citenamefont
  {Nikoli\ifmmode~\acute{c}\else \'{c}\fi{}}(2007)}]{PhysRevB.75.085328}%
  \BibitemOpen
  \bibfield  {author} {\bibinfo {author} {\bibfnamefont {R.~L.}\ \bibnamefont
  {Dragomirova}}\ and\ \bibinfo {author} {\bibfnamefont {B.~K.}\ \bibnamefont
  {Nikoli\ifmmode~\acute{c}\else \'{c}\fi{}}},\ }\href
  {https://doi.org/10.1103/PhysRevB.75.085328} {\bibfield  {journal} {\bibinfo
  {journal} {Phys. Rev. B}\ }\textbf {\bibinfo {volume} {75}},\ \bibinfo
  {pages} {085328} (\bibinfo {year} {2007})}\BibitemShut {NoStop}%
\bibitem [{\citenamefont {Lumbroso}\ \emph {et~al.}(2018)\citenamefont
  {Lumbroso}, \citenamefont {Simine}, \citenamefont {Nitzan}, \citenamefont
  {Segal},\ and\ \citenamefont {Tal}}]{Lumbroso}%
  \BibitemOpen
  \bibfield  {author} {\bibinfo {author} {\bibfnamefont {O.}~\bibnamefont
  {Lumbroso}}, \bibinfo {author} {\bibfnamefont {L.}~\bibnamefont {Simine}},
  \bibinfo {author} {\bibfnamefont {A.}~\bibnamefont {Nitzan}}, \bibinfo
  {author} {\bibfnamefont {D.}~\bibnamefont {Segal}},\ and\ \bibinfo {author}
  {\bibfnamefont {O.}~\bibnamefont {Tal}},\ }\href
  {https://doi.org/https://doi.org/10.1038/s41586-018-0592-2} {\bibfield
  {journal} {\bibinfo  {journal} {Nature}\ }\textbf {\bibinfo {volume} {562}},\
  \bibinfo {pages} {240} (\bibinfo {year} {2018})}\BibitemShut {NoStop}%
\bibitem [{\citenamefont {Rech}\ \emph {et~al.}(2020)\citenamefont {Rech},
  \citenamefont {Jonckheere}, \citenamefont {Gr\'emaud},\ and\ \citenamefont
  {Martin}}]{PhysRevLett.125.086801}%
  \BibitemOpen
  \bibfield  {author} {\bibinfo {author} {\bibfnamefont {J.}~\bibnamefont
  {Rech}}, \bibinfo {author} {\bibfnamefont {T.}~\bibnamefont {Jonckheere}},
  \bibinfo {author} {\bibfnamefont {B.}~\bibnamefont {Gr\'emaud}},\ and\
  \bibinfo {author} {\bibfnamefont {T.}~\bibnamefont {Martin}},\ }\href
  {https://doi.org/10.1103/PhysRevLett.125.086801} {\bibfield  {journal}
  {\bibinfo  {journal} {Phys. Rev. Lett.}\ }\textbf {\bibinfo {volume} {125}},\
  \bibinfo {pages} {086801} (\bibinfo {year} {2020})}\BibitemShut {NoStop}%
\bibitem [{\citenamefont {Rebora}\ \emph {et~al.}(2022)\citenamefont {Rebora},
  \citenamefont {Rech}, \citenamefont {Ferraro}, \citenamefont {Jonckheere},
  \citenamefont {Martin},\ and\ \citenamefont
  {Sassetti}}]{PhysRevResearch.4.043191}%
  \BibitemOpen
  \bibfield  {author} {\bibinfo {author} {\bibfnamefont {G.}~\bibnamefont
  {Rebora}}, \bibinfo {author} {\bibfnamefont {J.}~\bibnamefont {Rech}},
  \bibinfo {author} {\bibfnamefont {D.}~\bibnamefont {Ferraro}}, \bibinfo
  {author} {\bibfnamefont {T.}~\bibnamefont {Jonckheere}}, \bibinfo {author}
  {\bibfnamefont {T.}~\bibnamefont {Martin}},\ and\ \bibinfo {author}
  {\bibfnamefont {M.}~\bibnamefont {Sassetti}},\ }\href
  {https://doi.org/10.1103/PhysRevResearch.4.043191} {\bibfield  {journal}
  {\bibinfo  {journal} {Phys. Rev. Res.}\ }\textbf {\bibinfo {volume} {4}},\
  \bibinfo {pages} {043191} (\bibinfo {year} {2022})}\BibitemShut {NoStop}%
\end{thebibliography}%
\end{document}